\documentclass[a4paper,11pt]{article}

\usepackage{amsmath}
\usepackage{amssymb}
\usepackage{cite}
\usepackage{epsfig}
\usepackage{cancel}
\usepackage{feynmp}
\usepackage{xspace}
\usepackage{flafter}

\def\spa#1.#2{\left\langle#1\,#2\right\rangle}
\def\spb#1.#2{\left[#1\,#2\right]}
\def\spaa#1.#2.#3{\langle\mskip-1mu{#1}
                  | #2 | {#3}\mskip-1mu\rangle}
\def\spbb#1.#2.#3{[\mskip-1mu{#1}
                  | #2 | {#3}\mskip-1mu]}
\def\spab#1.#2.#3{\langle\mskip-1mu{#1}
                  | #2 | {#3}\mskip-1mu\rangle}
\def\spba#1.#2.#3{\langle\mskip-1mu{#1}^+
                  | #2 | {#3}^+\mskip-1mu\rangle}
\def\spav#1.#2.#3{\|\mskip-1mu{#1}
                  | #2 | {#3}\mskip-1mu\|^2}
\newcommand{\ca}{\ensuremath{C_{\!A}}}
\newcommand{\cf}{\ensuremath{C_{\!F}}}
\newcommand{\Nc}{\ensuremath{N_{\!C}}}
\newcommand{\gs}{\ensuremath{g}}
\newcommand{\as}{\ensuremath{\alpha_s}\xspace}
\newcommand\ve{\varepsilon}

\title{\vspace*{-4.cm}
\begin{normalsize}
\begin{flushright}
CERN-PH-TH/2009-153\\
DCPT/09/122\\
IPPP/09/61
\end{flushright}
\end{normalsize}
\vspace*{2cm}
Constructing All-Order Corrections to Multi-Jet Rates}% using Currents}
\author{Jeppe~R.~Andersen$^{a}$,
  Jennifer~M.~Smillie$^{b}$\\\mbox{}\\$^a$Theory Division, Physics
  Department, CERN, CH-1211 Geneva 23, Switzerland\\$^b$IPPP, University of
  Durham, Durham, DH1 3LE, UK.}

\begin{document}
\maketitle

\begin{abstract}
  We discuss the universal behaviour of scattering cross sections in the
  limit of infinite rapidity separation between all produced particles, and
  illustrate the behaviour explicitly for the production of $n$ jets,
  $W+n$~jets, $Z+n$~jets for $n=2,3,4$, and for $H+2,3$~jets. We give a set of rules
  for constructing scattering cross sections, which are exact in the given
  limit, and order-by-order reproduce well the full fixed order results when applied to LHC
  phenomenology. The approximation includes both real and virtual
  corrections, and is sufficiently simple to allow for the regulated
  all-order perturbative sum to be explicitly constructed. We test the
  expected accuracy by comparing results order-by-order with full fixed-order
  perturbation theory for the processes mentioned above.
\end{abstract}

\tableofcontents
\clearpage
\section{Introduction}
\label{sec:introduction}
% \item All resummations construct a factorisation and/or investigate a kinematical limit of
%   QCD such that the form of the higher order corrections simplify (after one checks that
%   the limit is in fact a relevant limit).
%\textbf{!! More references ?!!}

Achieving the full potential of the LHC in furthering our understanding of
particle physics will challenge our understanding and description of events
with multiple jets. Such events arise both within the Standard Model (SM),
and in many models of physics beyond the SM. Apart from furthering our
understanding of the solutions to the SM equations of motion, a better
understanding and description of the multi-jet predictions arising from the
SM is necessary in order to fully disentangle this contribution from that
which might arise from outside the SM.

The true complication of an observed jet in terms of its constituent hadrons
can currently only be described within the context of a ``General Purpose
Monte Carlo'', implementing a parton shower and hadronisation model as
described in e.g.~\cite{Sjostrand:2007gs,Bahr:2008pv,Gleisberg:2008ta}. The
parton shower is based on a resummation of soft and collinear
radiation. However, while obtaining a good description of the
\emph{structure} of each jet, this description severely underestimates the
\emph{rate} and \emph{hardness} ($p_\perp$-spectrum) of multi-jet
samples. Better predictions of such quantities in exclusive (meaning fixed
number) jet samples have so far been obtained in fixed order calculations;
for processes allowing more than two light partons in the final state, such
predictions are currently limited to NLO accuracy. However, putting aside
issues of hadronisation, many other questions which are important for the LHC
programme still cannot be answered satisfactorily in the relatively simple
description of the final states obtained at such low orders in the
perturbative expansion. Such questions range from the effects of central jet
vetos in samples of Higgs boson plus jets, to the simple question of the
transverse momentum spectrum of weak gauge bosons (both at large and small
transverse momenta) or the relative weight of various jet multiplicities in
inclusive samples (``inclusive'' here meaning sum over any number of jets).

A combination of fixed order calculations and further parton shower resummation has been
achieved, which so far corrects the approximation obtained in the parton shower to
full tree-level accuracy (for a fixed jet multiplicity~\cite{Catani:2001cc,Lonnblad:1992tz}) or NLO
accuracy~\cite{Frixione:2002ik,Frixione:2007nu} (so far in processes which at LO have just
zero or one light parton in the final state). Such matching schemes ensure that a given
process is described correctly at least at (N)LO, while all higher order
corrections are estimated from the parton shower.

All resummation schemes are built upon a particular kinematic limit in which
the perturbative corrections simplify, allowing for all-order approximations
to the full perturbative series. For the parton shower, this is the soft and
the collinear limit, i.e.~emissions under small invariant mass, and the
building blocks are e.g.~parton splitting functions. In the current study, we
will be guided by the simplifications of the perturbative corrections in the
case of large invariant mass between all emissions. We will seek to catch the
part of the perturbative corrections which controls additional jet
production, but ignore the collinear behaviour which gives rise to the
jet-sub-structure. We will obtain an approximation to both virtual and
real-emission corrections, which allows us to build an approximation to the
regularised, all-orders matrix element for each exclusive (resolved) parton
multiplicity. The focus of this paper is to obtain the relevant universal
building blocks for multi-jet predictions. Furthermore, we will investigate
the accuracy of this approach by comparing the lowest order predictions
obtained in our simplified approach to the full tree-level QCD perturbative
series for $n-$jet-production, $(n=2,3,4)$, both ``pure'' and in association
with a $W,Z$ or Higgs-boson.

The approximate or ``$t$-channel factorised'' matrix element will be
sufficiently fast to evaluate numerically that the resummation can be
constructed by explicit summation over the exclusive final states, and thus
any event analysis can be performed by simply imposing jet algorithms etc.~on
the exclusive final states (``exclusive'' meaning fully differential in the
momenta of all produced particles). We will leave the phenomenological
implications of the resummation of each of the four processes discussed here
to future studies.

In Section~\ref{sec:high-energy-appr} we will introduce the
\emph{Multi-Regge-Kinematic-limit}, which lies at the heart of our approach,
and discuss and illustrate the universal behaviour of scattering amplitudes
in this limit. In Section~\ref{sec:current-method} we will construct all the
building blocks for the relevant amplitudes. In
Section~\ref{sec:applications} we will check process-by-process,
multiplicity-by-multiplicity the level of accuracy at which the approach
reproduces the fixed order perturbative expansion in terms of not just cross sections but
also differential distributions, where these can be obtained with standard
tools implementing the full tree-level processes. By doing so, we hope to
instill trust in the all-order approximation which can be built using the
elements described in this paper.

%%% Local Variables: 
%%% mode: latex
%%% TeX-master: "jetcurrents"
%%% End: 

\section{The High Energy Limit of Scattering Amplitudes}
\label{sec:high-energy-appr}

We will start by studying the scattering matrix elements of a gluonic $2\to
n, n\ge 2,$ process in the so-called \emph{Multi--Regge-Kinematic} (MRK)
limit, where the scattering momenta $p_A,p_B\to p_1,\ldots, p_n$ in
terms of transverse momenta and rapidity
$y=\ln\left(\frac{E+p_z}{E-p_z}\right)$ fulfil the following conditions
\begin{align}
  \label{eq:MRKlimit}
  \begin{split}
    \forall i\in \{2,\ldots, n-1\}: y_{i-1}\gg y_i \gg y_{i+1}\\
    \forall i,j: |p_{i\perp}|\approx |p_{j\perp}|,
  \end{split}
\end{align}
or alternatively
\begin{align}
  \label{eq:MRKlimit2}
    \forall i,j: |p_{i\perp}|\approx |p_{j\perp}|, s_{ij}\to \infty,
\end{align}
where $s_{ij}=2\ p_i.p_j$ and $s=2\ p_A.p_B$. The notation $y_i \gg y_j$
really means $y_i-y_j\to\infty$. The transverse directions are with respect
to the incoming partons $p_A, p_B$ (i.e.~transverse to the beam line), and
the limit requires the transverse components to be kept fixed (i.e.~not
growing with $s$) as $s_{ij},|y_{i}-y_{j}|\to \infty$. We can take $p_A$ to
be the direction of positive light-cone momentum.

\begin{figure}[tb]
  \centering
  \input{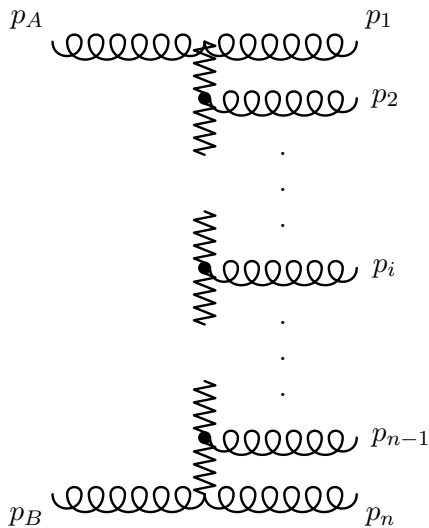}

  \vspace{0.3cm}
  \caption{Picture of effective vertices connected by $t$-channel exchanges}
  \label{fig:tchannelpicture}
\end{figure}

Explicit calculations of up to $2\to 4$ gluon scattering~\cite{Lipatov:1974qm}, and a
dispersive analysis of the $2\to n$ process~\cite{Fadin:1975cb,Kuraev:1977fs} showed that
in the MRK limit, the $2\to n$ scattering amplitude is dominated by the behaviour dictated
by the poles from $t$-channel gluon exchanges. $t$-channel here refers to the picture which arises of a
string of effective vertices for on-shell gluon production, which are connected by
off-shell ($t$-channel) gluon currents, see Fig.~\ref{fig:tchannelpicture}. The
$t$-channel momenta are here defined as
\begin{align}
  \label{eq:ti}
  t_i=p_A-\sum_{j=1}^{i} p_j.
\end{align}
In Section~\ref{sec:current-method} we will illustrate how the following
result arises, and re-derive the form of the building blocks of the
approximation. Here we will just quote the result for the LO colour and spin
summed and averaged scattering matrix element for $2\to n$ on-shell gluons in
the MRK limit:
\begin{align}
  \label{eq:Msqsumavg}
  \left|\overline{\mathcal{M}}_{gg\to g\cdots g}^{MRK}\right|^2 = \frac {4\ s^2}{\Nc^2-1}
  \ \frac{g^2\ \ca}{|p_{1\perp}|^2} \left( \prod_{i=2}^{n-1}\frac{4\ g^2
      \ca}{|p_{i\perp}|^2}\right) \frac{g^2\ \ca}{|p_{n\perp}|^2}.
\end{align}
% [discuss features]
One notices that obviously, after the MRK limit is taken, there is no
dependence left on the rapidities of the gluons (or rather, the dependence on
the rapidities is lost). The dependence on the centre-of-mass energy $s$ in
Eq.~\eqref{eq:Msqsumavg} is left in, although of course in the MRK limit,
$s\to\infty$. The partonic cross section is found by dividing the square of
the scattering matrix element by the flux factor, which is proportional to
$s^2$, so the MRK limit of the partonic cross section is a very simple
function of only the transverse momenta of the emitted gluons. 

% [summarise results]
The results of Ref.~\cite{Fadin:1975cb,Kuraev:1976ge,Kuraev:1977fs} extend
beyond just pure gluon scattering; the picture which arises is one in which
the description of a given $2\to n$ scattering process in the MRK limit
factorises into a product of effective vertices for particle emission, with
the vertices connected by propagators according to the ordering in rapidity
of the emitted particles. One immediate result is that in the MRK limit, the
$2\to n$ scattering process is dominated by rapidity orderings which allow
pure gluon exchanges (or more precisely: exchanges of the particle of highest
spin) between the scattering vertices. 

Another immediate result of the factorisation (arising from the assumption of a hierarchy
in the components of plus- and minus-momenta) is that in the MRK limit, the scattering
matrix element for e.g.~$gg\to ggg$, $qg\to qgg$ and $qQ\to qgQ$ differ only by colour
factors - this is a generalisation of the results leading to the ``effective
PDF''-approach to $2\to2$ scattering\cite{Combridge:1984jn}. The results equivalent to
Eq.~\eqref{eq:Msqsumavg} for the $qQ$ and $qg$-initiated scattering processes in the MRK
limit are then
\begin{align}
  \label{eq:qgMsqsumavg}
  \left|\overline{\mathcal{M}}_{qg\to qg\cdots g}^{MRK}\right|^2 = \frac {4\ s^2}{\Nc^2-1}
  \ \frac{g^2\ \cf}{|p_{1\perp}|^2} \left( \prod_{i=2}^{n-1}\frac{4\ g^2
      \ca}{|p_{i\perp}|^2}\right) \frac{g^2\ \ca}{|p_{n\perp}|^2},\\
  \label{eq:qQMsqsumavg}
  \left|\overline{\mathcal{M}}_{qQ\to qg\cdots Q}^{MRK}\right|^2 = \frac {4\ s^2}{\Nc^2-1}
  \ \frac{g^2\ \cf}{|p_{1\perp}|^2} \left( \prod_{i=2}^{n-1}\frac{4\ g^2
      \ca}{|p_{i\perp}|^2}\right) \frac{g^2\ \cf}{|p_{n\perp}|^2},
\end{align}
with the additional information that all final states not ordered in rapidity according to
the indication in the subscripts are suppressed (by powers of $s_{ij}$).  The difference
in colour factor when gluon 1 or $n$ is replaced by a quark is $\cf/\ca$, since the difference
in the summed and averaged colour factor is the replacement of
$f^{abc}f^{abd}/(\Nc^2-1)=\ca\delta^{cd}/(\Nc^2-1)$ with
$t^c_{ij}t^d_{ji}/\Nc=\delta^{cd}/(2\Nc)$.
Eqs.~\eqref{eq:Msqsumavg}--\eqref{eq:qQMsqsumavg} arise also as the expansions to fixed
order of the solution to the BFKL equation to leading logarithmic accuracy.

\begin{figure}[tb]
  \centering
  \epsfig{width=0.49\textwidth,file=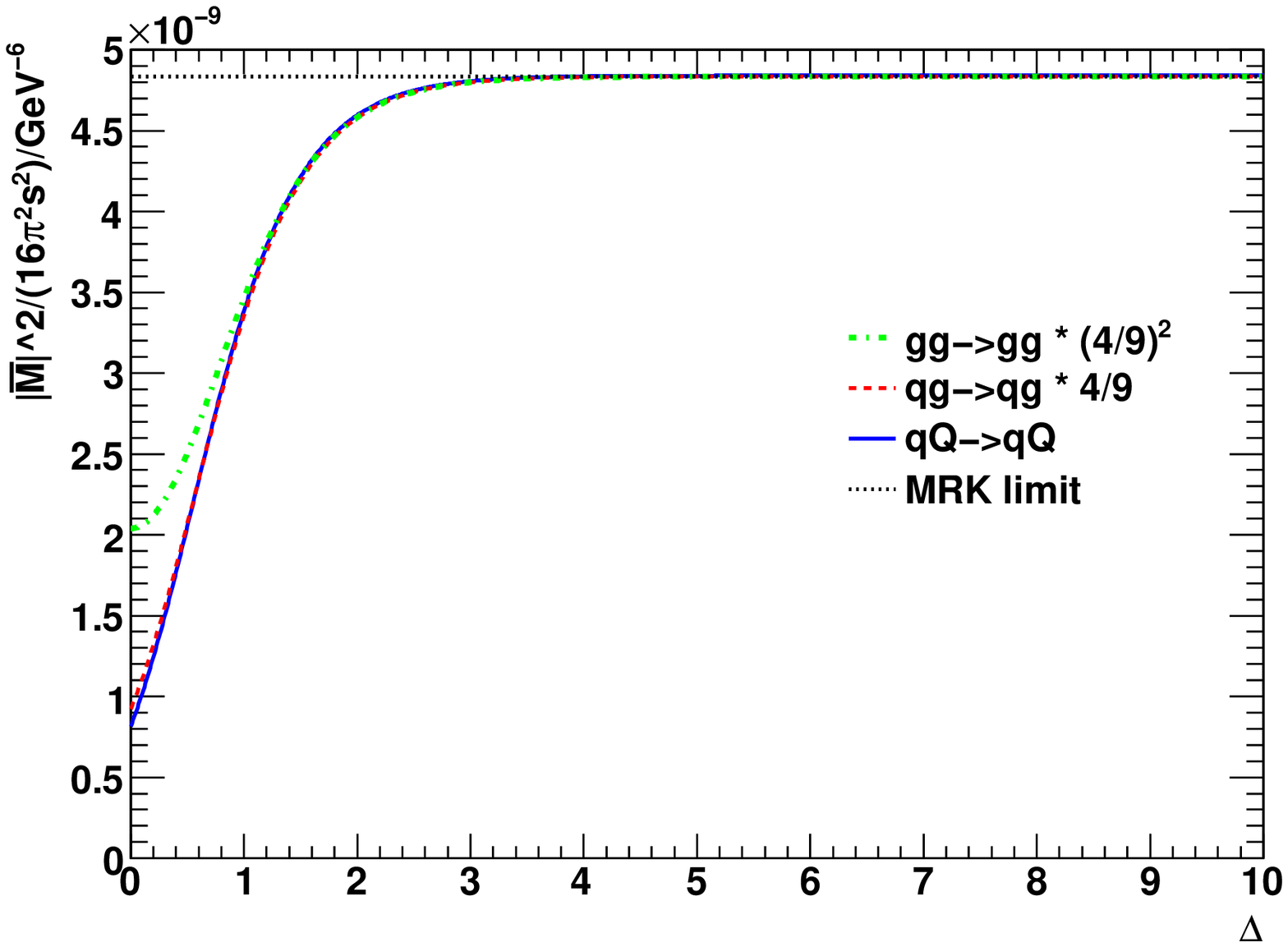}
  \epsfig{width=0.49\textwidth,file=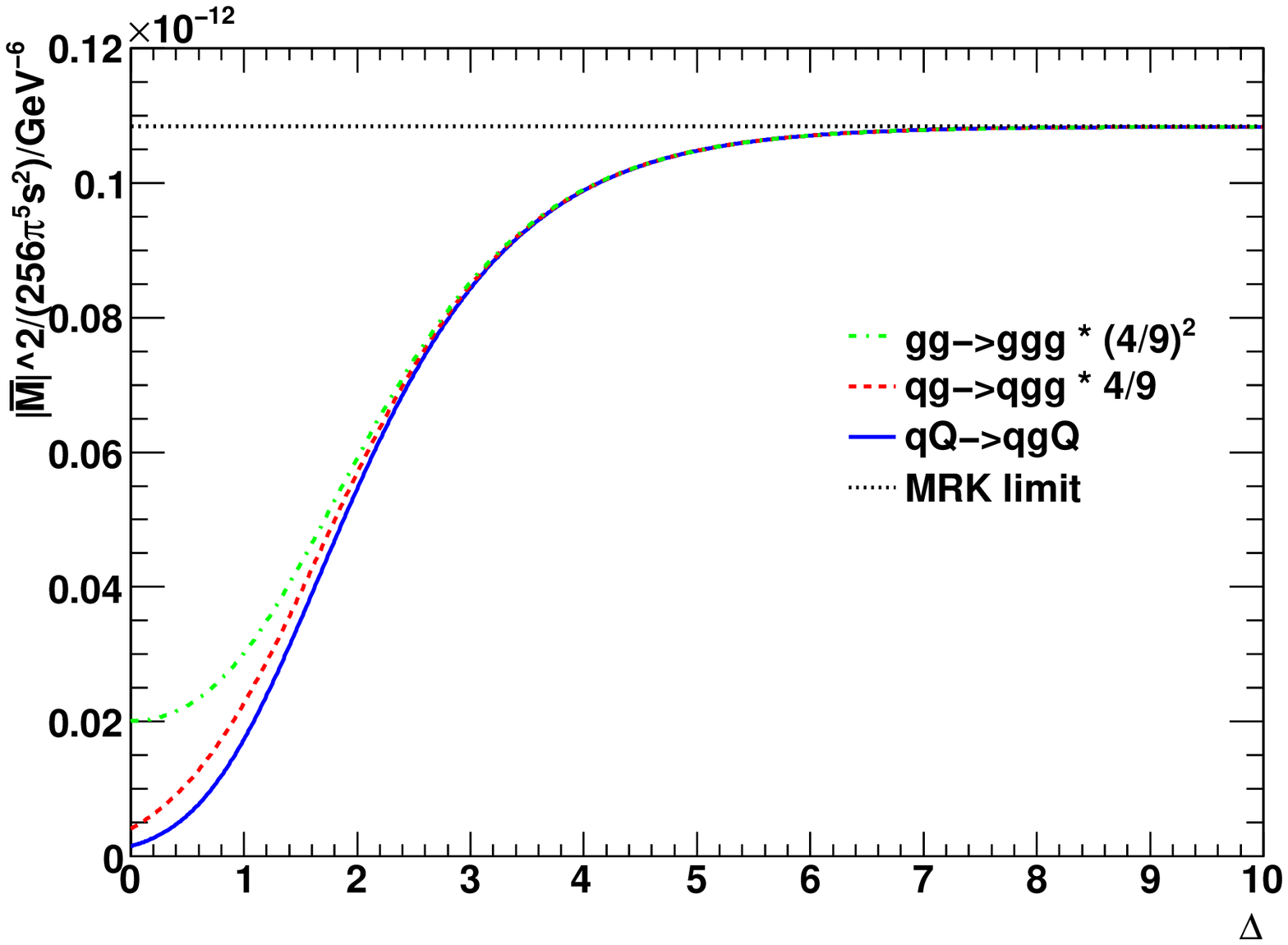}

  (a) \hspace{7.2cm}(b)\hspace{0.1cm}
  \epsfig{width=0.49\textwidth,file=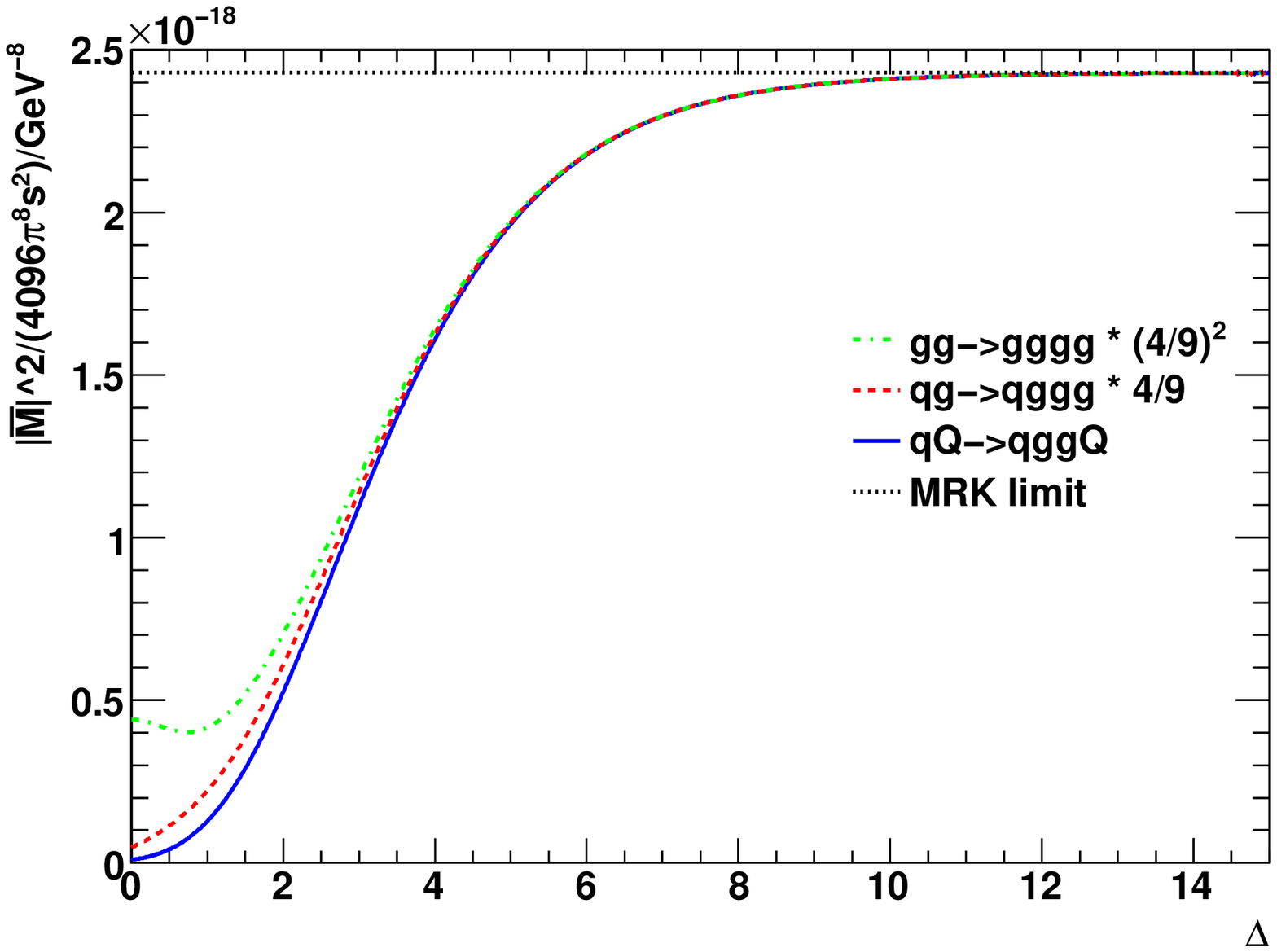}

  (c)
  \caption{The scattering matrix elements for (a) $qQ\to qQ$, $qg\to qg$ and $gg\to gg$,
    (b) $qQ\to qgQ$, $qg\to qgg$ and $gg\to ggg$
    and (c) $qQ\to qggQ$, $qQ\to qggQ$ and $gg\to gggg$, as a function of the rapidity
    difference $\Delta$ between the partons, defined in Eq.~\eqref{eq:3particlemomenta}.
    The MRK limit is the value obtained from Eq.~\eqref{eq:qQMsqsumavg}.}
  \label{fig:M3j}
\end{figure}
The results of Eqs.~\eqref{eq:Msqsumavg}--\eqref{eq:qQMsqsumavg} are
summarised in Fig.~\ref{fig:M3j}, which illustrates the behaviour of the
quantities
\begin{align}
  \label{eq:Mdivflux}
  \begin{split}
\mathrm{(a)}&\  \frac{\left| \mathcal{M}\right|^2}{16\ \pi^2\ s^2}\quad \mbox{for}\quad
  \left| \mathcal{M}\right|^2 \in \left\{ \left(\frac{\cf}{\ca}\right)^2
    \left|\overline{\mathcal{M}}_{gg\to gg}\right|^2,\ \left(\frac{\cf}{\ca}\right)
    \left|\overline{\mathcal{M}}_{qg\to qg}\right|^2,\ 
    \left|\overline{\mathcal{M}}_{qQ\to qQ}\right|^2 \right\}, \\
\mathrm{(b)}&\  \frac{\left| \mathcal{M}\right|^2}{256\ \pi^5\ s^2}\quad \mbox{for}\quad
  \left| \mathcal{M}\right|^2 \in \left\{ \left(\frac{\cf}{\ca}\right)^2
    \left|\overline{\mathcal{M}}_{gg\to ggg}\right|^2,\ \left(\frac{\cf}{\ca}\right)
    \left|\overline{\mathcal{M}}_{qg\to qgg}\right|^2,\ 
    \left|\overline{\mathcal{M}}_{qQ\to qgQ}\right|^2 \right\}, \\
  &\mathrm{and}\\
\mathrm{(c)}&\   \frac{\left| \mathcal{M}\right|^2}{4096\ \pi^8\ s^2}\quad \mbox{for}\quad
  \left| \mathcal{M}\right|^2 \in \left\{ \left(\frac{\cf}{\ca}\right)^2
    \left|\overline{\mathcal{M}}_{gg\to gggg}\right|^2,\ \left(\frac{\cf}{\ca}\right)
    \left|\overline{\mathcal{M}}_{qg\to qggg}\right|^2,\ 
    \left|\overline{\mathcal{M}}_{qQ\to qggQ}\right|^2 \right\}.
  \end{split}
\end{align}
The full tree-level matrix elements are extracted from
MadGraph\cite{Alwall:2007st}, and the momenta of the final state particles are chosen as
\begin{align}
  \begin{split}
    \label{eq:3particlemomenta}
    p_i&=(k_\perp \cosh(y_i), k_\perp \cos(\phi_i), k_\perp \sin(\phi_i) ,
    k_\perp \sinh(y_i) ),\ k_\perp=40\mathrm{GeV},\\
    \mathrm{(a):}&\ y_1=\Delta,\ y_2=-\Delta,\quad \phi_1=0,\ \phi_2=\pi,\\
    \mathrm{(b):}&\ y_1=\Delta,\ y_2=0,\ y_3=-\Delta, \quad \phi_1=0,\
    \phi_2=\frac{2\pi}3,\ \phi_3=-\frac{2\pi}3,\\
    \mathrm{(c):}&\ y_1=\Delta,\ y_2=\frac{\Delta}3,\
    y_3=-\frac{\Delta}3,\ y_4=-\Delta, \quad \phi_1=0,\ \phi_2=\frac{\pi}2,\
    \phi_3=-\frac{\pi}2,\ \phi_4=\pi .
  \end{split}
\end{align}

The same universal behaviour is seen in processes where a $W,Z$ or $H$ boson
is produced in association with jets, as shown in
Figs.~\ref{fig:WMRKlimit}--\ref{fig:HMRKlimit}.  The momentum configurations
for these plots are given in Appendix~\ref{sec:moment-conf}.  The rapidities
of the jets in each case are as in Eq.~\eqref{eq:3particlemomenta}, except
for the $H+3j$ process, where the jet rapidities are $\Delta$, $-\Delta/3$
and $-\Delta$ respectively.  The MRK limit for the $W$ processes is taken
from Ref.~\cite{Andersen:2001ja} while that for the Higgs processes is taken
from Ref.~\cite{DelDuca:2003ba}.

\begin{figure}[!p]
  \centering
  \epsfig{width=0.35\textwidth,file=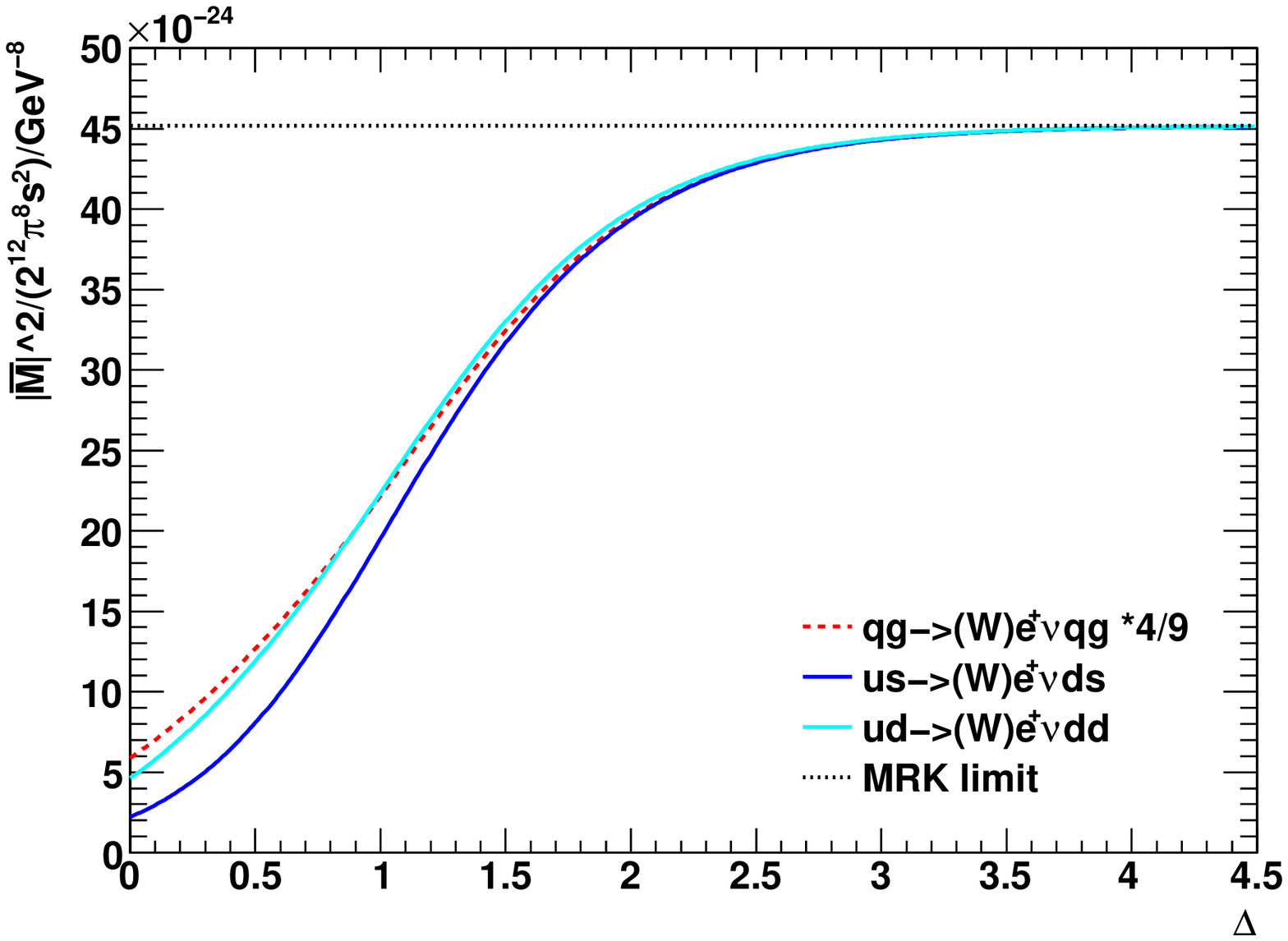} \hspace{-0.7cm}
  \epsfig{width=0.35\textwidth,file=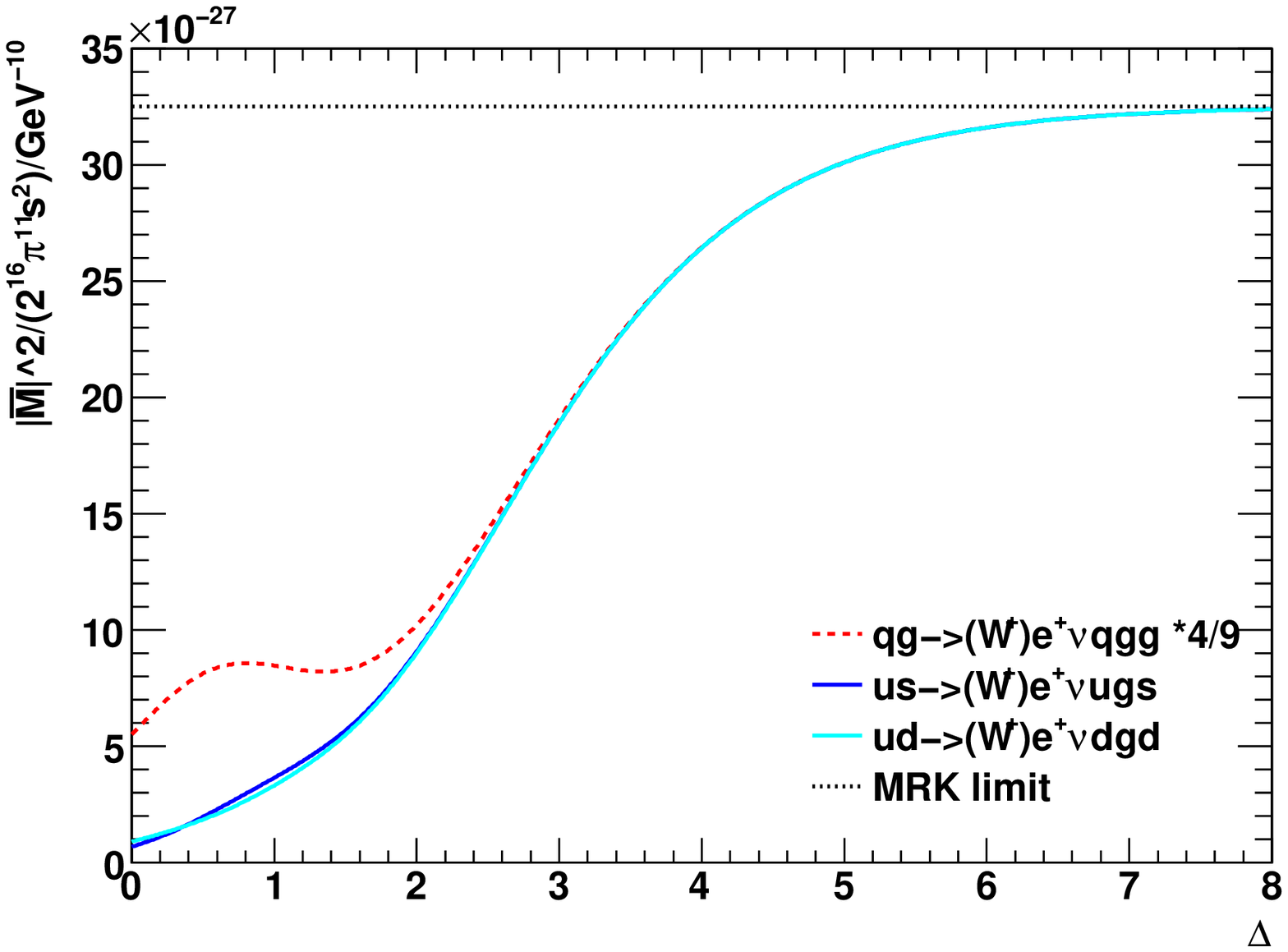} \hspace{-0.7cm}
  \epsfig{width=0.35\textwidth,file=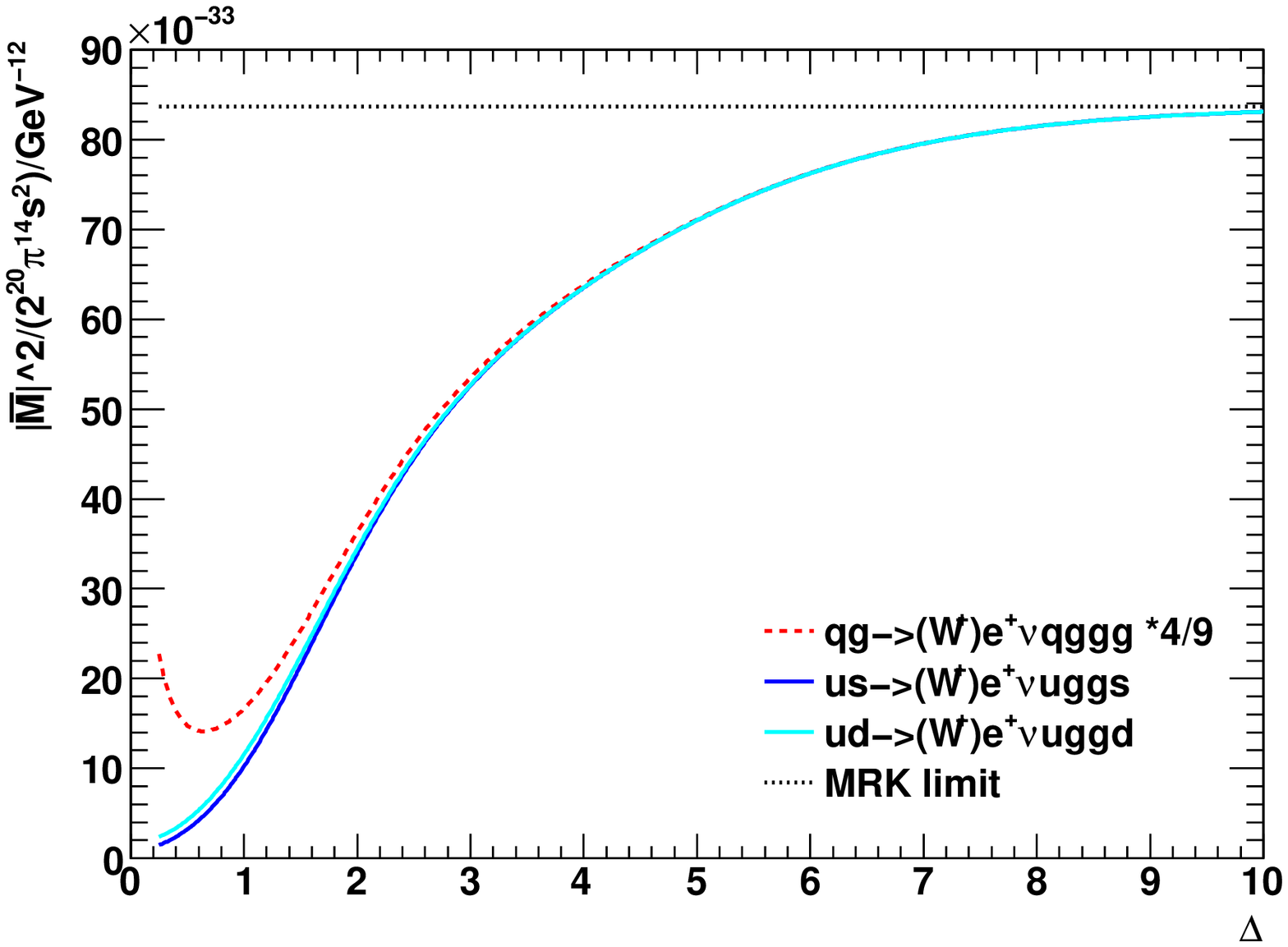}

  (a) \hspace{4.5cm} (b) \hspace{4.5cm} (c)
  \caption{The scattering matrix elements for jet production in association with a $W$
    boson, for (a) 2, (b) 3 and (c) 4 jets.}
  \label{fig:WMRKlimit}
\end{figure}
\begin{figure}[tbp]
  \centering
  \epsfig{width=0.35\textwidth,file=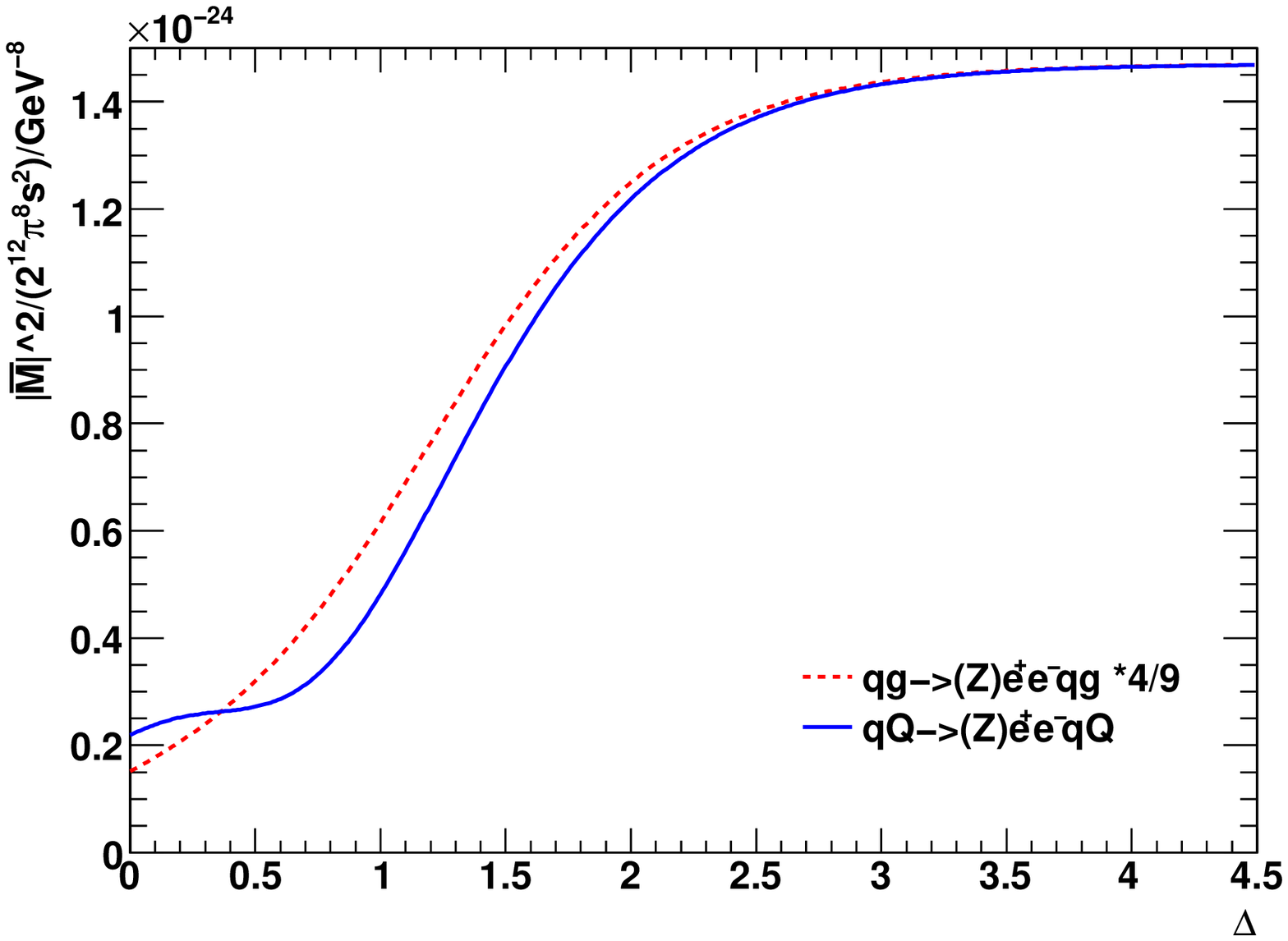} \hspace{-0.7cm}
  \epsfig{width=0.35\textwidth,file=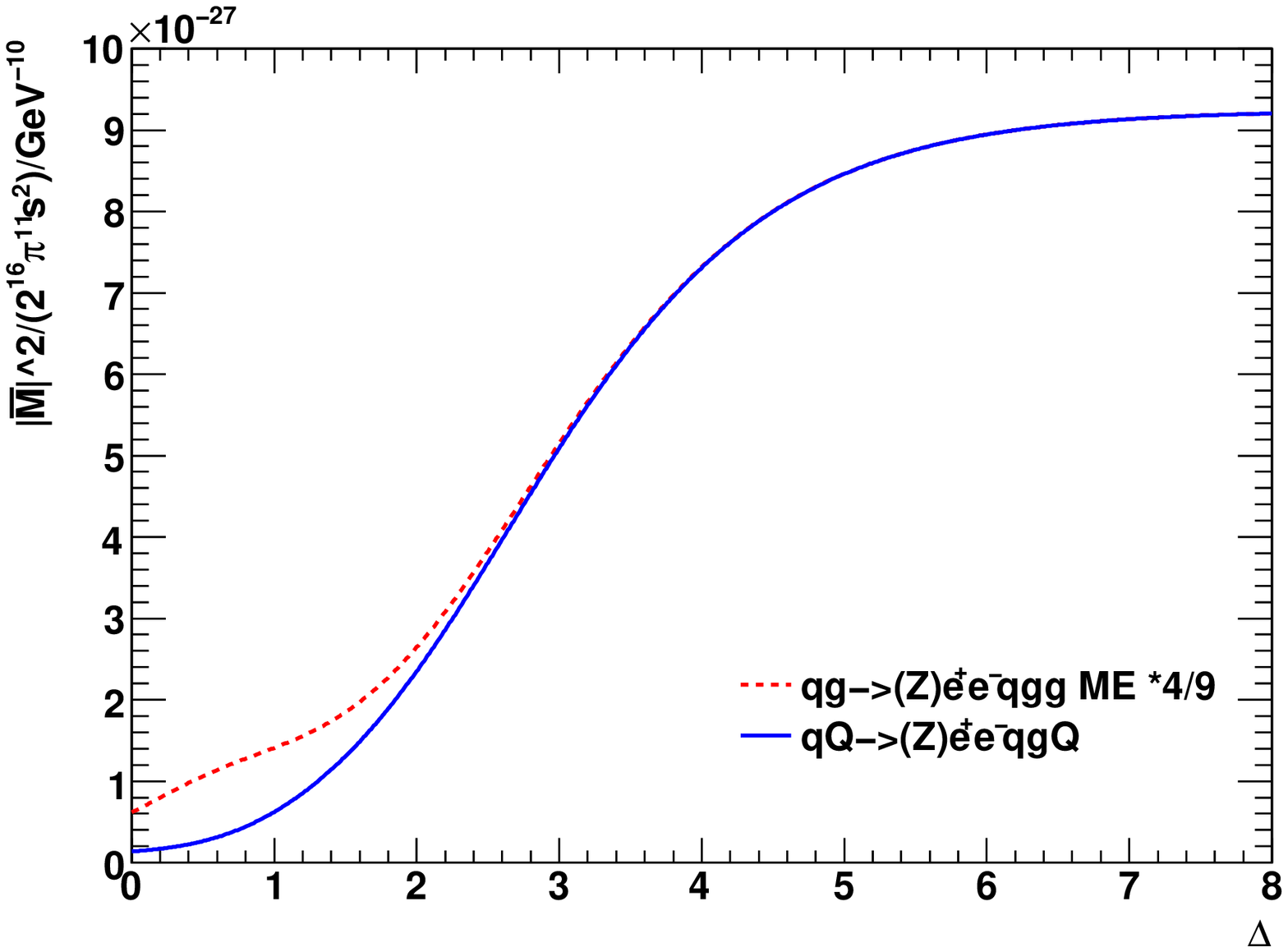} \hspace{-0.7cm}
  \epsfig{width=0.35\textwidth,file=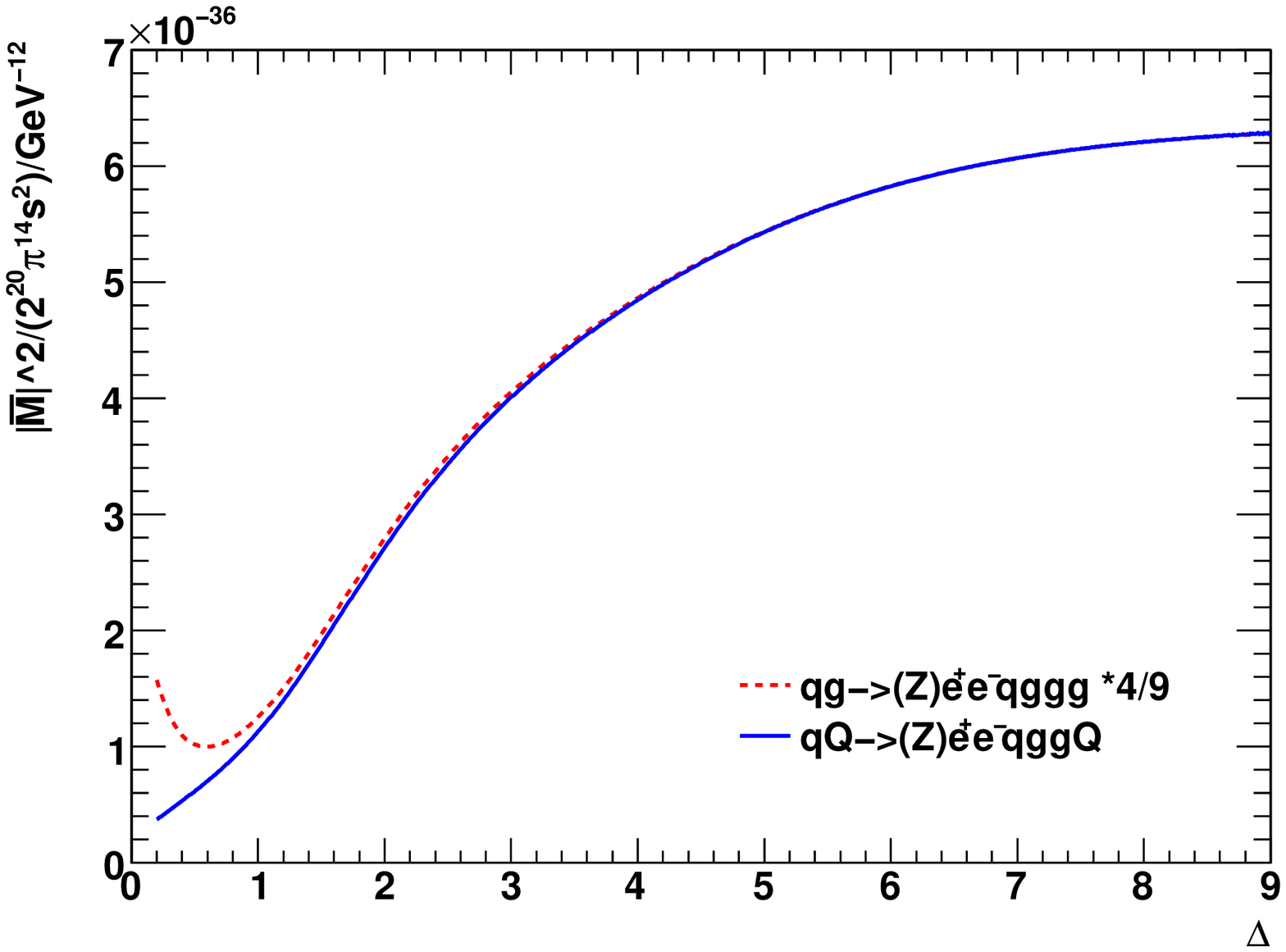}
  \caption{The scattering matrix elements for jet production in association with a $Z$
    boson, for (a) 2, (b) 3 and (c) 4 jets.}
  \label{fig:ZMRKlimit}
\end{figure}
\begin{figure}[tbp]
  \centering
  \epsfig{width=0.35\textwidth,file=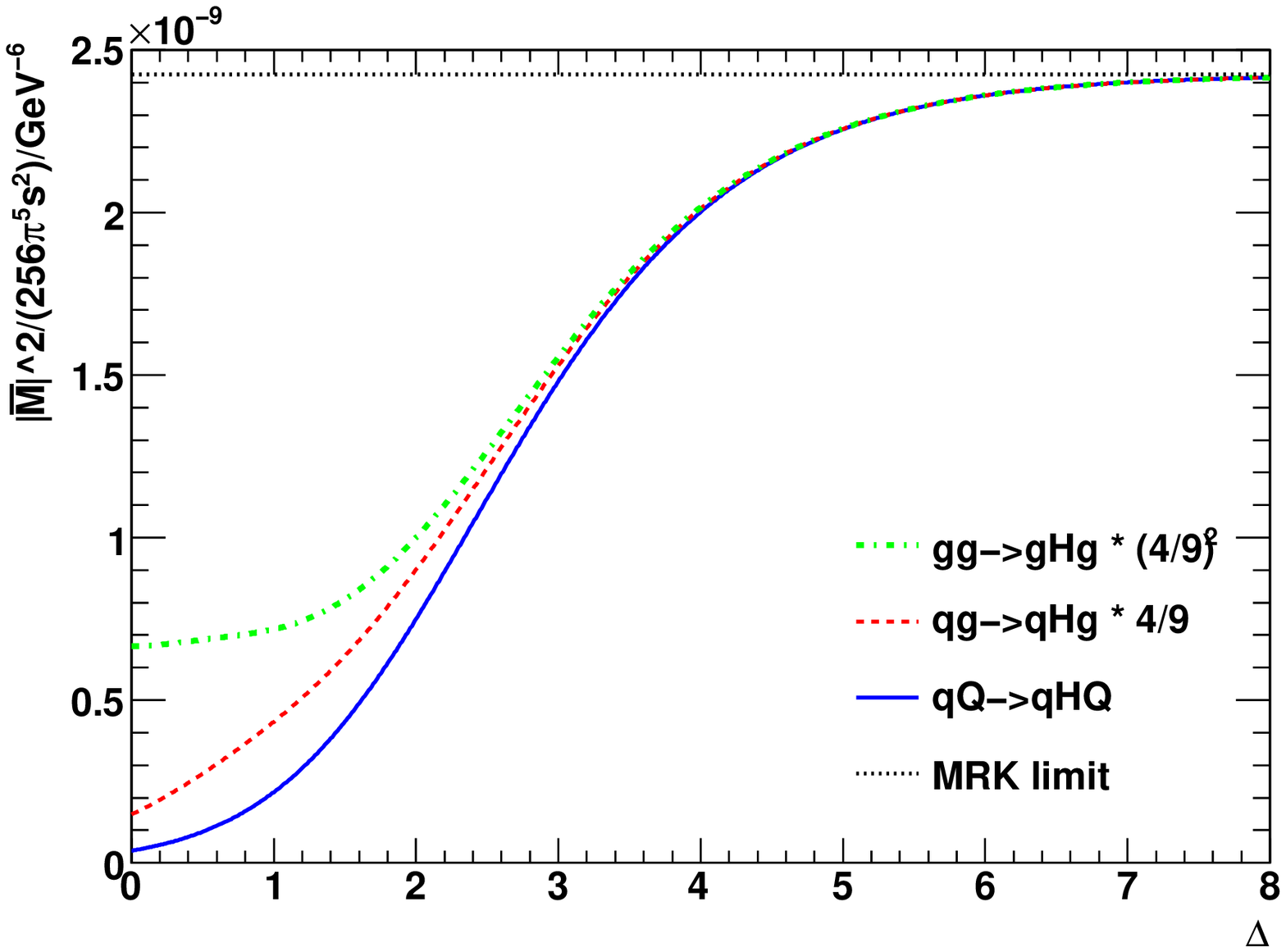} 
  \epsfig{width=0.35\textwidth,file=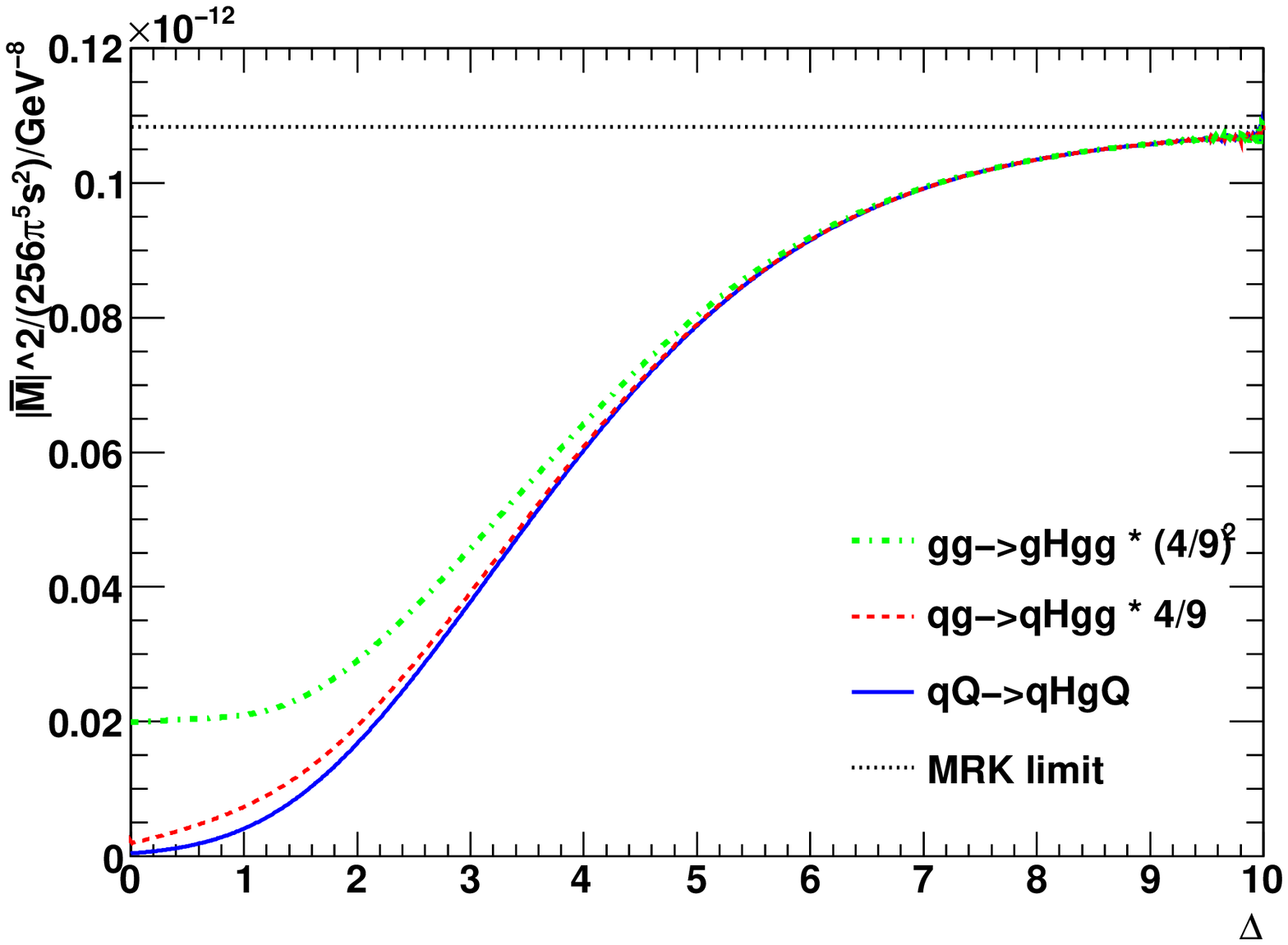} 

  (a) \hspace{5cm} (b) 
  \caption{The scattering matrix elements for jet production in association with a $H$
    boson, for (a) 2 and (b) 3 jets.}
  \label{fig:HMRKlimit}
\end{figure}
As can be seen on the plots, the correct MRK limit is obtained for all the
channels. However, while the simple formulae of
Eqs.~\eqref{eq:Msqsumavg}--\eqref{eq:qQMsqsumavg} do indeed describe the MRK limit of the
full amplitudes, this limit is approached only outside the region of relevance for the
LHC. The power vested in this implementation of the factorised picture in terms of ability
to calculate a specific limit of the $n$-gluon amplitude is turning out not to be relevant
for LHC phenomenology, if implemented according to Eq.~\eqref{eq:Msqsumavg} or the BFKL
equation.
% [MRK description not correction until very large rapidities - borderline
%  whether reachable by LHC]

% [Universality of channels indication of factorised picture? Try to obtain
%  building-blocks using fewer kinematical approximations than what is needed to
%  obtain the simple formula Eq.~\eqref{eq:Msqsumavg}-~\eqref{eq:qgMsqsumavg}. ]

% [matching corrections to describe the difference between gg,qg,qQ, and then
%  take qQ as model]

In contrast, the \emph{universality} of the $qQ,qg$ and $gg$-channels, which
arises naturally within and to some extent even implies the picture indicated
in Fig.~\ref{fig:tchannelpicture}, \emph{is} reached in the region relevant
for the LHC phenomenology. In fact, the $qQ$ and $qg$ channels behave very
similarly over a very wide range of rapidities. This poses the question
whether one can construct better building blocks still within a factorised
picture, which would allow one to get estimates for the all-order cross
sections, but crucially in a wider kinematic region which would be relevant
for the LHC.  This is the focus of the next section.

%%% Local Variables: 
%%% mode: latex
%%% TeX-master: "jetcurrents"
%%% End: 

\section{The Method Of Currents}
\label{sec:current-method}
The aim of this section is to construct a method for approximating the hard
scattering matrix element for jet production (pure or in association with a
$W/Z/H$-boson) to any order in $\alpha_s$. The method should have the
following characteristics :-
\begin{enumerate}
\item Inclusivenes: The approach should capture both real and virtual
  corrections to all orders for a given process. 
\item Simplicity:
  \begin{enumerate}
  \item All orders exclusively: In order to allow for arbitrary analyses of
    the process, the all-order result must be constructed as an explicit sum
    over $n$-particle final states, with access to the momenta of all emitted
    particles. This obviously requires that the evaluation of the scattering
    amplitude for any number (necessary) of particles is sufficiently fast to
    allow for the $n$-body phase space integration to be performed
    explicitly.
  \item Cancellation of IR poles: The formalism has to be sufficiently simple
    that the cancellation of IR poles between real and virtual corrections
    can be organised while keeping the all-order summation simple.
  \end{enumerate}
\item Accuracy: 
  \begin{enumerate}
  \item The obtained results must reproduce the full perturbative result
    order by order in $\alpha_s$ in the limit of infinite invariant mass
    between all partons.
  \item At the same time, the result must maintain relevance in the kinematic regime of
    TeV-scale colliders; we will calculate the cross section and kinematic distributions
    for $2,3,4$-jet production alone and in association with a $W,Z$ or $H$ boson within
    the simple procedure allowing an all-order construction, and compare the results to
    those obtained with full, fixed-order calculations.
  \end{enumerate}
\end{enumerate}
%% Based on dominance of behaviour dictated by poles in $t$-channel momenta
%% only 'FKL' configurations in this paper
%% Matching corrections
Our approach will be based on capturing the behaviour dictated by poles in
the $t$-channel momenta of the scattering amplitude (as defined in
Eq.~\eqref{eq:ti}, where the momenta $p_i$ are rapidity
ordered). Furthermore, in this paper we will focus on the rapidity ordering
of particles which allows colour octet exchanges between all neighbouring (in
rapidity) particles. It is
well-known\cite{Kuraev:1977fs,Fadin:2006bj,Bogdan:2006af} that for a given
$2\to n$ process, these rapidity-orderings will form the leading contribution
to $n$-jet production, and all other orderings will be suppressed by powers
of the invariant mass between jets of flavours which cannot be connected by a colour
octet exchange. The relevance of this property of the scattering amplitudes
for inclusive cross sections (i.e.~not just as an asymptotic argument) was
investigated for Higgs boson production in association with at least two jets
in Ref.\cite{Andersen:2008ue,Andersen:2008gc}. 

\subsection{Current-Current Scattering} 
\label{sec:curr-curr-scatt}
Based on the close resemblance between the various partonic channels in $2\to2, 2\to 3,
2\to 4$ scattering, we choose to model all channels on $qQ\to q(g\ldots
g)Q$-scattering. The similarity in the behaviour of the various partonic channels in the
MRK limit, where the $t$-channel dominance has set in, is further supported by the behaviour
of the three-gluon vertex and the quark current (Fig.~\ref{fig:quarkTGV}) in the limit
$p_A\sim p_1$ (using the standard spinor formalism):
\begin{figure}[tbp]
  \centering
  \input{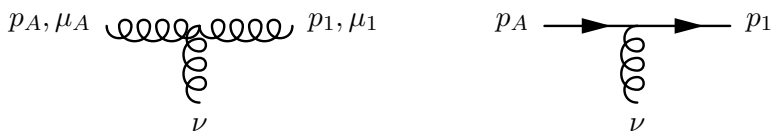}
  \caption{The three-gluon vertex and the quark current shown are identical in the limit
    $p_A\sim p_1$, Eq.~\eqref{eq:HEcurrentandTGV}.}
  \label{fig:quarkTGV}
\end{figure}
\begin{align}
  \begin{split}
    \label{eq:HEcurrentandTGV}
    &\langle 1|\nu|A\rangle\ \to\ 2 p_{A/1}^\nu\\
    &\ve_{A\mu_A}\ve_{1\mu_1}^* \left( (p_A+p_1)^\nu\ g^{\mu_A\mu_1}\ +\ (q-p_1)^{\mu_A}\
      g^{\mu_1\nu}\ +\ (-q-p_A)^{\mu_1}g^{\mu_A\nu}\right)\ \to\
    \ve_{A}\cdot\ve_{1}^*\times 2 p_{A/1}^\nu.
  \end{split}
\end{align}
%% Instead of applying kinematic limits, we will choose to study a specific
%% channel which has only t-channel gluon exchanges.

The building blocks of the traditional (B)FKL formalism are termed
\emph{impact factors} (describing the coupling of e.g.~two on-shell partons
to an off-shell ($t$-channel) gluon) and the \emph{kernel} (describing
the evolution of the off-shell gluon under the emission of partons). These
building blocks can be obtained by taking the MRK limit of specific
processes, see e.g.~Ref.\cite{DelDuca:1999ha} and references
therein. However, with building blocks obtained in this way, the resulting
approximation for the scattering amplitude will have lost all dependence on
the rapidity difference between particles (since this has been taken to
infinity), and only the limits in Fig.~\ref{fig:M3j} are obtained. When the
BFKL equation is used, this approximation is then applied in all of phase
space, resulting in a gross over-estimate order-by-order of the scattering amplitude.

Instead of relying on a kinematic limit to ensure the $t$-channel dominance
of the scattering, we will instead obtain the building blocks from the
process $qQ\to qQ$, which consists only of a $t$-channel gluon exchange. This
means that the lowest order results for the $qQ\to qQ$ process will be
identical in full QCD and with the effective rules. The benefits are even
greater for e.g.~W or Z plus jets, where in the standard approach (see
e.g.~Ref.\cite{Andersen:2001ja}), kinematical assumptions on also the weak
boson (and its decay products) as well as the jets had to be applied in order to obtain the building
blocks for a resummation. We will see that the results for Higgs boson
production in association with
jets~\cite{DelDuca:2003ba,Andersen:2008ue,Andersen:2008gc} can also be
improved using this new formalism. 

When the effective Feynman rules we derive here are used in constructing
all-order results, the differences between the various partonic channels at
small rapidities (and the corrections to the approximations for multi-parton
production) will be taken into account order-by-order by matching to the full
fixed-order perturbative
result using the same approach as in
Ref.\cite{Andersen:2008ue,Andersen:2008gc}. 

In the spinor notation, the (colour and coupling stripped) matrix element for
the process $qQ\to qQ$ for negative helicity quarks reads
\begin{align}
  \label{eq:MqQ}
  M_{q^-Q^-\to q^-Q^-} &=\ \spab1.\mu.a \frac{g^{\mu\nu}}{t}\spab 2.\nu.b\\
 &=\ \frac{2\spb a.b\spa2.1}{t}.
  \label{eq:MqQ2}
\end{align}
We will use the scattering of quark currents as the basis for our framework,
since these consist of $t$-channel gluon exchanges only, and explicitly
exhibit the factorisation into two components (spinor strings), each
depending only on the momenta along each quark line. This is also obviously
true for all other helicity configurations.

Let us denote the spinor string (for helicities $h_a,h_1,h_b,h_2$ of the
quarks) appearing in the amplitude as
\begin{align}
  \label{eq:spinorstring}
  S^{h_ah_b\to h_1h_2}_{qQ\to qQ}=\langle 1\ h_1|\mu|a\ h_a\rangle\
  g^{\mu\nu}\ \langle 2\ h_2|\nu| b\ h_b\rangle.
\end{align}
This complex number will be calculated using an explicit representation for
the spinors (see Appendix~\ref{sec:spin-repr}), and we will denote the sum
over helicities of the absolute square of this number by
\begin{align}
  \label{eq:spinorsqsum}
  \left\|S_{qQ\to qQ}\right\|^2=\sum_{h_a,h_a,h_b,h_2} \left |S^{h_ah_b\to h_1h_2}_{qQ\to qQ}\right|^2.
\end{align}
Of course in this case non-zero contributions arise only when $h_a=h_1$ and $h_b=h_2$.

The colour and helicity summed and averaged matrix element for the scattering
process $qQ\to qQ$ is then
\begin{align}
  \begin{split}
    \label{eq:MqQqQchsa}
    \left|\overline{\mathcal{M}}^t_{qQ\to qQ}\right|^2\ =\ &\frac 1 {4\
      (\Nc^2-1)}\ \left\|S_{qQ\to qQ}\right\|^2\\
    &\cdot\ \left(g^2\ \cf\ \frac 1 {t_1}\right)\\
    &\cdot\ \left(g^2\ \cf\ \frac 1 {t_2}\right).
  \end{split}
\end{align}
with $t_1=(p_a-p_1)^2$ and $t_2=(-p_b+p_2)^2$ ($t_1=t_2$ in this case of a
$2\to2$-process).  The superscript $t$ is meant to indicate that this is the ``$t$-channel
factorised''-approximation, but this is of course exact in the channel $qQ\to qQ$. The
approximations for the gluon channels are obtained by multiplying $\ca/\cf$ for each pair
of quarks replaced by gluons.  Fig.~\ref{fig:M3j} shows that this is an extremely good
approximation.

The spinor formalism clearly displays the factorisation in the $t$-channel of the
scattering, i.e.~within the spinor formalism, the ``impact factors'' are clearly
identified as the quark currents; in contrast, the ``spinor product'' rewriting of
Eq.~\eqref{eq:MqQ2} mixes momenta from the two quark lines.  The standard procedure for
extracting impact factors using the helicity formalism\cite{DelDuca:1999ha} applies the
kinematic approximations valid in the MRK limit. In terms of invariants, the square of the
colour and spin averaged and summed scattering matrix element for $qQ\to qQ$ is
\begin{align}
  \label{eq:qQqQinv}
  g^4\ \frac 4 9\ \frac{s^2+u^2}{t^2}.
\end{align}
The $u^2$-terms arise from scattering of quark currents of different helicities, and
spoil the factorisation implied in Eq.~\eqref{eq:qQMsqsumavg} from being exact,
necessitating the consideration of kinematic limits of the squared scattering matrix
element. This despite the fact that for processes which proceed only through a $t$-channel
gluon exchange, the starting expression in terms of spinor strings is already factorised.

In the MRK limit (of infinite rapidity separation between the scattered
partons), the expression for the colour and helicity summed and averaged
matrix element simplifies to the 2-jet part of Eq.~\eqref{eq:qQMsqsumavg}
since all allowed helicity scatterings give the same result in the limit and
\begin{align}
  \label{eq:simplifications}
  \left |\spb a.b\spa2.1\right| = s,\quad t\to -|p_\perp|^2.
\end{align}
This is the lowest order results in Eqs.~\eqref{eq:Msqsumavg}--\eqref{eq:qQMsqsumavg}.

\subsection{Multi-Parton Production}
\label{sec:multi-part-prod}

\begin{figure}[tbp]
  \centering

  \vspace{0.3cm}
  \input{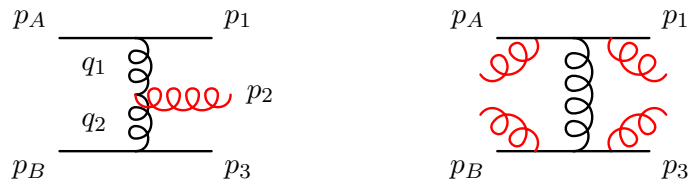}  
  \caption{We add contributions where the 3rd jet (red) is emitted from the $t$-channel
    gluon, and from each of the four external quark lines.}
  \label{fig:2plusgemission}
\end{figure}

In this section we will develop the picture of the scattering of two quark currents to
take into account the emission of additional gluons.  We first consider adding one extra
gluon to the $qQ\to qQ$ scattering we have taken as our model so far; this may be emitted
from the $t$-channel gluon or from each of the external quark lines,
Fig.~\ref{fig:2plusgemission}.

Firstly the three gluon vertex in the $t$-channel emission gives a contribution of
\begin{align}
  \label{eq:tgluonfrom2}
  \mathcal{A}_g=\frac{-\mathcal{C}_gg_s^3}{t_1t_2} \ \bar u(p_1)\gamma_\mu u(p_A)\ \bar
  u(p_3) \gamma_\nu u(p_B)  \ \ve^{*}_{\rho}\ 
  \left( (q_1+q_2)^\rho g^{\mu \nu} + (p_2-q_2)^\mu g^{\nu \rho} - (q_1+p_2)^\nu
    g^{\mu\rho} \right),
\end{align}
where $\mathcal{C}_g=T^w_{a_1a_A}f^{wi_2v}T^v_{a_3a_B}$.  In the MRK limit, we can use
Eq.~\eqref{eq:HEcurrentandTGV} for the spinor strings, and $q_1=p_3-p_B+p_2$ and $q_2=p_a-p_1-p_2$ to get
\begin{align}
  \label{eq:tgluonlim}
  \mathcal{A}_g \to \frac{-2g_s^3}{t_1t_2}\ \ve^{*}_{\rho}\ \left( -p_A^\rho
    (s_{3B}+2s_{2B})+p_B^\rho(s_{1A}+2s_{2A}) + (q_1+q_2)^\rho \hat s \right).
\end{align}
The MRK limit (Eq.~\eqref{eq:MRKlimit}) gives $s_{2B}\gg s_{3B}$ and $s_{2A}\gg s_{1A}$ so
we are left with
\begin{align}
  \label{eq:tgluonfinallim}
  \mathcal{A}_g^{\mathrm{MRK}} = \frac{-2g_s^3\hat s}{t_1t_2}\ \ve^{*}_{\rho}\ \left(
    -2p_A^\rho \frac{s_{2B}}{\hat s} + 2p_B^\rho\frac{s_{2A}}{\hat s} + (q_1+q_2)^\rho
  \right).
\end{align}
We now add the contributions from the emissions of gluons from each external quark
line.  We treat these emissions as soft, and use Eikonal factors which are
valid for a gluon emitted between
the quark jets in the MRK limit.  These add to give
\begin{align}
  \label{eq:quarkcontribs}
  \mathcal{A}_q=\mathcal{A}_{qQ\to qQ}\times (ig_s)\ \ve^{*}_{\rho}\ \left(
    \mathcal{C}_1\frac{p_1^\rho}{p_1\cdot p_2} - \mathcal{C}_A\frac{p_A^\rho}{p_A \cdot
      p_2} + \mathcal{C}_3 \frac{p_3^\rho}{p_3\cdot p_2} -
    \mathcal{C}_B\frac{p_B^\rho}{p_B \cdot p_2} \right).
\end{align}
The $\mathcal{C}_i$ are the relevant colour factors:
\begin{align}
  \label{eq:Ddefn}
  \mathcal{C}_1=T^{i_2}_{a_1b} T^w_{b a_A} T^w_{a_3a_B},\
  \mathcal{C}_A=T^w_{a_1b}T^{i_2}_{ba_A}T^w_{a_3a_B},\
  \mathcal{C}_3=T^w_{a_1a_A}T^{i_2}_{a_3b}T^w_{ba_B},\ 
  \mathcal{C}_B=T^w_{a_1a_A}T^w_{a_3b}T^{i_2}_{ba_B}.
\end{align}
Now in the MRK limit, $p_A\to p_1$, $p_B\to p_3$ and $\mathcal{A}_{qQ\to qQ}=S_{qQ\to qQ}/ t$, so
\begin{align}
  \label{eq:nearlylimq}
  \mathcal{A}_q \to \frac{S_{qQ\to qQ}}{q_1^2 q_2^2}\ (-ig_s^3)\ \ve^{*}_{\rho}\ \left(
    (\mathcal{C}_1-\mathcal{C}_A)\ q_1^2\ \frac{p_A^\rho}{p_A \cdot p_2} +
    (\mathcal{C}_3-\mathcal{C}_B)\ q_2^2\ \frac{p_B^\rho}{p_B \cdot p_2}
  \right).
\end{align}
However, $(\mathcal{C}_1-\mathcal{C}_A)=i\mathcal{C}_g$ and
$(\mathcal{C}_3-\mathcal{C}_B)=-i\mathcal{C}_g$ so that all five contributions give the
\emph{same} overall colour factor.  (The same pairing up is seen when a quark line is
replaced by a gluon line, only here it is the Jacobi Identity which can be used to give
a single overall factor.)

\begin{figure}[htbp]
  \centering
  \epsfig{width=0.49\textwidth,file=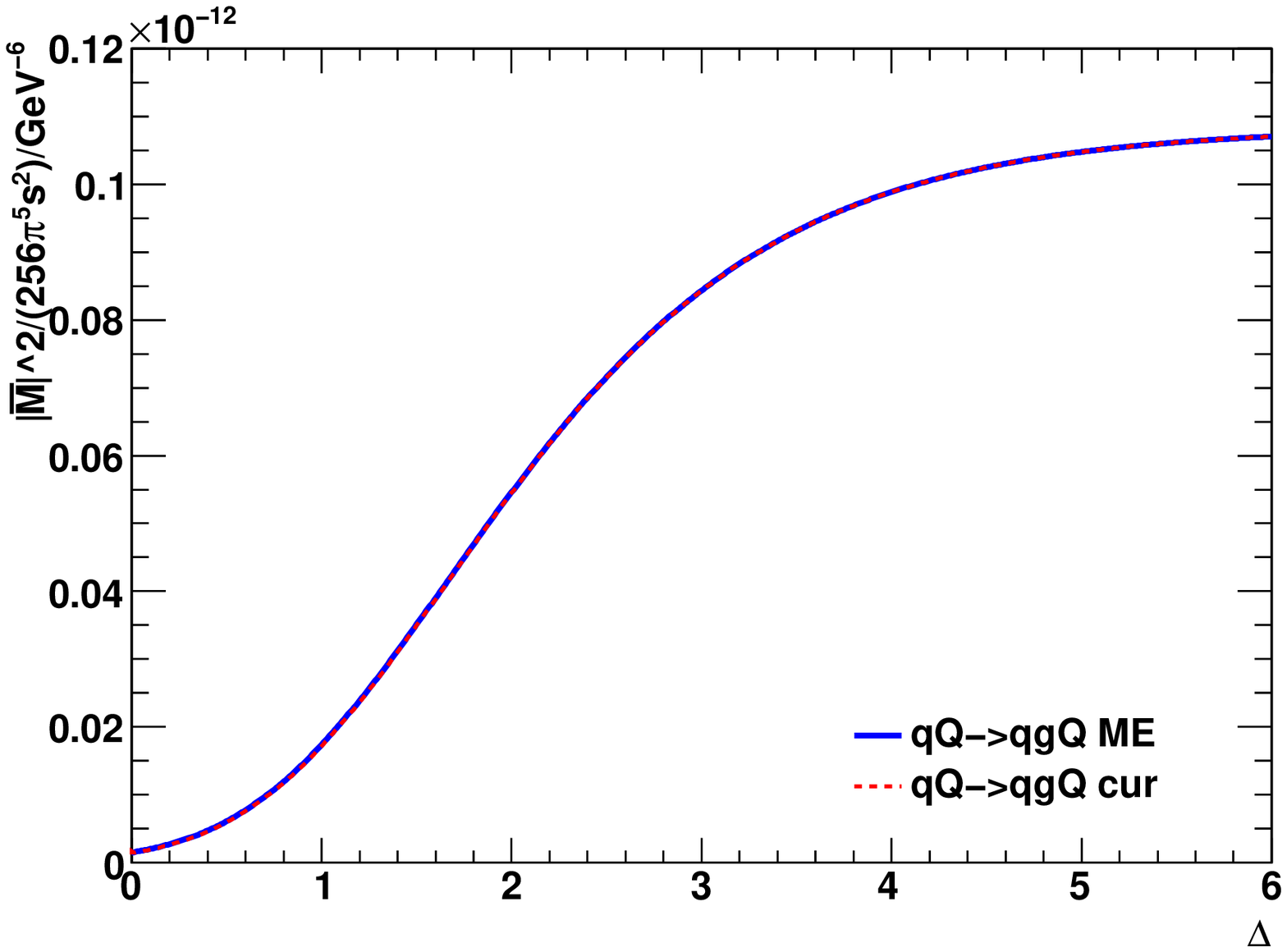}
  \epsfig{width=0.49\textwidth,file=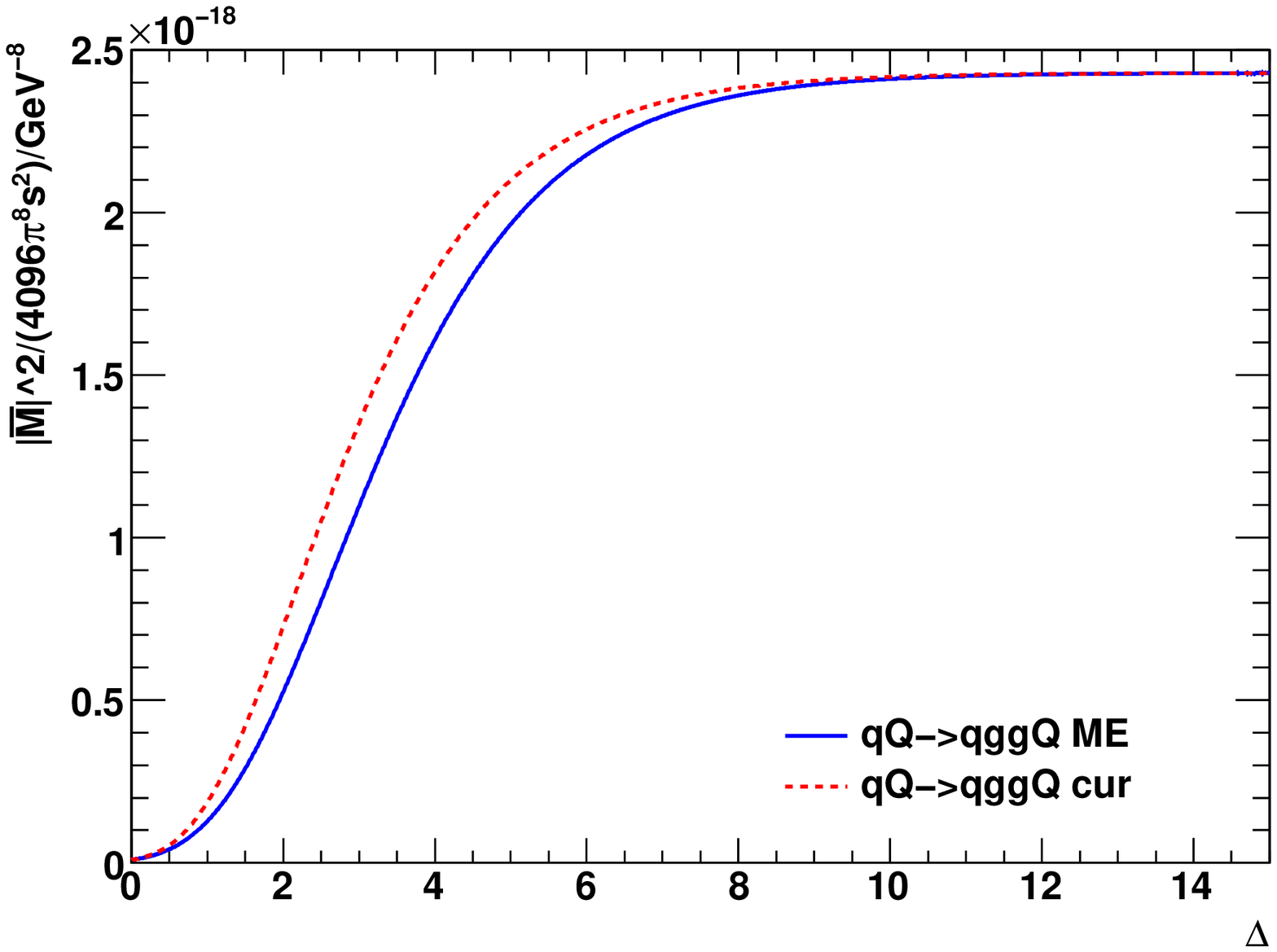}

  (a) \hspace{7.2cm}(b)\hspace{0.1cm}
  \caption{Results for the matrix elements of (a) $qQ\to qgQ$ and (b) $qQ\to qggQ$
    processes respectively.  The momentum configurations are as for Fig.~\ref{fig:M3j}(b)
    and (c) respectively.}
  \label{fig:2jMEs}
\end{figure}

We could then add Eq.~\eqref{eq:nearlylimq} to Eq.~\eqref{eq:tgluonfinallim}; this gives
the form used in \cite{Andersen:2008ue,Andersen:2008gc}.
However we choose to first reinstate the full kinematic structure of
Eq.~\eqref{eq:quarkcontribs} in order to capture more of the original process.  We also
adapt Eq.~\eqref{eq:tgluonfinallim} to take account of both \{$p_A,p_B$\} and
\{$p_1,p_3$\} to give the following approximation for the $qQ\to
qgQ$-scattering matrix element
\begin{align}
  \label{eq:3jstruct}
  \mathcal{A}_{qQ\to qgQ} = g_s^3\ \mathcal{C}_g\ \ve^{*}_{\rho}\ \frac{S_{qQ\to qQ}}{q_1^2 q_2^2}\ V^\rho(q_1,q_2)
\end{align}
where
\begin{align}
  \label{eq:EmissionV}
  \begin{split}
  V^\rho(q_1,q_2)=&-(q_1+q_2)^\rho \\
  &+ \frac{p_A^\rho}{2} \left( \frac{q_1^2}{p_2\cdot p_A} +
  \frac{p_2\cdot p_B}{p_A\cdot p_B} + \frac{p_2\cdot p_3}{p_A\cdot p_3}\right) + p_A
\leftrightarrow p_1 \\ 
  &- \frac{p_B^\rho}{2} \left( \frac{q_2^2}{p_2 \cdot p_B} + \frac{p_2\cdot
      p_A}{p_B\cdot p_A} + \frac{p_2\cdot p_1}{p_B\cdot p_1} \right) - p_B
  \leftrightarrow p_3.
  \end{split}
\end{align}
This form of the effective vertex is gauge invariant; the Ward Identity, $p_g\cdot V=0$
can easily be checked.  The spinor structure is exactly as for $qQ\to qQ$ and so we find 
\begin{align}
  \label{eq:3jetVs}
  \begin{split}
    \left|\overline{\mathcal{M}}^t_{qQ\to qgQ}\right|^2\ =\ &\frac 1 {4\
      (\Nc^2-1)}\ \left\|S_{qQ\to qQ}\right\|^2\\
    &\cdot\ \left(g^2\ \cf\ \frac 1 {t_1}\right) \cdot\ \left(g^2\ \cf\ \frac 1
      {t_{2}}\right)\\
    & \cdot \left( \frac{-g^2 C_A}{t_1t_2}\ V^\mu(q_1,q_2)V_\mu(q_1,q_2) \right).
  \end{split}
\end{align}
The results of this formalism are compared to the full tree-level matrix element for $qQ\to qgQ$ in
Fig.~\ref{fig:2jMEs}(a).  The results from the two methods are indistinguishable by eye.

It can be checked that in the MRK limit, Eq.~\eqref{eq:EmissionV} gives
\begin{align}
  \label{eq:MRKLipatov}
  -V^\mu(q_1,q_{2})V_\mu(q_1,q_{2}) \longrightarrow 4\frac{|q_{1\perp}|^2
    |q_{2\perp}|^2}{|p_{2\perp}|^2}.
\end{align}
The $|q_{j\perp}|^2$ terms in the numerator cancel with the factors of $t_j$ in
Eq.~\eqref{eq:3jetVs} to give an MRK limit which agrees with Eq. \eqref{eq:qQMsqsumavg}.
Motivated by the structure of Eq.~\eqref{eq:qQMsqsumavg}, we describe a general
process with $n$ jets in the final state (Fig.~\ref{fig:tchannelpicture}) as
\begin{align}
  \label{eq:multijetVs}
  \begin{split}
    \left|\overline{\mathcal{M}}^t_{qQ\to qg\ldots gQ}\right|^2\ =\ &\frac 1 {4\
      (\Nc^2-1)}\ \left\|S_{qQ\to qQ}\right\|^2\\
    &\cdot\ \left(g^2\ \cf\ \frac 1 {t_1}\right) \cdot\ \left(g^2\ \cf\ \frac 1
      {t_{n-1}}\right)\\
    & \cdot \prod_{i=1}^{n-2} \left( \frac{-g^2 C_A}{t_it_{i+1}}\
      V^\mu(q_i,q_{i+1})V_\mu(q_i,q_{i+1}) \right),
  \end{split}
\end{align}
where $V^\mu(q_i,q_{i+1})$ is the obvious generalisation of Eq.~\eqref{eq:EmissionV}:
\begin{align}
  \label{eq:GenEmissionV}
  \begin{split}
  V^\rho(q_i,q_{i+1})=&-(q_i+q_{i+1})^\rho \\
  &+ \frac{p_A^\rho}{2} \left( \frac{q_i^2}{p_{i+1}\cdot p_A} +
  \frac{p_{i+1}\cdot p_B}{p_A\cdot p_B} + \frac{p_{i+1}\cdot p_n}{p_A\cdot p_n}\right) +
p_A \leftrightarrow p_1 \\ 
  &- \frac{p_B^\rho}{2} \left( \frac{q_{i+1}^2}{p_{i+1} \cdot p_B} + \frac{p_{i+1}\cdot
      p_A}{p_B\cdot p_A} + \frac{p_{i+1}\cdot p_1}{p_B\cdot p_1} \right) - p_B
  \leftrightarrow p_n.
  \end{split}
\end{align}
Each emission vertex comes with a factor of $\ca$ in $|\mathcal{M}|^2$ because (by the
discussion above) there is a $Tr(f^{abc}f^{abc})$ in place of a $\delta^{bc}\delta^{bc}$
compared to the process with one fewer jet.  

The results for 4 jet final states are shown in Fig.~\ref{fig:2jMEs}(b) compared to the
full matrix element, and show that the formalism performs well.  A more detailed study of
the results obtained in this formalism for kinematic distributions is
presented in Section~\ref{sec:pure-jets}.

% It is important to check that including a string of these vertices in this way reproduces
% the expected form in the MRK limit.  In this limit, 
% \begin{align}
%   \label{eq:MRKamp}
%   \left|\overline{\mathcal{M}}^t_{qQ\to qg\ldots gQ}\right|^2\ \longrightarrow\
%   &\frac{4s^2}{(\Nc^2-1)}\cdot\ \left(\frac{g^2\cf}{|p_{1\perp}|^2} \right)
%   \cdot\ \left(\frac{g^2\cf}{|p_{n\perp}|^2}\right) \cdot \prod_{i=1}^{n-2} \left(
%     \frac{4g^2 C_A}{|p_{i+1\perp}|^2} \right)
% \end{align}

\subsection{W and Z Boson Production in Association with Jets}
\label{sec:w-boson-production}

\begin{figure}[btp]
  \centering
  \input{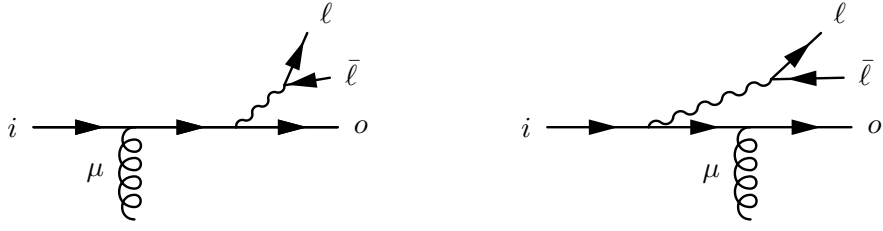}
  \caption{The two diagrams which contribute when a $W$ or $Z$ boson is emitted from a
    quark current line joined to the rest of the diagram.}
  \label{fig:Wurrent}
\end{figure}

To apply this formalism to $W$ and $Z$ production with jets, we
follow the same structure as the pure jets case: modelling both $qQ$ and $qg$ channels
on $qQ\to q(g\ldots g)Q + (V\to )\ell \bar \ell$.  When a $W$ or $Z$ is emitted from a
quark line, the two diagrams shown in Fig.~\ref{fig:Wurrent} need to be considered.
The gluon line with free index $\mu$ links this
current to the rest of the diagram.  The two diagrams add to give
\begin{align}
  \label{eq:Zsum}
    J_V^\mu(i,\ell, \bar \ell,,o)=\left( \frac{ \bar u_o \gamma^\alpha (\cancel{p}_o +
        \cancel{p}_\ell +
        \cancel{p}_{\bar \ell}) \gamma^\mu u_i}{t_{o\ell \bar \ell}} + \frac{\bar u_o
        \gamma^\mu(\cancel{p}_i - \cancel{p}_\ell - \cancel{p}_{\bar \ell}) \gamma^\alpha
        u_i}{t_{i\ell \bar \ell}} \right) \bar u_\ell \gamma_\alpha v_{\bar \ell}.
\end{align}
We can rewrite this in terms of manageable $\overline{\rm{spinor}}$-matrix-spinor pieces
as:
\begin{align}
  \label{eq:finalZ}
  \begin{split}
    J_V^\mu(i,\ell, \bar \ell,,o)=&\left( \frac{2p_o^\alpha [\bar u_o \gamma^\mu P_i u_i]
        + [\bar u_o \gamma^\alpha P_o
        u_{\bar \ell}][\bar u_{\bar \ell} \gamma^\mu P_i u_i] + [\bar u_o \gamma^\alpha
        P_o u_\ell][\bar u_\ell \gamma^\mu P_i u_i]}{t_{o\ell \bar \ell}} \right. \\
    &\quad \left. + \frac{2p_i^\alpha [\bar u_o \gamma^\mu P_o u_i] - [\bar u_o
        \gamma^\mu P_o u_\ell][\bar u_\ell \gamma^\alpha P_i u_i] - [\bar u_o \gamma^\mu
        P_o u_{\bar \ell}] [\bar u_{\bar \ell} \gamma^\alpha P_i u_i]}{t_{i\ell \bar
          \ell}} \right) \\ & \qquad \times [\bar u_\ell \gamma_\alpha P_\ell u_{\bar
      \ell}],
  \end{split}
\end{align}
where $P_x$ is the projection operator $P_\pm=(1\pm\gamma_5)/2$ according to the helicity
of particle $x$.  The current is zero unless $P_o=P_i$.

There are two classes of subprocess to consider, those where
\begin{enumerate}
\item the $W$ or $Z$ boson can only be emitted from one quark line, which is the case in
  all $qg$ channels and certain $qQ$ channels for $W$ production, or
\item the $W$ or $Z$ boson can be emitted from either quark line.
\end{enumerate}
In the following, without loss of generality, we pick a specific example subprocess for
each case: for case 1, $ud\to dd(W\to)\nu_\ell \bar \ell$ and for case 2, $ud\to ud(Z\to )
\ell \bar \ell$.

Case 1 is straightforward and the analogue of equation \eqref{eq:spinorstring} is
\begin{align}
  \label{eq:Wspinorstring}
  S^{-h_b\to -h_2--}_{ud\to dd\nu_\ell \bar \ell}=J_{V\mu}(i,\ell,\bar \ell,
  o)\  g^{\mu\nu}\ \langle 2\ h_2|\nu| b\ h_b\rangle.
\end{align}
Again we use double vertical line notation to symbolise the sum over helicities of each $|
S|^2$.  We write
\begin{align}
  \label{eq:combinecouplings}
    ||S_{ud\to d \nu_\ell \bar \ell d}||^2 &= \frac{g_W^4}{4} \sum_{h_b,h_2} |S_{ud\to d
      \nu_\ell \bar \ell d}^{- h_b \to - h_2 - -}|^2, \\
  \begin{split}
     \left|\overline{\mathcal{M}}^t_{ud\to d\nu_\ell \bar
        \ell d}\right|^2\ &=\ \frac 1 {4\
      (\Nc^2-1)}\ \left\|S_{ud\to d\nu_\ell \ell d}\right\|^2\\
    &\cdot\ \left(g^2\ \cf\ \frac 1 {t_1}\right)\\
    &\cdot\ \left(g^2\ \cf\ \frac 1 {t_2}\right)
  \end{split}
\end{align}
where now $t_1=(p_a-p_1-p_\nu-p_\ell)^2$ and $t_2=(-p_b+p_2)^2$ (still $t_1=t_2$ in this
$2j$ case).

In case 2, it is similar but slightly more complicated as one must divide by $t$s before
squaring so we get
\begin{align}
  \label{eq:case2M}
  \begin{split}
    \left|\overline{\mathcal{M}}^t_{ud\to u\ell \bar \ell d}\right|^2\ &=\ \frac 1 {4\
      (\Nc^2-1)}\ \cdot\ \left(g^2\ \cf\right)\cdot\ \left(g^2\ \cf\right) \\
    & \times \sum_{
      \begin{array}{c}
        h_a,h_b,h_1, \\ h_2,h_\ell,h_{\bar \ell}
      \end{array} } k_{Z\ell \bar \ell}^2\ \left| k_u\
    J_{V\mu}(a,\ell,\bar \ell, 1)\ g^{\mu\nu}\ \langle 2\
    h_2|\nu| b\ h_b\rangle \cdot \frac{1}{t_{t}} \right. \\
  &\hspace{4cm} \left. +k_d\ \langle 1\ h_1|\nu| a\ h_a\rangle\ g^{\mu\nu}
    J_{V\nu}(b,\ell,\bar \ell,2)\cdot \frac{1}{t_{b}}\right|^2,
  \end{split}
\end{align}
where $k_{u,d}$ are the different couplings of top and bottom emission, $k_{Z\ell \bar
  \ell}$ is the $Z$ coupling to the leptons, $t_{t}=(p_a-p_1-p_\ell-p_{\bar \ell})^2$ and
$t_{b}=(p_b-p_2-p_\ell-p_{\bar \ell})^2$.

\begin{figure}[tbp]
  \centering
  \epsfig{width=0.49\textwidth,file=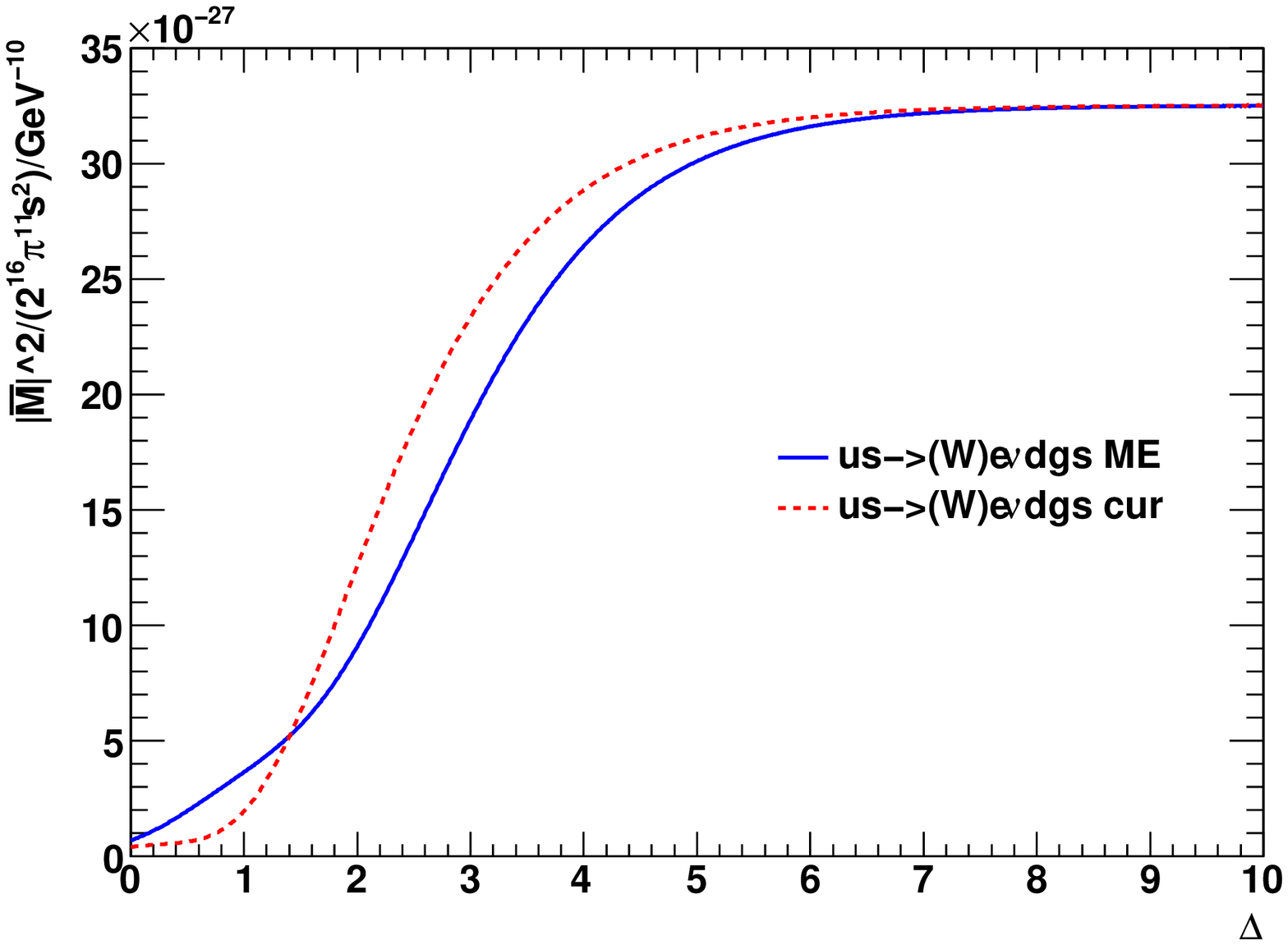}
  \epsfig{width=0.49\textwidth,file=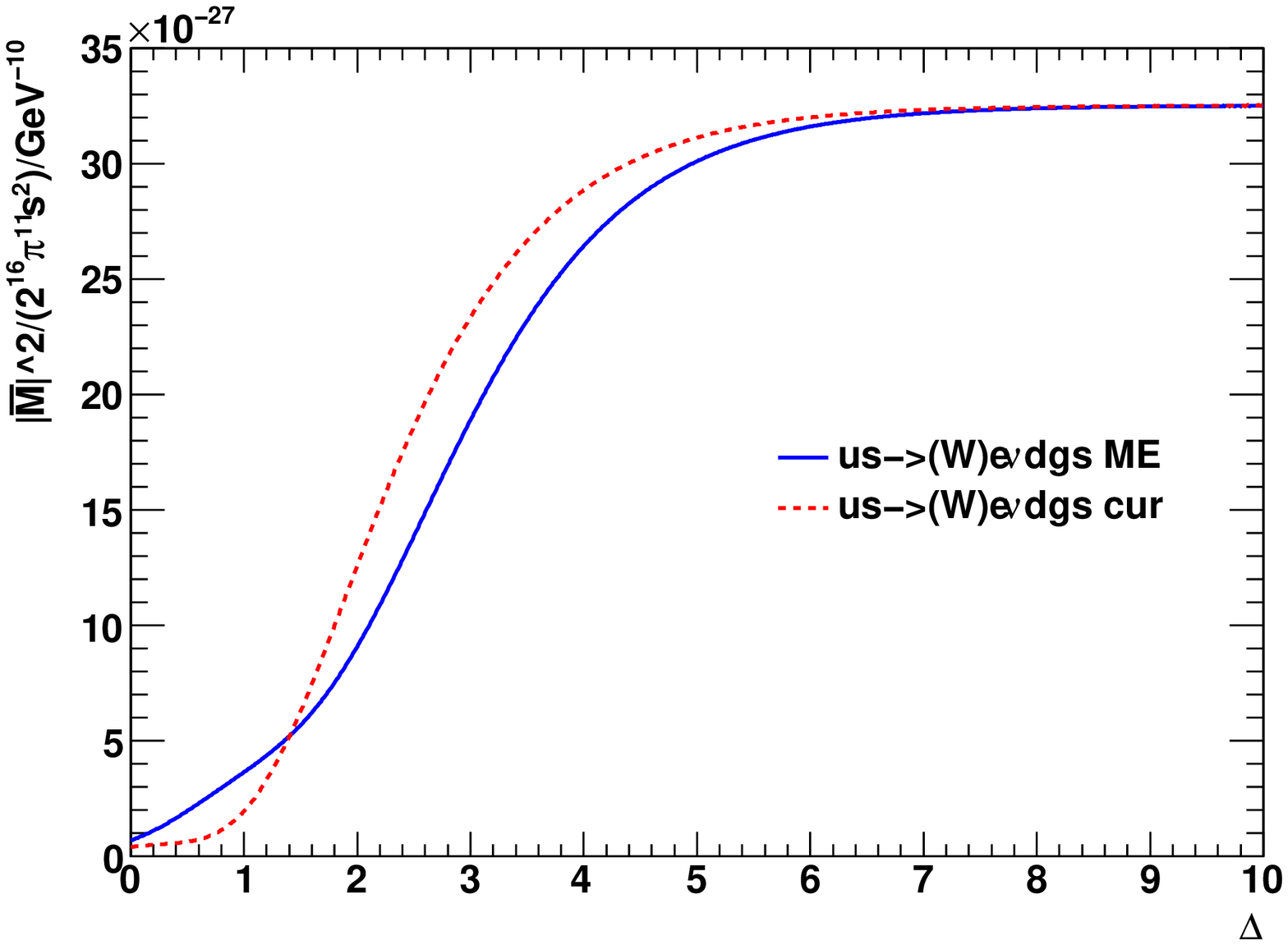}

  (a) \hspace{7.2cm}(b)\hspace{0.1cm}
  \caption{A comparison of the matrix elements obtained from Madgraph and from our
    formalism for (a) $W+3j$ and (b) $W+4j$.  The jets have rapidities $\Delta, 0,
    -\Delta$ in (a), and $\Delta, \Delta/3, -\Delta/3, -\Delta$ in (b).  The full momentum
  configurations are in Appendix \ref{sec:moment-conf}.}
  \label{fig:WMEs}
\end{figure}

As in the case of pure jets, these results are exact for $qQ\to qQ + \ell \bar \ell$.  We
now want to extend this to $2j\to nj+\ell \bar \ell$ for $n>2$.  In case 1, the inclusion
of further gluon emissions (ordered in rapidity) is done in exactly the same way as for
the pure jet case, outlined in section \ref{sec:multi-part-prod}.  The results are shown
in Fig.~\ref{fig:WMEs} for $W+3j$ and $W+4j$.

\begin{figure}[tbp]
  \centering
  \epsfig{width=0.49\textwidth,file=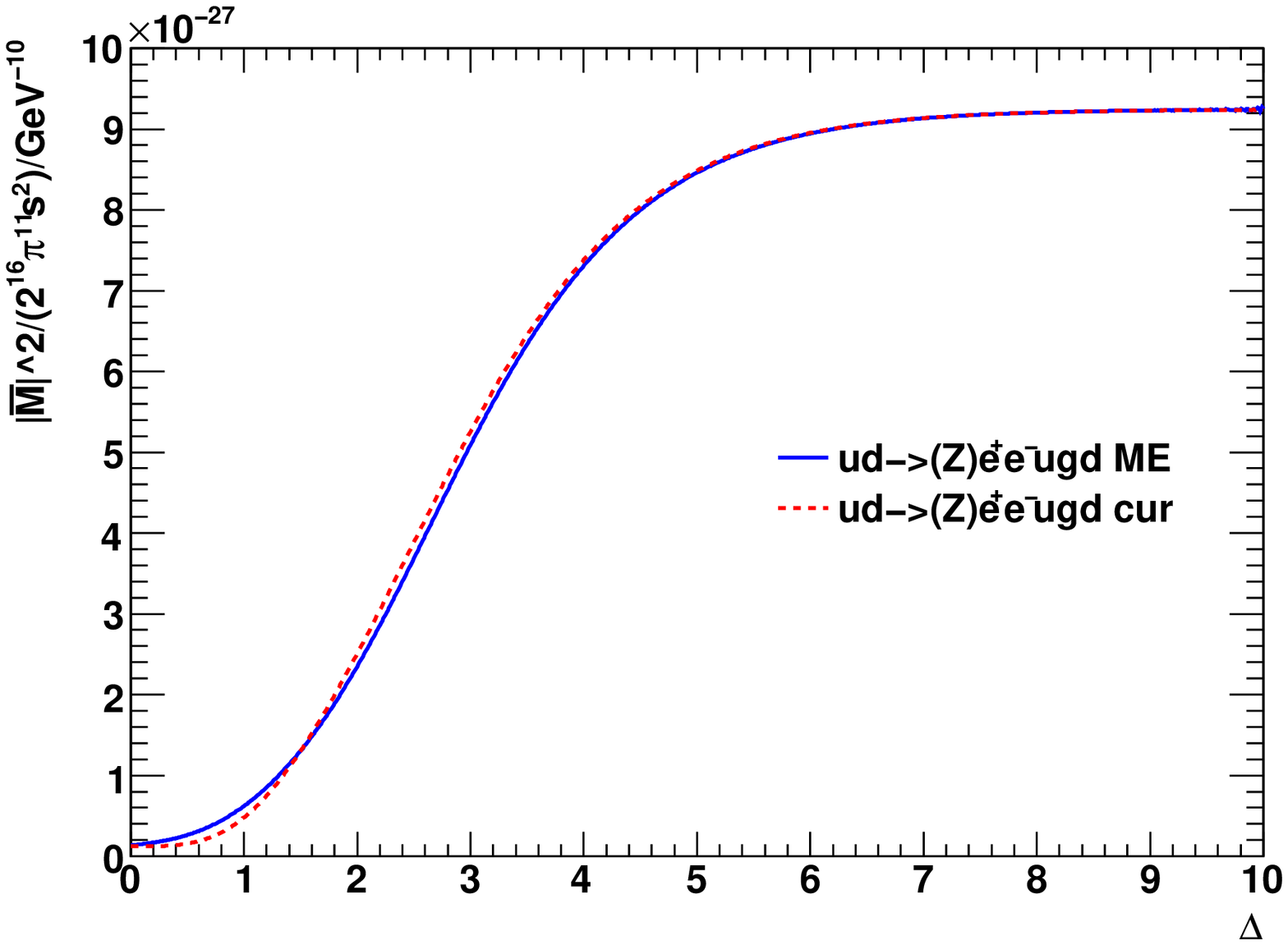}
  \epsfig{width=0.49\textwidth,file=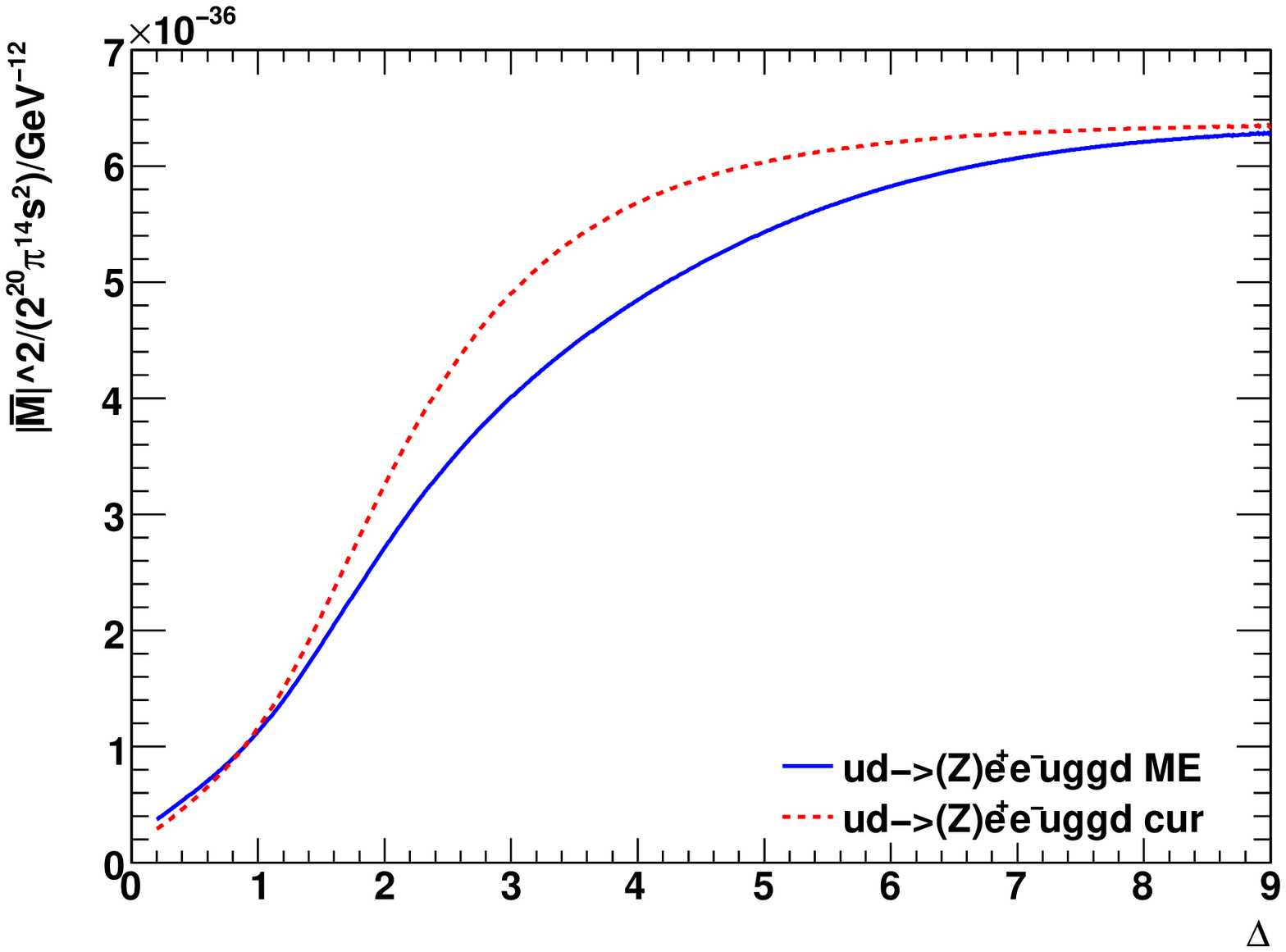}

  (a) \hspace{7.2cm}(b)\hspace{0.1cm}
  \caption{A comparison of the matrix elements obtained from Madgraph and from our
    formalism for (a) $Z+3j$ and (b) $Z+4j$ (Case 2).  The rapidities are as in
    Fig.~\ref{fig:WMEs}.  The full momentum configurations are in Appendix
    \ref{sec:moment-conf}.}
  \label{fig:ZMEs}
\end{figure}

Case 2 is very similar, except the Lipatov vertices must be included in the analogue of
equation \eqref{eq:case2M} before squaring.
\begin{figure}[tbp]
  \centering
  \input{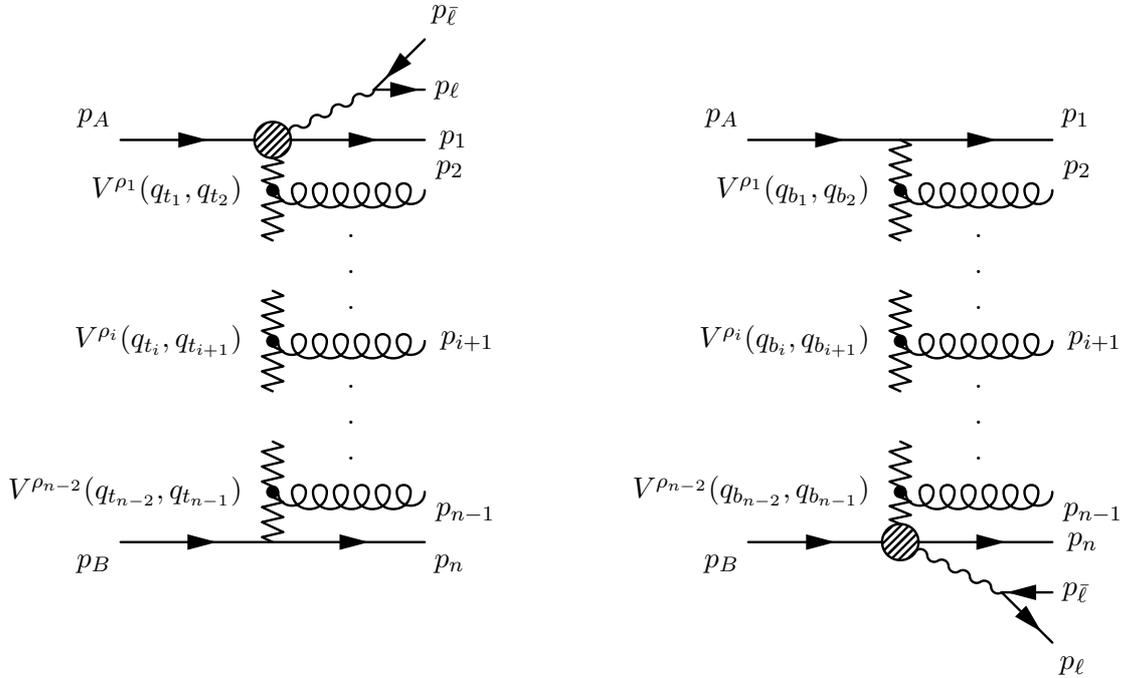}
  \caption{The two diagrams which are added at amplitude level for a Case 2 process with
    an $n$ jet final state.}
  \label{fig:Wplusn}
\end{figure}
For example, consider our sample $ud$ process with $n$ jets in the final state,
Fig.~\ref{fig:Wplusn}.  To make the (simple) structure clear, we introduce the shorthands
\begin{align}
  \label{eq:defa}
  a_t= J_{V\mu}(a,\ell,\bar \ell, 1)\ g^{\mu\nu}\ \langle 2\
    h_2|\nu| b\ h_b\rangle\qquad \mathrm{and}\qquad a_b = \langle 1\ h_1|\nu| a\ h_a\rangle\
    g^{\mu\nu} J_{V\nu}(b,\ell,\bar \ell,2)
\end{align}
for the current parts.  Then the $| \ldots |^2$ expression in equation \eqref{eq:case2M}
becomes
\begin{align}
  \label{eq:3jZ}
  \begin{split}
    & \left( k_u\ a_t \cdot \frac{V^{\rho_1}(q_{t_1},q_{t_2})\ldots
        V^{\rho_{n-2}}(q_{t_{n-2}},q_{t_{n-1}})}{t_{t_1}\ldots t_{t_{n-1}}} +\ k_d\
    a_b\cdot \frac{V^{\rho_1}(q_{b_1},q_{b_2})\ldots V^{\rho_{n-2}}}{t_{b_1}\ldots
      t_{b_{n-1}}} \right) \\ 
  & \quad \times (-g_{\rho_1 \sigma_1})\ldots (-g_{\rho_{n-2}\sigma_{n-2}})\\
  & \quad \times\left(  k_u\ a_t^* \cdot \frac{V^{\sigma_1}(q_{t_1},q_{t_2})\ldots
        V^{\sigma_{n-2}}(q_{t_{n-2}},q_{t_{n-1}})}{t_{t_1}\ldots t_{t_{n-1}}} +\ k_d\
    a_b^*\cdot \frac{V^{\sigma_1}(q_{b_1},q_{b_2})\ldots V^{\sigma_{n-2}}}{t_{b_1}\ldots
      t_{b_{n-1}}}\right) 
  \end{split} \nonumber \\ \\ \nonumber 
  \begin{split}
  =&\ (-1)^n \left( k_u^2|a_t|^2 \frac{V(q_{t_1},q_{t_2})^2\ldots
    V(q_{t_{n-2}},q_{t_{n-1}})^2}{t_{t_1}^2\ldots  t_{t_{n-1}}^2} +k_d^2|a_b|^2
  \frac{V(q_{b_1},q_{b_2})^2\ldots V(q_{b_{n-2}},q_{b_{n-1}})^2}{t_{b_1}^2\ldots
    t_{b_{n-1}}^2} \right. \\ 
  &\qquad \qquad \left. + 2k_uk_d(a_t a_b^*+a_t^* a_b)
  \frac{V(q_{t_1},q_{t_2}) \cdot V(q_{b_1},q_{b_2})\ldots V(q_{t_{n-2}},q_{t_{n-1}})\cdot
    V(q_{b_{n-2}},q_{b_{n-1}})}{t_{t_1}\ldots t_{t_{n-1}}\ t_{b_1}\ldots t_{b_{n-1}}} \right),
  \end{split}
\end{align}
where $t_{ti}$ and $t_{bi}$ are the analogues of equation \eqref{eq:ti} for top
and bottom line emission respectively.  What is particularly appealing about this form is
that there are only ever 3 terms, independent of the number of final jets.  The results
are shown in Fig.~\ref{fig:ZMEs} for $Z+3j$ and $Z+4j$.

% \subsection{Z Boson Production in Association with Jets}
% \label{sec:z-boson-production}

%%% Local Variables: 
%%% mode: latex
%%% TeX-master: "jetcurrents"
%%% End: 
 
\subsection{Higgs Boson Production in Association with Jets}
\label{sec:higgs-boson-prod}
In this study we will consider just the production of a Higgs boson
in-between (in rapidity) jets, since this will nicely illustrate the
increased predictive power of the factorised current formalism over the
formalism based on kinematic limits of amplitudes. The production of a Higgs
boson ``outside'' jets could be fully included by obtaining results for
``Higgs+gluon''-impact factors (see Ref.\cite{DelDuca:2003ba} for details)
also within the current formalism.

Let us start by considering the scattering $q^-Q^-\to q^- Q^- H$ with the
scattering matrix element
\begin{align}
  \label{eq:MqmQmqmQmH}
  M_{q^-Q^-\to q^-Q^-H} = \spab 1.\mu.a\ \frac {g^{\mu\sigma_1}} {q_1^2}\
  V^H_{\sigma_1\sigma_2}(q_1,q_2)\ \frac {g^{\sigma_2\nu}} {q_2^2}\ \spab2.\nu.b,
\end{align}
where the effective coupling of two off-shell gluons to the Higgs boson field through a
top-quark triangle is described by the vertex obtained in the infinite
top-mass limit (please note however that the picture of factorised
amplitudes does not rely on taking the infinite top-mass limit)
\begin{align}
  \label{eq:Vh}
  V_H^{\sigma_1\sigma_2}(q_1,q_2)=g^{\sigma_1\sigma_2}\ q_1.q_2 - q_1^{\sigma_2}q_2^{\sigma_1}.
\end{align}
Here, $q_1=p_a-p_1$, $q_2=q_1-p_H$. Again, we will use the sum of all allowed
helicity configurations as the model for all partonic channels. Let us define
the spinor/Lorentz string
\begin{align}
  \label{eq:Higgsjetsspinorstring}
  S^{h_ah_b\to h_1h_2}_{qQ\to qHQ}(q_1,q_2)\ =\ \langle 1\ h_1|\mu|a\ h_a\rangle\
  g^{\mu\sigma_1}\ V^H_{\sigma_1\sigma_2}(q_1,q_2)\ {g^{\sigma_2\nu}}\
  \langle 2\ h_2|\nu| b\ h_b\rangle.
\end{align}
Setting again
\begin{align}
  \label{eq:Higgsjetsspinorstringsq}
  \|S_{qQ\to qHQ}(q_1,q_2)\|^2=\sum_{h_a,h_a,h_b,h_2} \left |S^{h_ah_b\to h_1h_2}_{qQ\to qHQ}(q_1,q_2)\right|^2,
\end{align}
the helicity and colour summed and averaged square of the matrix element is
given by
\begin{align}
  \begin{split}
  \label{eq:qhQME2}
  \left|\overline{\mathcal{M}}^t_{qQ\to qHQ}\right|^2\ =\ &\frac 1 {4\
      (\Nc^2-1)}\ \left\|S_{qQ\to qHQ}(q_1,q_2)\right\|^2\\
    &\cdot\ \left(g^2\ \cf\ \frac 1 {t_1}\right)\\
    &\cdot\ \left(\frac 1 {t_1}\ \left(\frac{\alpha_s}{6\ \pi\ v} \right)^2\
      \frac 1 {t_2}\right)\\
    &\cdot\ \left(g^2\ \cf\ \frac 1 {t_2}\right).
  \end{split}
\end{align}
Compared to the $2\to2$ current scattering of Eq.~\eqref{eq:spinorstring}, an
extra Lorentz-tensor has been inserted in the spinor/Lorentz structure,
contracted with the external currents
(Eq.~\eqref{eq:Higgsjetsspinorstring}). In the square of the amplitude, an
extra inverse power of the square of the momentum either side of the Higgs
boson vertex has appeared. This is the same situation as with the emission of
a gluon by use of the effective vertex Eq.~\eqref{eq:EmissionV}. Therefore,
the approximation for Higgs production in association with three jets will be
based on the physical picture of (high energy) current-current scattering
with rapidity ordered $t$-channel couplings of the effective gluon emission
vertex (Eq.~\eqref{eq:EmissionV}) and the $g^*g^*H$-vertex of
Eq.~\eqref{eq:Vh}, in line with the four-jet formalism. Therefore, for the
multi-gluon emissions, we will take $q_1,q_2$ in the argument of the Higgs
boson vertex to be the $t$-channel momentum of the propagators either side of
the Higgs boson-vertex.

The approximation for the colour and helicity summed and averaged square of
the matrix element for the rapidity ordering implied by the subscript
\begin{align}
  \begin{split}
    \label{eq:hjjjsumaveraged}
    \left|\overline{\mathcal{M}}^{t}_{qQ\to qHgQ}\right|^2\ =\ &\frac 1 {4\
      (\Nc^2-1)}\ \left\|S_{qQ\to qHQ}(q_1,q_2)\right\|^2\\
    &\cdot\ \left(g^2\ \cf\ \frac 1 {t_1}\right)\\
    &\cdot\ \left(\frac 1 {t_1}\ \left(\frac{\alpha_s}{6\ \pi\ v} \right)^2\
      \frac 1 {t_2}\right)\\
    &\cdot\ \left(g^2\ \ca\ \frac 1 {t_2}\ \left(-V.V\right)\ \frac 1 {t_3}\right)\\
    &\cdot\ \left(g^2\ \cf\ \frac 1 {t_3}\right),
  \end{split}
\end{align}
where $V^{\mu}=V^\mu(q_2,q_3)$ of Eq.~\eqref{eq:EmissionV}, $q_1=p_a-p_1,
q_2=q_1-p_H, q_3=q_2-p_2$, and $t_i=q_i^2$, with the momentum of the quark
currents $(p_a,p_1)$ and $(p_b,p_3)$ respectively. Once again, the factorised
formalism for the scattering is extremely simple, and the generalisation to
multiple emissions is straightforward.  Again, the approximations for the
gluon channels differ only by colour factors (multiplication by $\ca/\cf$ for
each parton line replaced). 

\begin{figure}[tbp]
  \centering
  \epsfig{width=0.5\textwidth,file=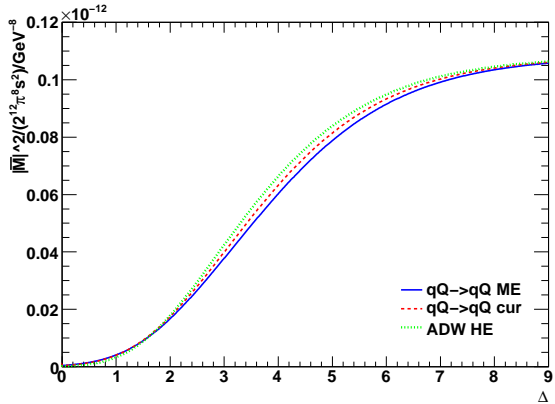}
  \caption{The full matrix element for $qQ\to qHgQ$, the result of
    Eq.~\eqref{eq:hjjjsumaveraged} (``cur'') and the approximation
    of~\cite{Andersen:2008ue,Andersen:2008gc} (``ADW HE'').} 
  \label{fig:HME}
\end{figure}

We show in Fig.~\ref{fig:HME} the results of this formalism compared to the full matrix
element.  We also show the approximation of Ref.~\cite{Andersen:2008ue,Andersen:2008gc}.
We will in Section~\ref{sec:higgsjetsrevisited}
investigate the degree to which this formalism captures the physics of the
full scattering amplitude.

%%% Local Variables: 
%%% mode: latex
%%% TeX-master: "jetcurrents"
%%% End: 

\subsection{Virtual Corrections and Regularisation}
\label{sec:virt-corr-regul}
So far, we have discussed how the real emission is approximated to all
orders. In this subsection we will discuss how the MRK limit of the virtual
corrections can be implemented according to the \emph{Lipatov Ansatz}. This
will facilitate the construction of a regularised, all-order scattering
matrix element for each $n$-parton exclusive final state. We emphasise that:-
a) in the current study, we construct scattering matrix elements only for
rapidity orderings which allow colour octet-exchanges between each pair of
rapidity ordered particles, b) the contribution from other rapidity orderings
are systematically suppressed by powers of invariant masses; some of these
configurations would arise when considering the next-to-leading
contributions to the factorisation, but can still be included in a
$t$-channel effective description\cite{Bogdan:2006af}. The degree to which
the leading flavour assignments dominate the $n$-jet cross sections was
discussed in e.g.~Ref.\cite{Andersen:2008ue,Andersen:2008gc} and are repeated
in Section~\ref{sec:higgsjetsrevisited}. We leave the
treatment of the sub-leading configurations to a later study.

The \emph{Lipatov Ansatz} states that the order by order, the virtual
corrections to the full $n$-parton scattering amplitude in the MRK limit can
be obtained by replacing in the scattering amplitudes
\begin{align}
  \label{eq:LipatovAnsatz}
  \frac 1 {t_i}\ \to\ \frac 1 {t_i}\ \exp\left[\hat \alpha (q_i)(y_{i-1}-y_i)\right]
\end{align}
with
\begin{align}
\hat{\alpha}(q_i)&=\alpha_s\ \ca\
t_i\int\frac{d^2k_\perp}{(2\pi)^2}\frac{1}{k_\perp^2(q_i-k)_\perp^2}\label{eq:alpha}
\\
&\to-\gs^2\ \ca\ \frac{\Gamma(1-\varepsilon)}{(4\pi)^{2+\varepsilon}}\frac
  2 \varepsilon\left({\bf q}^2/\mu^2\right)^\varepsilon.
\label{eq:ahatdimreg}
\end{align}
This ansatz for the exponentiation of the virtual corrections in the
appropriate limit of the $n$-parton scattering amplitude has been proved to
even the sub-leading
level\cite{Bogdan:2006af,Fadin:2006bj,Fadin:2003xs,Fadin:2005pt}.

Our discussion of the regularisation of the soft divergences will follow
closely the discussion in Ref.\cite{Andersen:2008gc}. We will show that order
by order in $\alpha_s$, the soft divergence for the emission of gluons
cancels with the soft divergence from the virtual corrections implemented
according to the Lipatov Ansatz for the resummed $t$-channel
propagator. Consider the limit where the transverse momentum of the $i$'th
emitted gluon is vanishing. In this limit,
\begin{align}
  \label{eq:kisoftlimit}
  \left|\overline{\mathcal{M}}^t_{p_a\ p_b\to p_1\ \cdots\ p_{i-1}\ p_i\ p_{i+1}\ \cdots\ 
    p_n}\right|^2\ \stackrel{{{\bf p}_{i}}^2\to 0}{\longrightarrow}\
\left(\frac{4\ \gs^2\ \ca}{{{\bf p}_{i}}^2}\right)
  \left|\overline{\mathcal{M}}^t_{p_a\ p_b\ \to\ p_1\ \cdots\ p_{i-1}\
      p_{i+1}\ \cdots\ p_n}\right|^2,
\end{align}
where the matrix element on the RHS has $n-1$ final state particles, and
${\bf p}_i^2$ is the sum of the squares of the transverse components of $p_i$
in the Euclidean metric. By integrating over the soft region ${\bf
  p}_i^2<\lambda^2$ of phase space in $D=4+2\varepsilon$ dimensions we find
\begin{align}
  \begin{split}
    \label{eq:realdiv1}
    \int_0^\lambda& \frac{\mathrm{d}^{2+2\varepsilon}{\bf p}\
      \mathrm{d}y_i}{(2\pi)^{2+2\varepsilon}\ 4\pi} \left(\frac{4 \gs^2
        \ca}{{\bf p}^2}\right)\mu^{-2\varepsilon}\\
    &=\frac{4\gs^2\ca}{(2\pi)^{2+2\varepsilon}4\pi}\Delta y_{{i-1},{i+1}}
    \frac{\pi^{1+\varepsilon}}{\Gamma(1+\varepsilon)} \frac 1 \varepsilon (\lambda^2/\mu^2)^\varepsilon.
  \end{split}
\end{align}

The square of the matrix element on the left hand side of
Eq.~(\ref{eq:kisoftlimit}) contains the exponential $\exp(2\alpha(q_i)\Delta
y_{{i-1},{i+1}})$. By expanding the exponential to first order in \as and in
$\varepsilon$, the resulting pole in $\varepsilon$ does indeed cancel that of
Eq.~(\ref{eq:realdiv1}), and the combined effect of one soft real emission
and the first term in the expansion of the Reggeised propagator is a factor
\begin{align}
  \label{eq:virtualexponent}
  \Delta y_{{i-1},{i+1}}\frac{\as\Nc}{\pi} \ln\left(\frac{\lambda^2}{{\bf q}^2}\right)
\end{align}
multiplying the $(n-1)$-particle matrix element.

It is clear that the nested rapidity integrals of additional soft radiation in the
$t$-channel factorised multi-parton amplitudes will build up the exponential needed to
cancel the poles from the virtual corrections to all orders in \as. The divergence arising
from a given real emission is therefore cancelled by that arising from the virtual
corrections in the Reggeised $t$-channel propagator of the matrix element without the
relevant real emission. Furthermore, the organisation of the cancellation of infra-red
poles can be achieved with a simple phase-space slicing. Since the $t$-channel factorised
matrix elements are very fast to evaluate and the regularisation procedure does not add
any complexity (because of the simple IR structure of the $t$-channel factorised matrix
elements), the radiative corrections to all orders can be constructed as an explicit phase
space integral over each number of resolved ($|{\bf k}_{i}|^2>\lambda^2, i=2,\ldots n-1$)
gluons emitted.  The cancellation of the poles in $\varepsilon$ ensures that the
logarithmic dependence on $\lambda$ generated by the lower limit on the transverse
momentum integrals cancels with the logarithmic $\lambda$-dependence of the virtual +
unresolved-real correction. This is similar to the explicit construction of the solution
to the BFKL evolution, where the very weak dependence of the solution on $\lambda$ at
leading logarithmic accuracy was studied in Ref.~\cite{Schmidt:1996fg,Orr:1997im}, and in
Ref.~\cite{Andersen:2003wy} at next-to-leading logarithmic accuracy.

The construction of an explicit integration over emissions to all orders relies on an
efficient phase-space generator~\cite{Andersen:2008gc,Andersen:2008ue}, which should
sample final states with the number of particles varying by more than an order of
magnitude. The problem is significantly different to that of a so-called general purpose
Monte Carlo (Pythia\cite{Sjostrand:2007gs}, Herwig\cite{Bahr:2008pv},
SHERPA\cite{Gleisberg:2008ta}) where also the number of final state particles varies,
since in these the approximation to the virtual corrections is \emph{defined} such that
the emission of particles is \emph{unitary}, i.e.~does not change the total cross section,
which allows for a simple probabilistic interpretation. In the problem at hand, (an
approximation to) the virtual corrections are also calculated, and introduce a
\emph{suppression} of the regularised matrix element for any final state with a finite
number of partons as the rapidity length of the event is increased. This is countered by
the (positive) contribution from the emission of additional gluons, and introduces a
correlation between the number of final state partons and the typical rapidity length of
an event, which depends also on the transverse momenta. This probabilistic correlation is
absolutely crucial to incorporate in the phase space generator in order to obtain
numerical stability in finite amount of time. Such a phase space integrator can be
efficiently implemented by following the ideas of Ref.\cite{Andersen:2006sp}.

%%% Local Variables: 
%%% mode: latex
%%% TeX-master: "jetcurrents"
%%% End: 

\section{Applications}
\label{sec:applications}
The framework developed here is only useful if it is relevant to the current
generation of colliders.  In this section, we compare the results of using
the $t$-channel factorised matrix elements order by order with those obtained
with full tree-level matrix elements (extracted from
MadGraph~\cite{Alwall:2007st}) for a 10~TeV proton-proton-collider. We will
compare both rapidity and transverse momentum distributions for a
representative set of processes with a minimal set of cuts. Since the only
difference between the two calculations presented is the evaluation of the
hard scattering matrix element, the choice of coupling or parton distribution
functions is largely irrelevant for the purpose of comparison and judging the
accuracy of the simple picture for higher order corrections. In the later
construction of the resummation, we envisage using a NLO set of
pdfs. Therefore, we choose also in this study to apply the MSTW2008NLO
set~\cite{Martin:2009iq} of pdfs, so that one can judge already now the level
of corrections which will arise from matching the resummation to fixed order
accuracy. We will use the $k_t$-algorithm as implemented in
Ref.\cite{Cacciari:2005hq} to define jets.

In the following, we will only integrate over final states in which
the partons are arranged in rapidity such that colour octets can be exchanged
between all rapidity-neighbours; this constraint is relevant only for some
partonic channels. Other contributions are systematically suppressed by the
invariant mass between the non-ordered partons, and the relevance of this
suppression was checked in an earlier publication\cite{Andersen:2008gc}. We
will repeat this analysis for the updated set of cuts in
Section~\ref{sec:higgsjetsrevisited}. The non-leading contributions can
also be included in a $t$-channel factorised formalism\cite{Bogdan:2006af},
but we have not yet done so, and so we will compare with the results from
full tree-level QCD only of the final states which we claim to reproduce.

We emphasise that the following results will form the basis of a resummation
procedure, in which the discrepancies between the full tree-level results and
those obtained using the approximation of a $t$-channel factorised scattering
can be corrected for order-by-order, where the full result is known. However,
the close resemblance between the results obtained order-by-order in the
simple picture (allowing all-order constructions) and with full QCD should
instill trust in the results which will be obtained from the resummation
procedure.

\subsection{Pure Jets}
\label{sec:pure-jets}
\begin{figure}[tbp]
  \centering
  \epsfig{width=0.49\textwidth,file=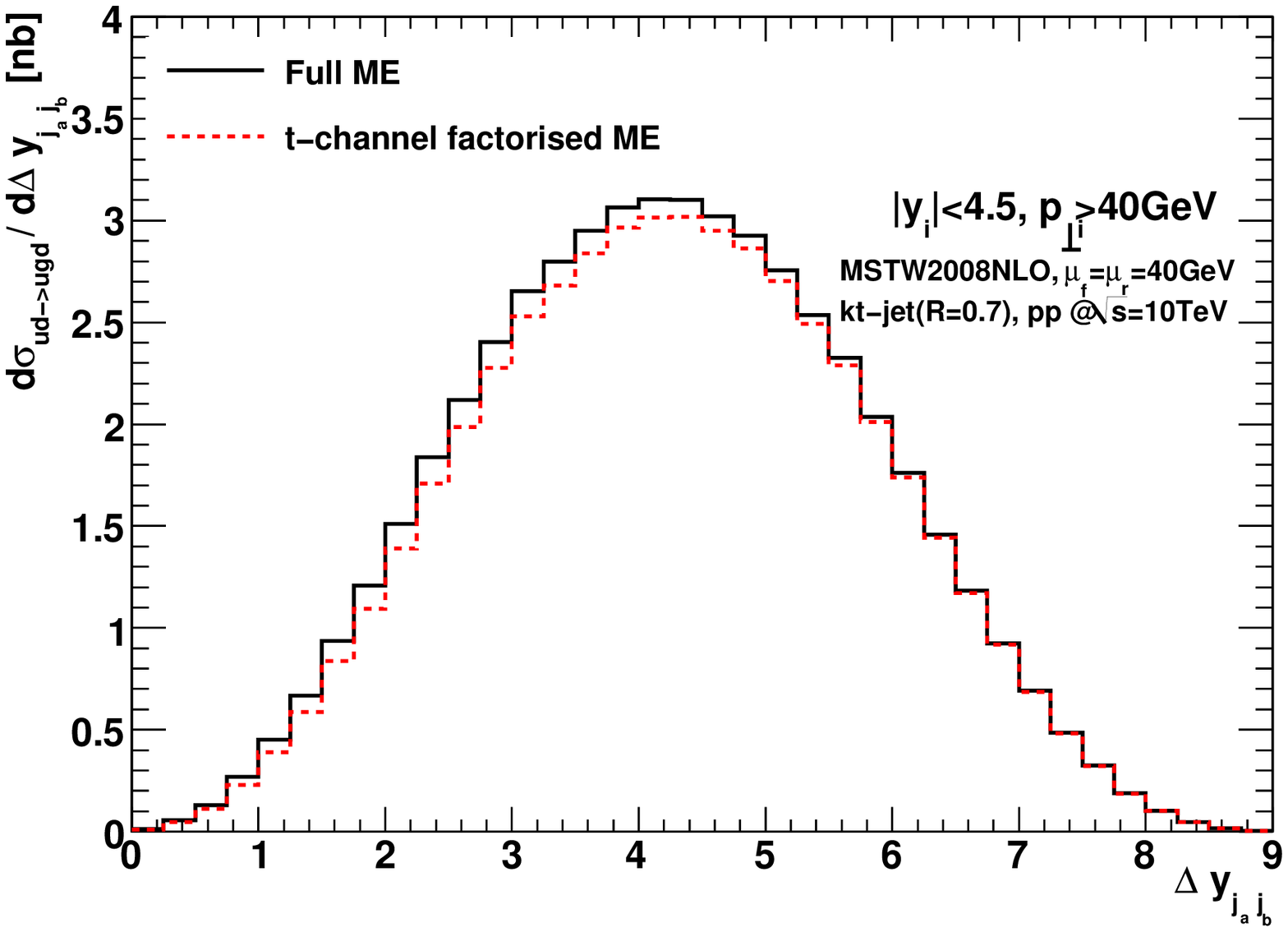}
  \epsfig{width=0.49\textwidth,file=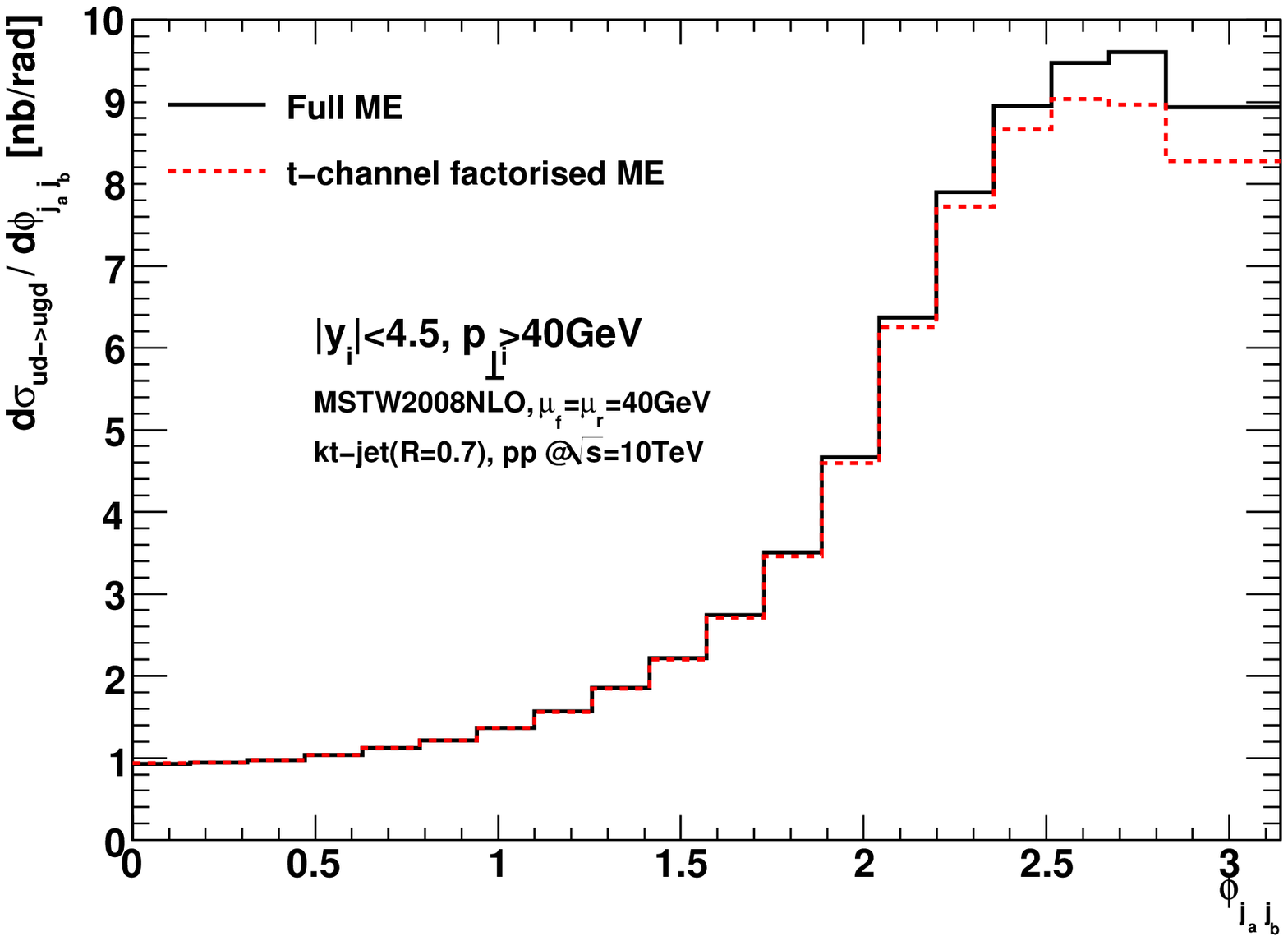}
  (a) \hspace{7.2cm}(b)\hspace{0.1cm}\\
   \epsfig{width=0.49\textwidth,file=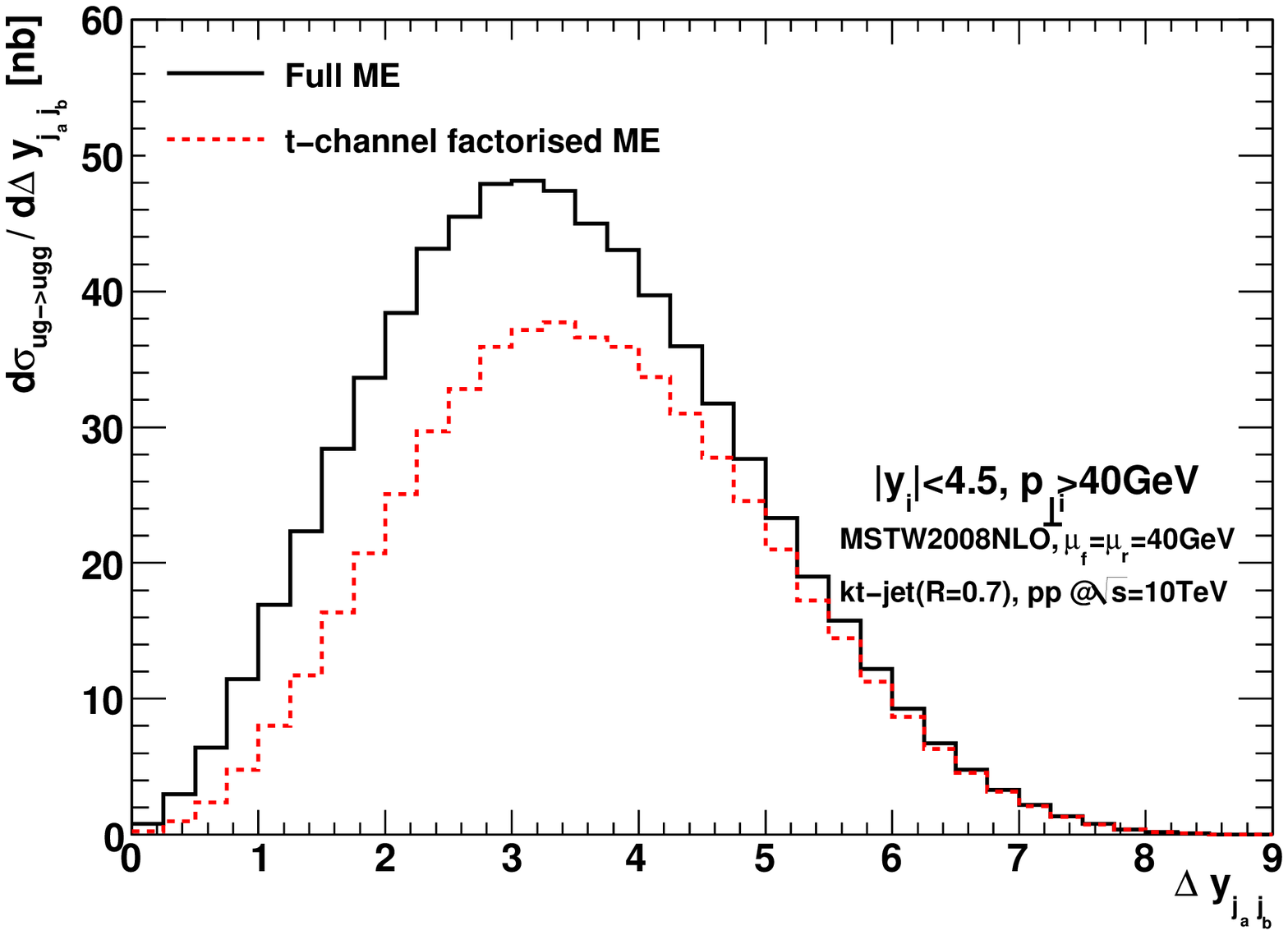}
   \epsfig{width=0.49\textwidth,file=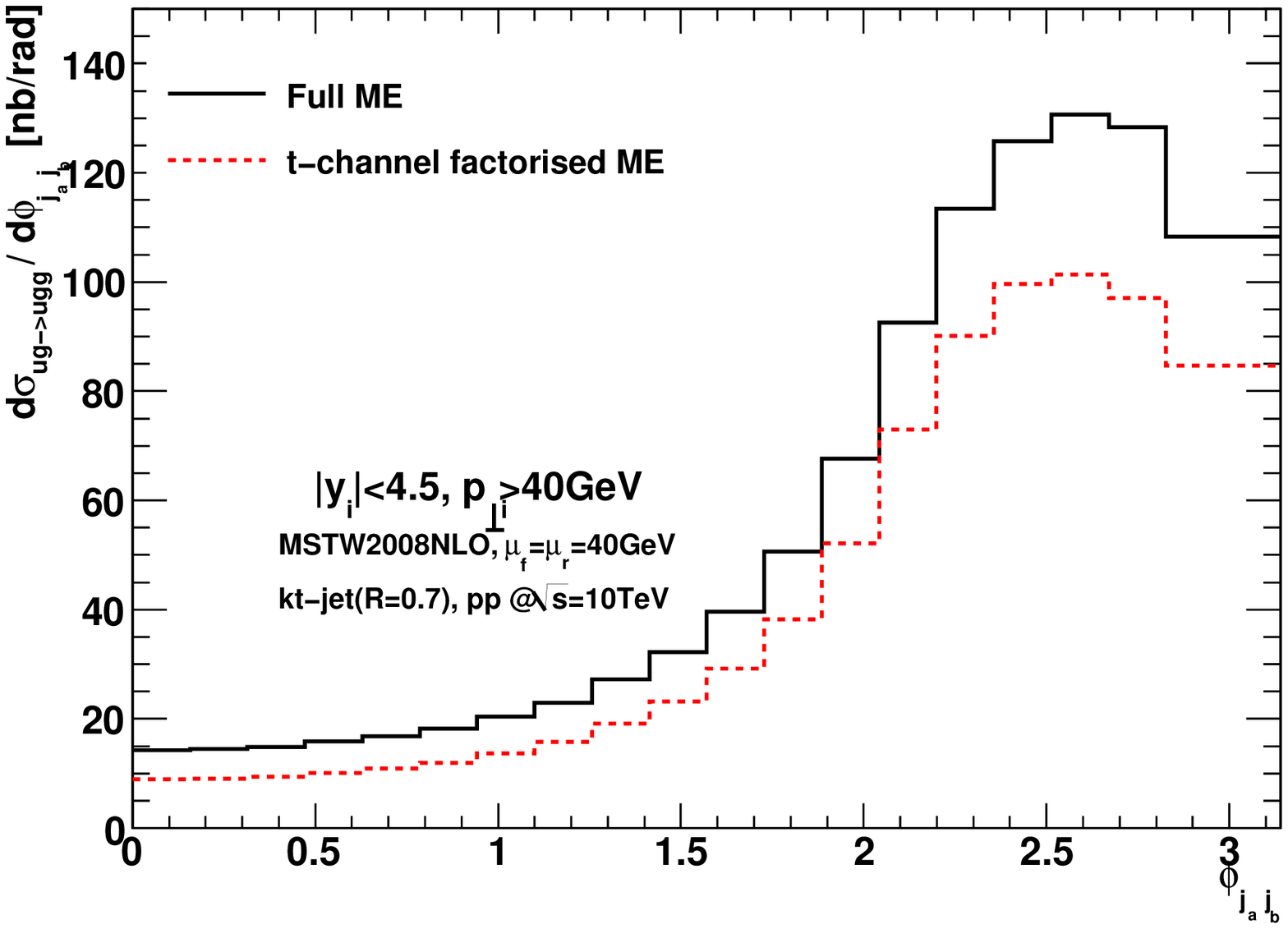}
  (c) \hspace{7.2cm}(d)\hspace{0.1cm}\\
   \epsfig{width=0.49\textwidth,file=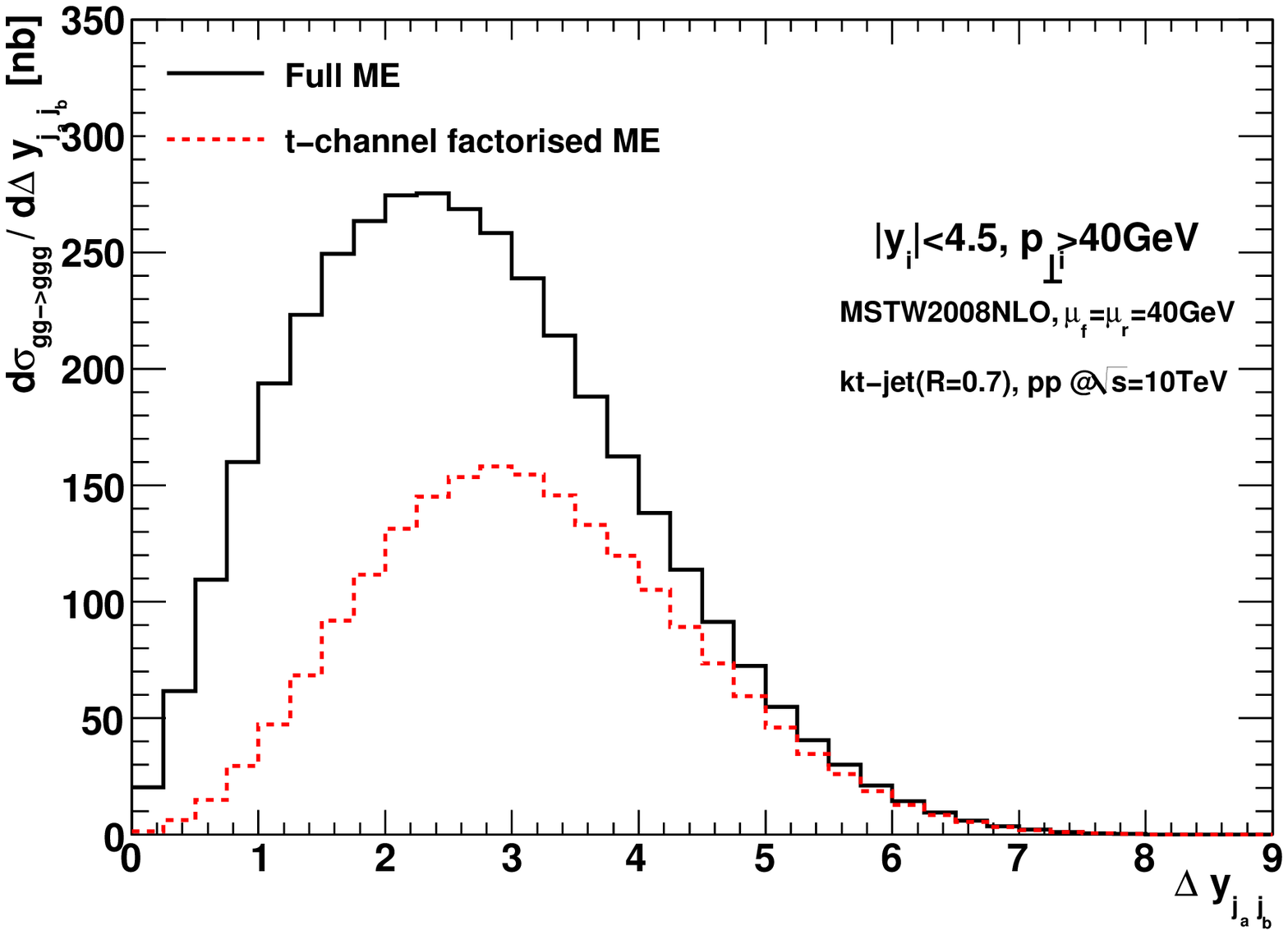}
   \epsfig{width=0.49\textwidth,file=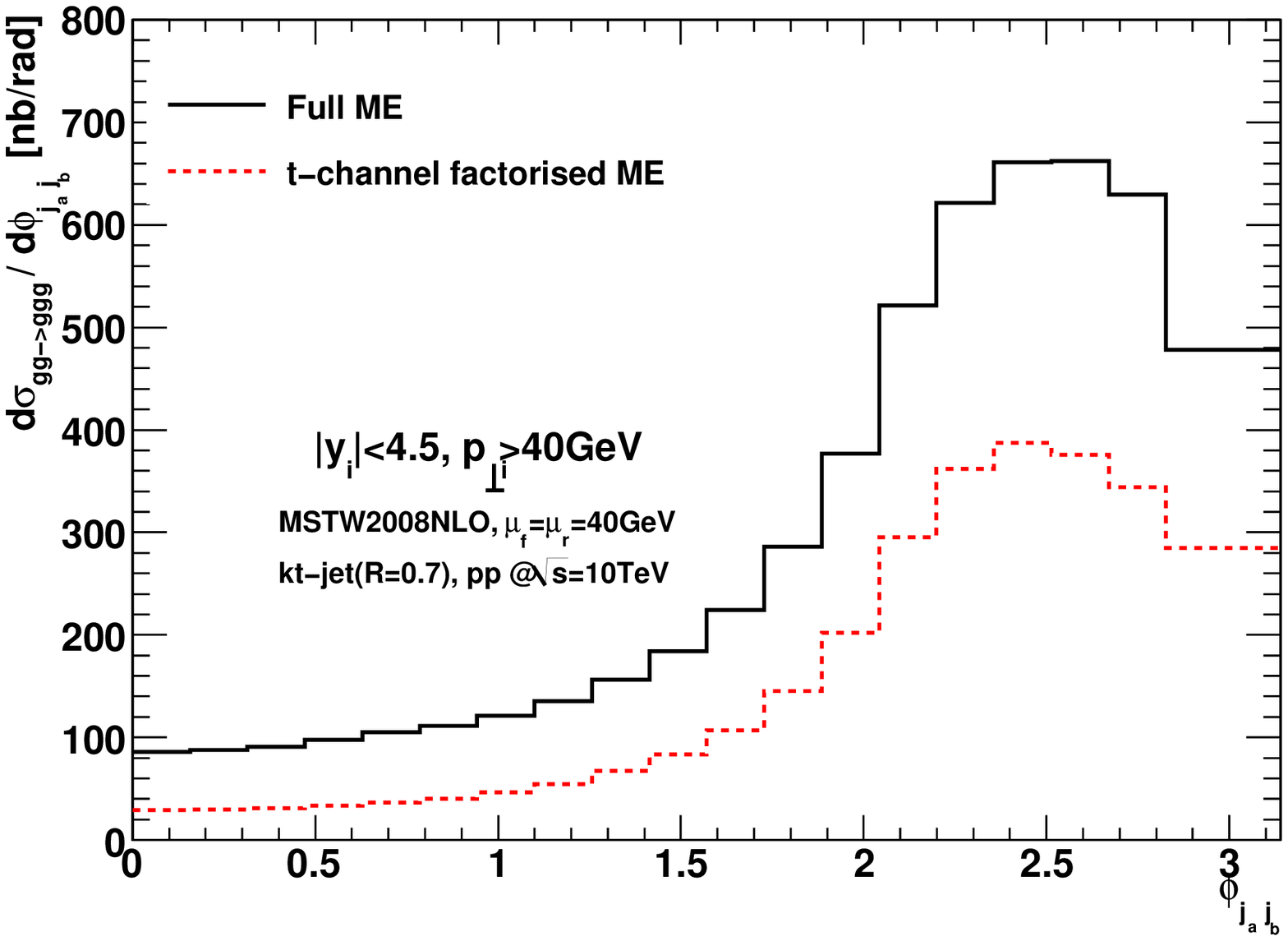}
  (e) \hspace{7.2cm}(f)\hspace{0.1cm}\\
  \caption{A comparison between the differential 3-jet cross section obtained
    using lowest order matrix elements obtained from Madgraph and the
    $t$-channel factorised formalism for the process (a)-(b) $ud\to ugd$ with the
    $u$-quark incoming in the positive $z$-direction. (a) the
    rapidity difference between the most forward and backward jet (b) the
    distribution on the azimuthal angle between the most forward and backward
    jet. Similarly for $ug\to ugg$ (c)-(d) and $gg\to ggg$ (e)-(f).}
  \label{fig:3j}
\end{figure}

Since the two-jet cross section is reproduced exactly for the pure quark
channels, we will present comparisons only for the three and four-jet
channels. We will use the following minimal set of cuts
\begin{center}
  \begin{tabular}{|rl||rl|}
    \hline
    $p_{j_\perp}$& $> 40$ GeV & $|y_j|$ & $<$ 4.5\\\hline
  \end{tabular}
\end{center}
In Fig.~\ref{fig:3j} we compare the differential cross section
with respect to the rapidity difference and the azimuthal angle between the
most forward and backward jet between the full leading order matrix element
and the approximation of this obtained in the $t$-channel factorised
approach. In the case of $ud$-scattering $(a)-(b)$, we note that because of the
valence quark contribution, the rapidity-distribution is peaked at rapidities
around 4, and a very good agreement between the result obtained within the
full tree-level and the simpler $t$-channel factorised approach; we emphasise
that no constraints on the minimal separation between jets have been
applied. Neither a constraint on the similarity of transverse scales in
accordance with Eq.~\eqref{eq:MRKlimit}.

In the $qQ$-channel, the $t$-channel factorised picture obviously works
particularly well across all rapidities. Also the azimuthal distribution is
described very well. Indeed, the (small) discrepancy is isolated to the
collinear region, where the forward and backward jet is back-to-back in
azimuth, and thus the third hard jet is radiated in the same direction as one
of the other jets. A discrepancy in this region is hardly surprising, since
the collinear singularity is not included in our $t$-channel factorised
description.
For the $gq$ and $gg$-channel, the peak in the rapidity distribution obtained
with the full LO matrix element is at ever decreasing rapidities; slightly
above 2 for the channel $gg\to ggg$. As expected, in these channels the
$t$-channel dominance is only reached for larger rapidities as compared to
the $ud$-channels. However, we emphasise that matching corrections can be
taken into account in the resummation procedure which will be built upon
these approximations, such that for smaller rapidities, where the first few
orders of the perturbative expansion is sufficient (i.e. the effects of the
resummation are small), full LO results will be used, and for larger
rapidities where the resummation effects are important, the approximations
fare better and so the approximations can be better trusted. This is achieved
following the matching procedure discussed in Ref.\cite{Andersen:2008gc}.

\begin{figure}[tbp]
  \centering
  \epsfig{width=0.49\textwidth,file=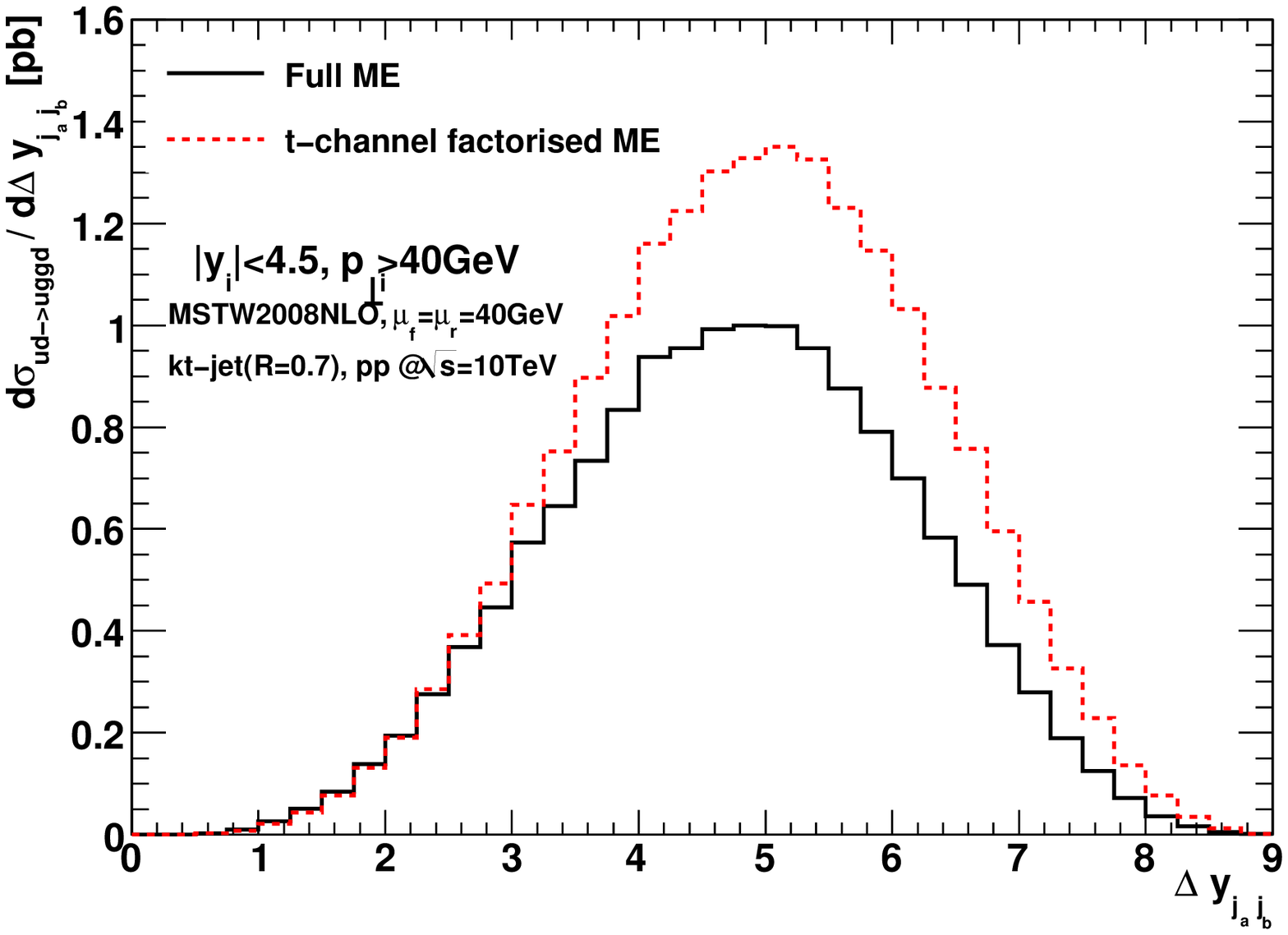}
  \epsfig{width=0.49\textwidth,file=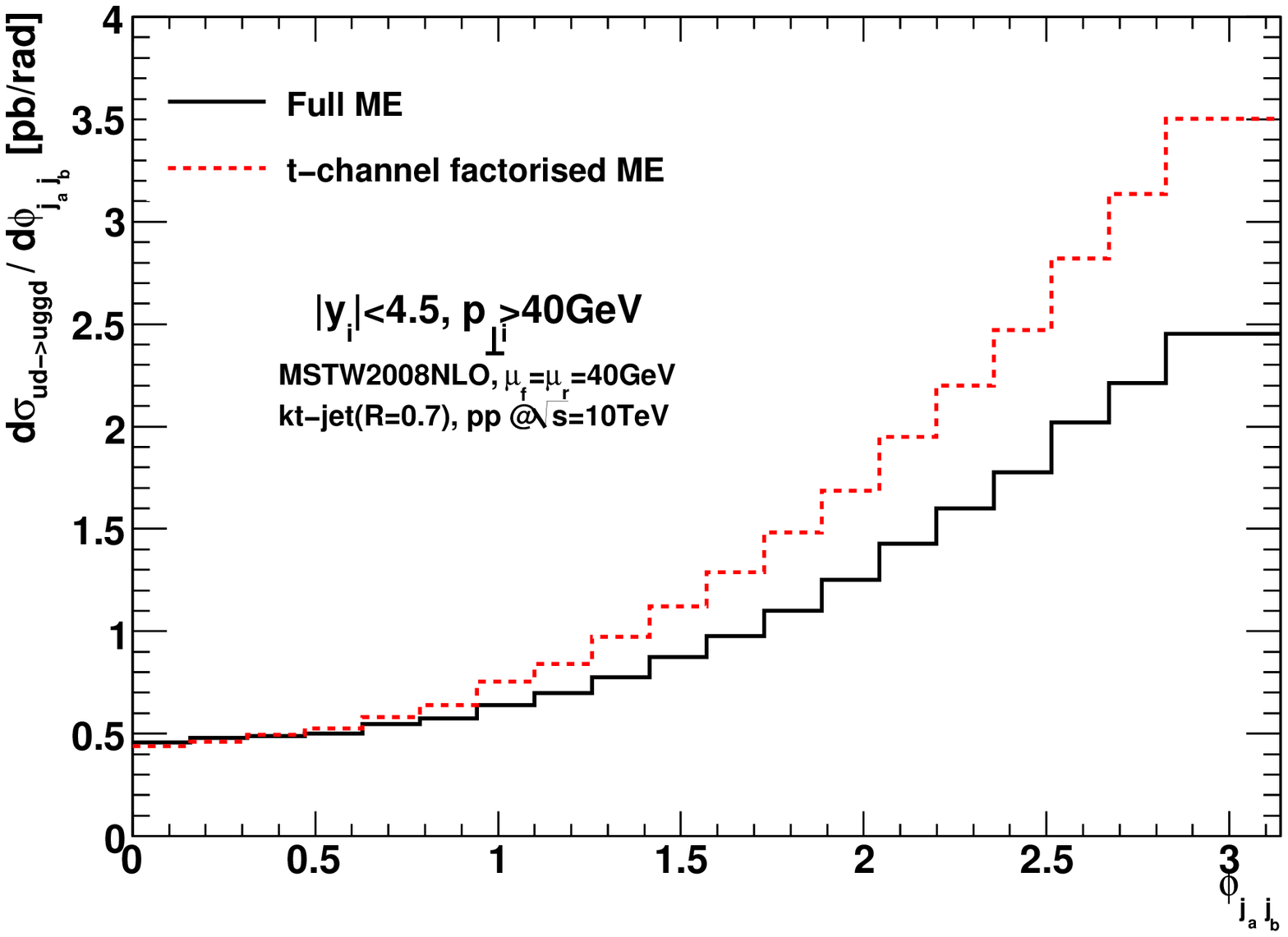}
  (a) \hspace{7.2cm}(b)\hspace{0.1cm}\\
   \epsfig{width=0.49\textwidth,file=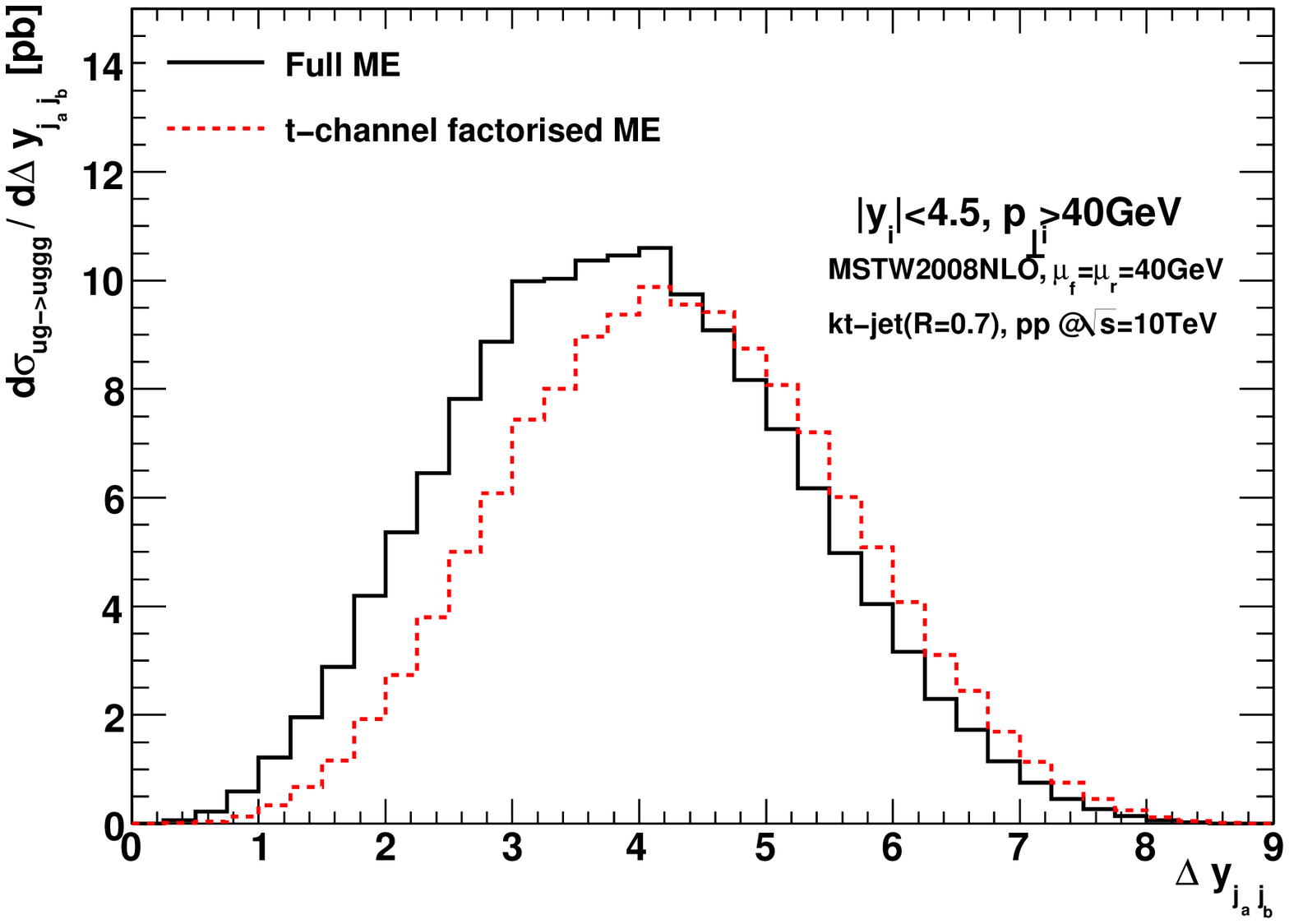}
   \epsfig{width=0.49\textwidth,file=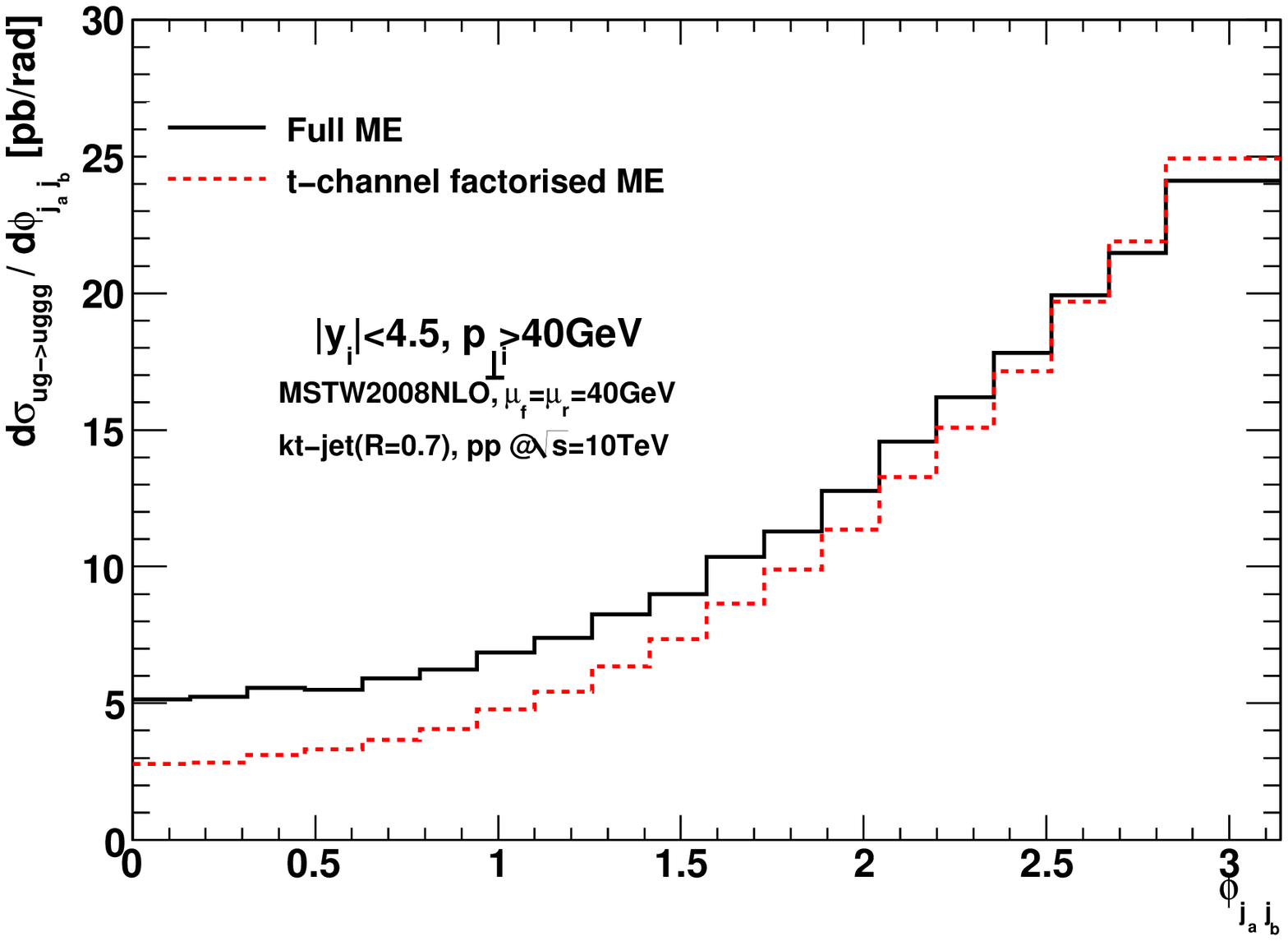}
  (c) \hspace{7.2cm}(d)\hspace{0.1cm}\\
   \epsfig{width=0.49\textwidth,file=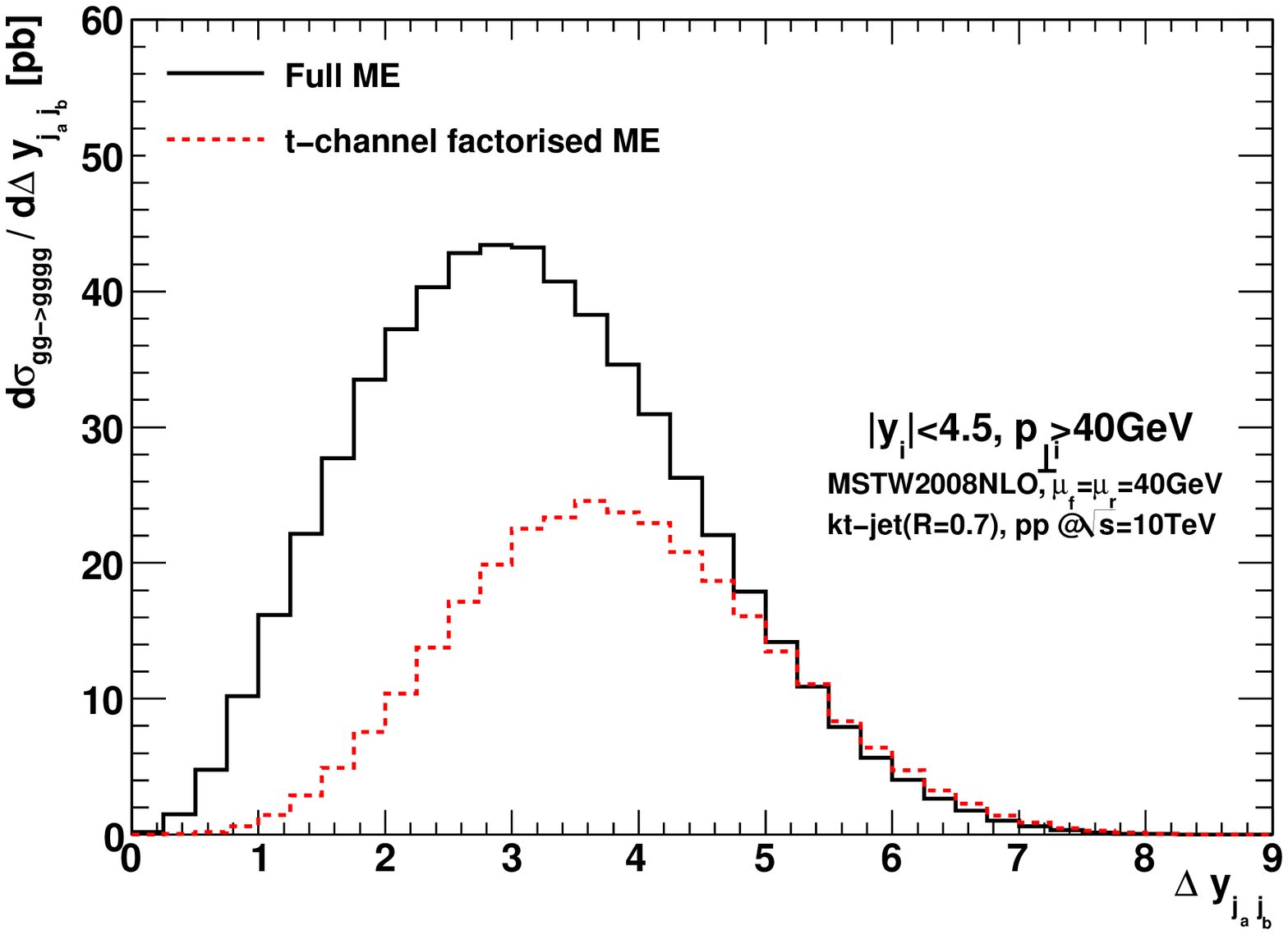}
   \epsfig{width=0.49\textwidth,file=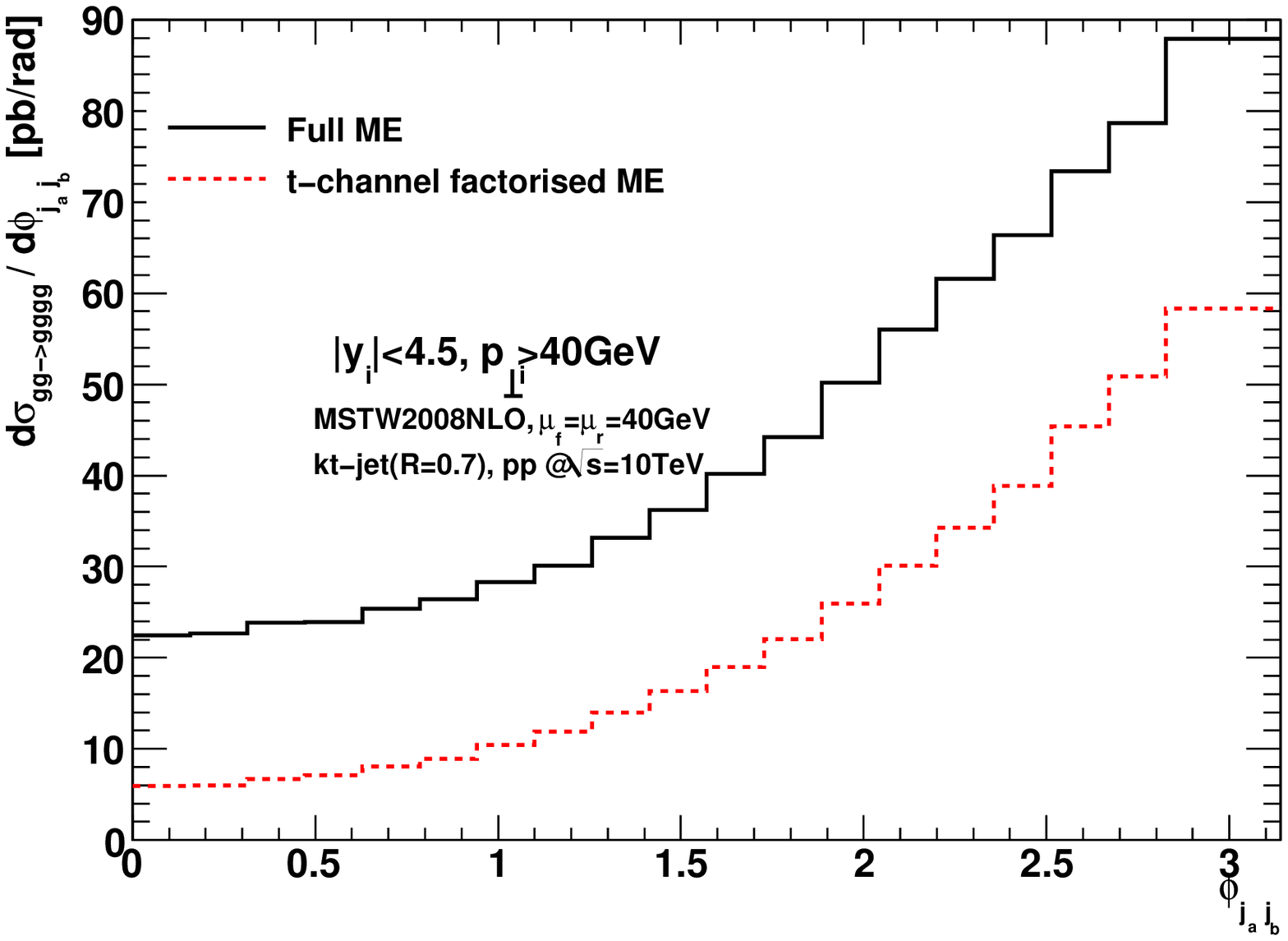}
  (e) \hspace{7.2cm}(f)\hspace{0.1cm}\\
  \caption{A comparison between the differential 4-jet cross section obtained
    using lowest order matrix elements obtained from Madgraph and the
    $t$-channel factorised formalism for the process (a)-(b) $ud\to uggd$ with the
    $u$-quark incoming in the positive $z$-direction. (a) the
    rapidity difference between the most forward and backward jet (b) the
    distribution on the azimuthal angle between the most forward and backward
    jet. Similarly for $ug\to uggg$ (c)-(d) and $gg\to gggg$ (e)-(f).}
  \label{fig:4j}
\end{figure}
In Fig.~\ref{fig:4j} we have plotted the similar distributions in the 4-jet cases. We
emphasise that these (and those in the subsequent sections) are not normalised
distributions - the total rates and distributions are really approximated by the simple
$t$-channel factorised framework as well as indicated in the figures. We note that for all
the channels, the peak in the rapidity distributions have moved to slightly larger
rapidities (roughly one unit) than in the equivalent three-jet channels. This is simply
because of two effects:- 1) the opening of phase space and 2) that the partonic cross
section reaches the asymptotic (and large) values at increasing rapidity spans for an
increasing number of final state particles (see Fig.~\ref{fig:M3j}). We also note that the
importance of the $qg$ channel compared with the $gg$-one increases with the jet count.

\subsection{W + Jets}
\label{sec:wjets}

We now compare the results for the production of a (leptonically decaying) $W$ boson in
association with three and four jets.  For brevity, we only show plots for the dominant $ug\to
e^+\nu_e dg(g)gg$ channel; the $t$-channel factorised matrix elements reproduce the channels
with an initial state of only quarks even better.

We apply the following set of cuts
\begin{center}
  \begin{tabular}{|rl||rl|}
    \hline
    $p_{j_\perp}$ & $> 40$~GeV &  $p_{e\perp}$& $> 20$~GeV\\
    $|y_j|$ & $<$ 4.5 &  $|y_e|$& $< 2.5$ \\
    &&$p_{\nu\perp}$ & $> 20$~GeV \\\hline
  \end{tabular}
\end{center}
The tree-level description of the channels $q\ Q\ \to\ q'\ Q\ (W\to)\ e\ \nu$
are reproduced exactly; this is an improvement over the earlier
approximations of Ref.\cite{Andersen:2001ja}, where kinematical limits of
also the $W$ (and its decay products) had to be applied in order to extract the relevant impact
factor. In particular, the $W$ had to be assumed to be produced in the forward
region, with its rapidity increasing with that of the emitting quark. This
assumption is in contradiction with the requirement of the charged lepton of
the $W$ decay products to be central for detection. The formalism developed
in Section~\ref{sec:w-boson-production} requires no kinematical constraints
to be placed on the $W$ or its decay products in order to extract the
building blocks for the resummation.

\begin{figure}[tbp]
  \centering
  \epsfig{width=0.49\textwidth,file=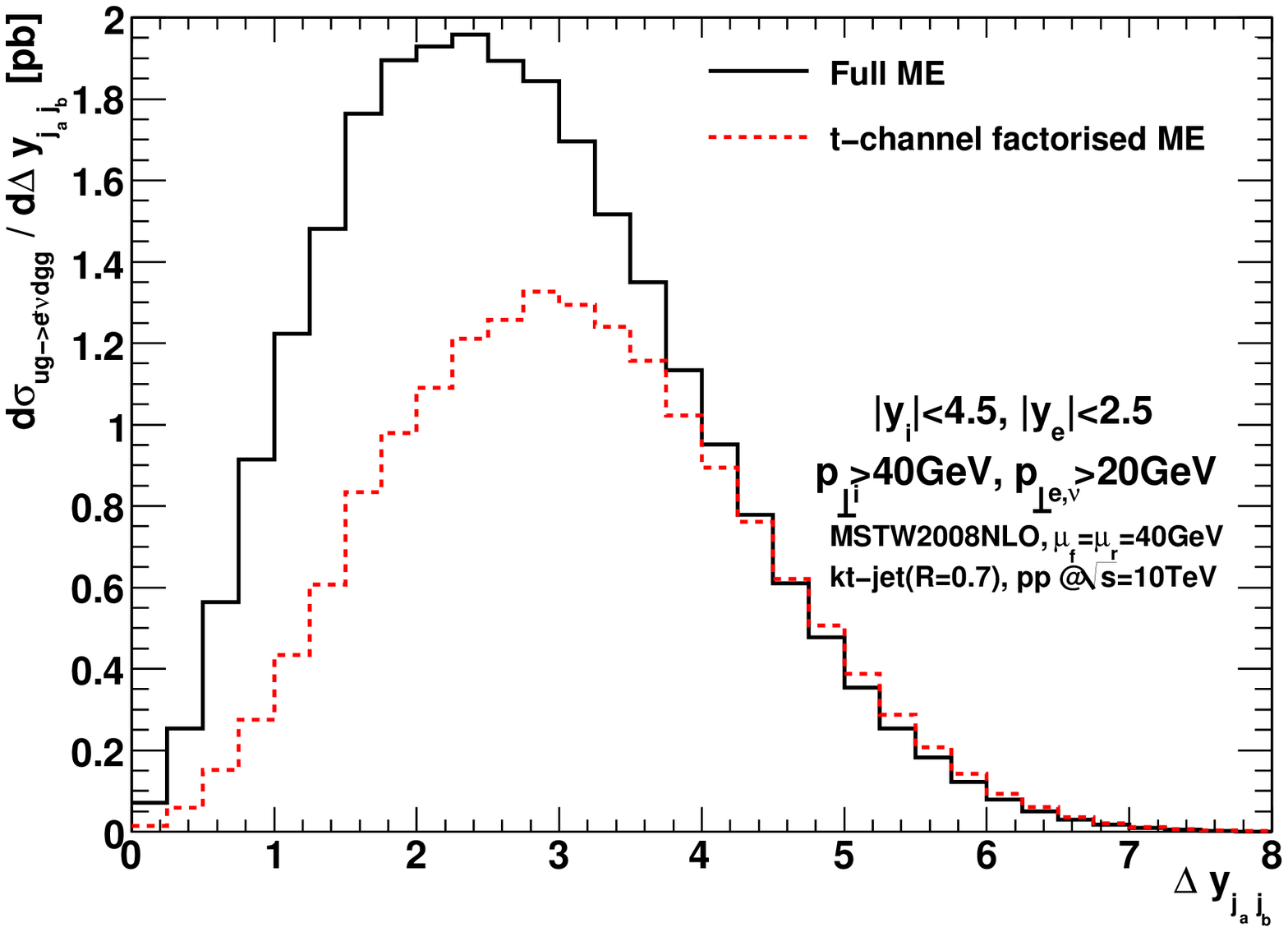}
  \epsfig{width=0.49\textwidth,file=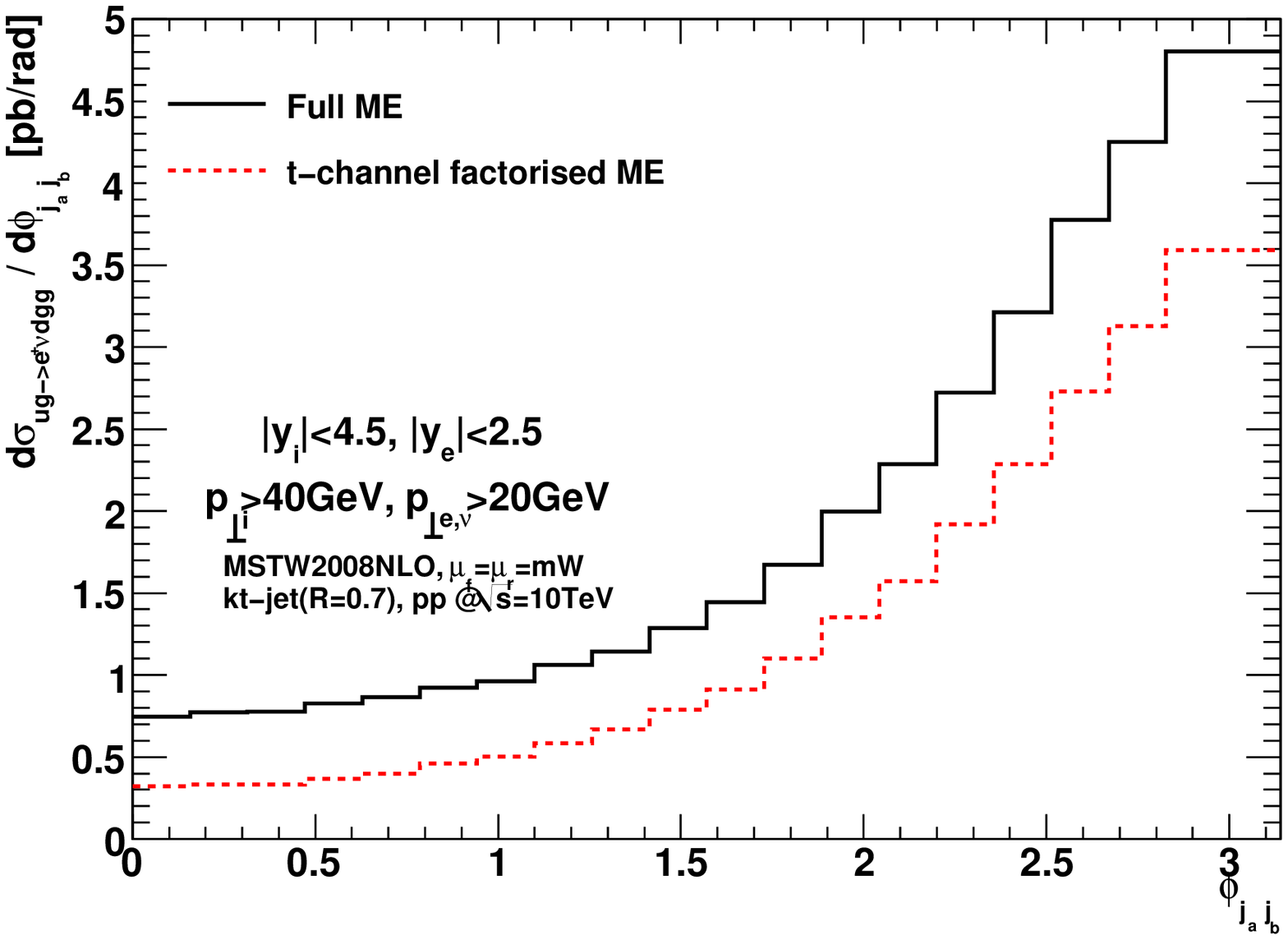}

  (a) \hspace{7.2cm}(b)\hspace{0.1cm}

  \epsfig{width=0.49\textwidth,file=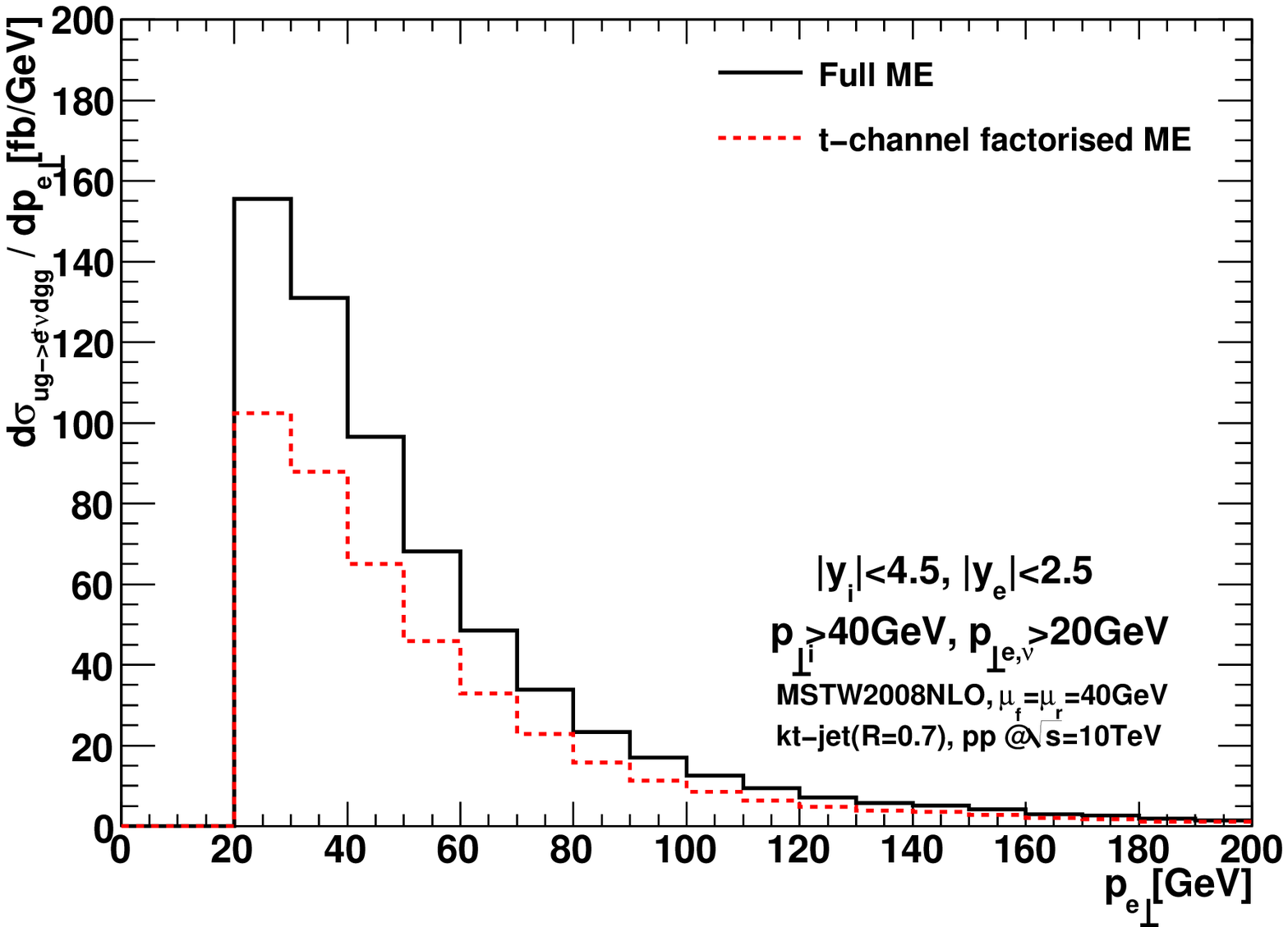}
  \epsfig{width=0.49\textwidth,file=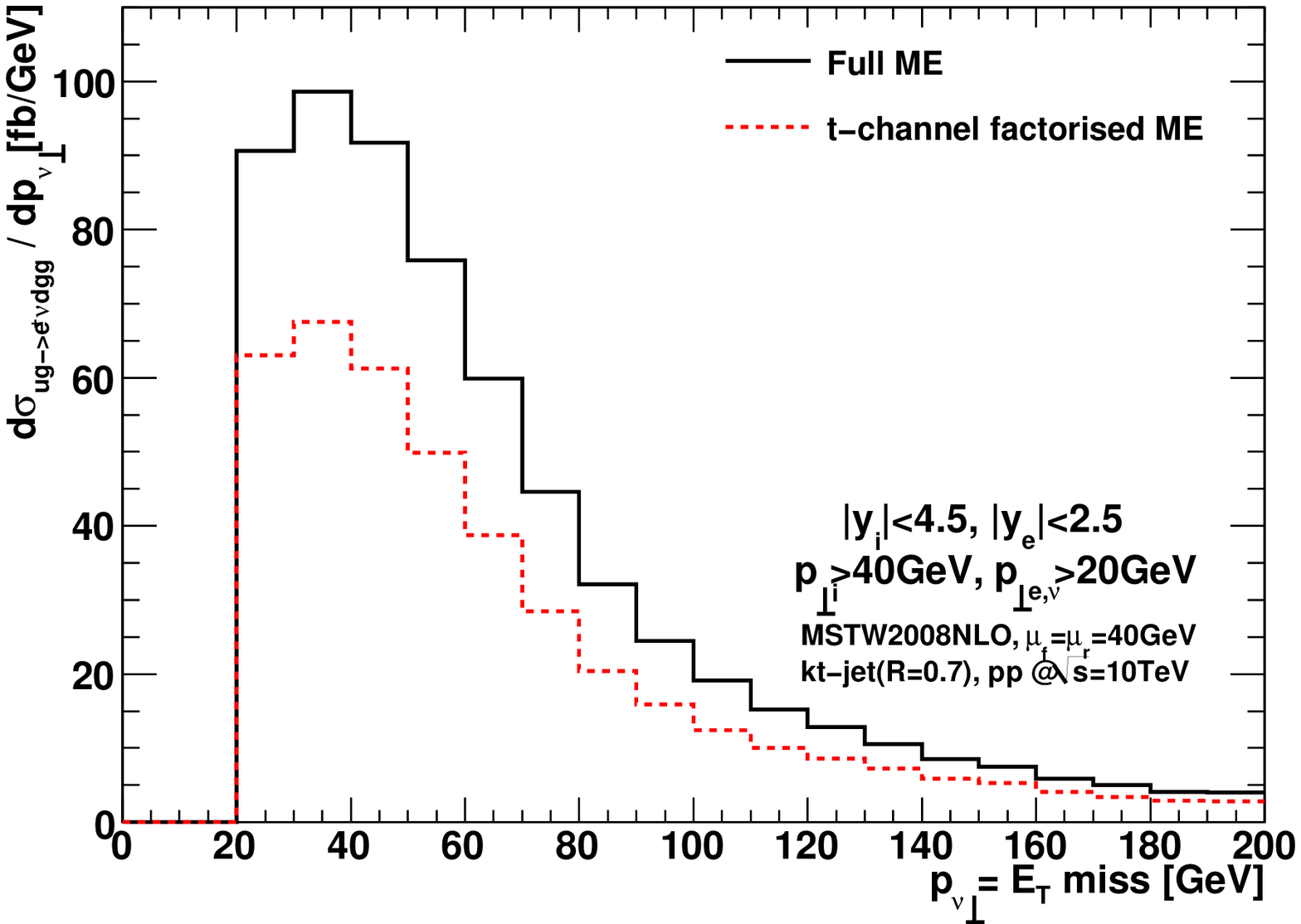}

  (c) \hspace{7.2cm}(d)\hspace{0.1cm}
  \caption{A comparison between the differential $W$ + 3-jet cross section obtained
    using lowest order matrix elements obtained from Madgraph and the
    $t$-channel factorised formalism for the process $ug\to e^+\nu dgg$ with the
    $u$-quark incoming in the positive $z$-direction. (a) the rapidity
    difference between the most forward and backward jet (b) the distribution
    on the azimuthal angle between the most forward and backward jet (c) the $p_\perp$
    distribution of the electron and (d) the $p_\perp$ distribution of the neutrino, which
    is the missing transverse energy.}
  \label{fig:qg3jenu}
\end{figure}

We again illustrate the performance of the approximations by analysing the differential cross section with respect to the
rapidity difference and azimuthal angle between the most forward and most backward jets in
Fig.~\ref{fig:qg3jenu}.  Also shown are the $p_\perp$ distributions of both the electron
and the neutrino. The latter is the distribution of missing transverse energy for these
events.

The rapidity distribution is peaked slightly to the right in our formalism, as for the
jets, and we still see that at larger rapidity differences the results converge.  The
shapes of the azimuthal angle, $p_{e\perp}$ and $p_{\nu\perp}$ distributions are very good
although in each case, our formalism slightly underestimates the full matrix element, as
was the case for the pure jet events.  This is probably again due to collinear
enhancements in the full matrix element, and the general underestimate of the
$qg$-channel for smaller rapidities.

% \begin{figure}[tbp]
%   \centering
%   \epsfig{width=0.49\textwidth,file=qg3jenu_pte.eps}
%   \epsfig{width=0.49\textwidth,file=qg3jenu_ptnu.eps}

\begin{figure}[tbp]
  \centering
  \epsfig{width=0.49\textwidth,file=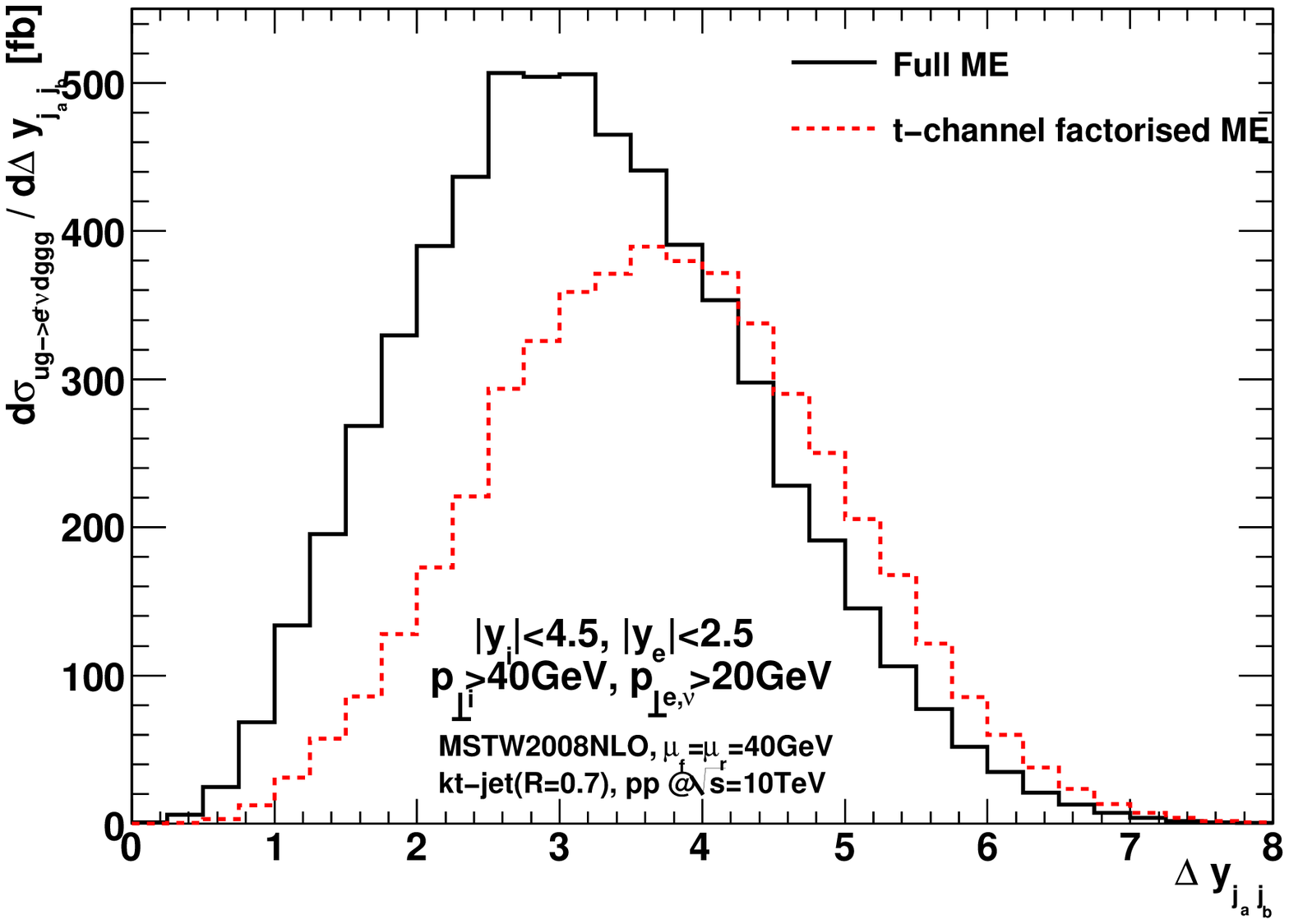}
  \epsfig{width=0.49\textwidth,file=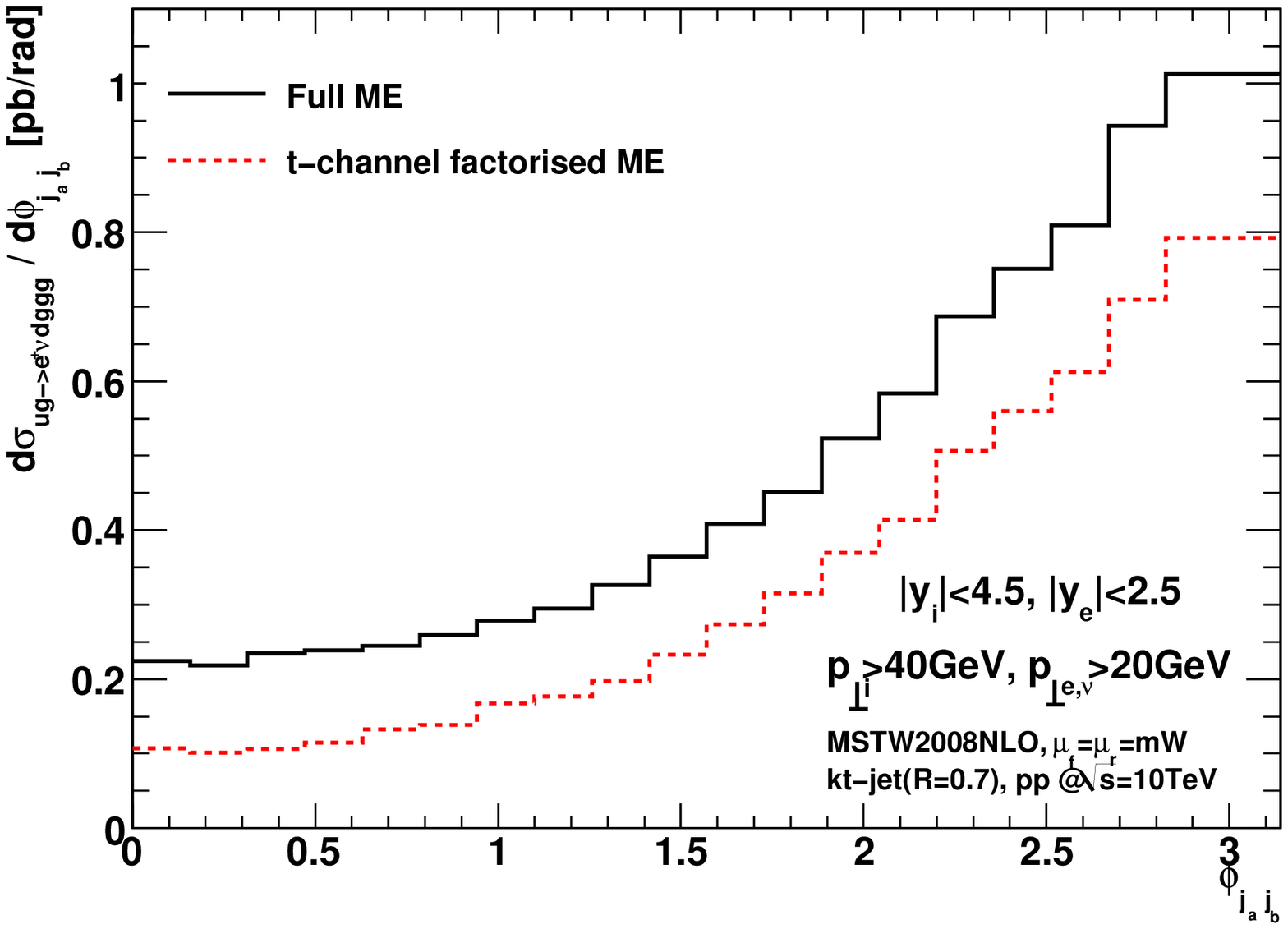}

  (a) \hspace{7.2cm}(b)\hspace{0.1cm}

  \epsfig{width=0.49\textwidth,file=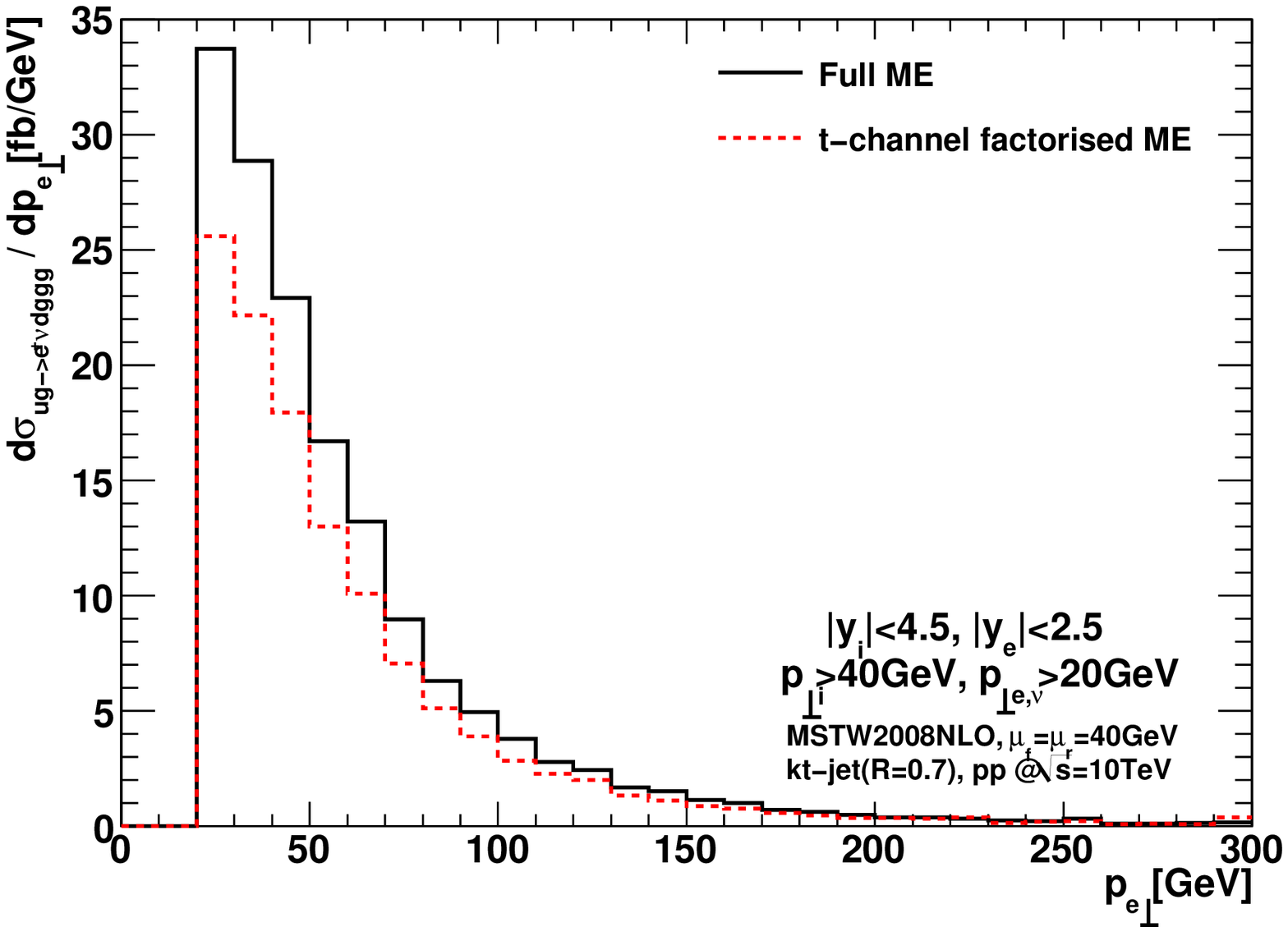}
  \epsfig{width=0.49\textwidth,file=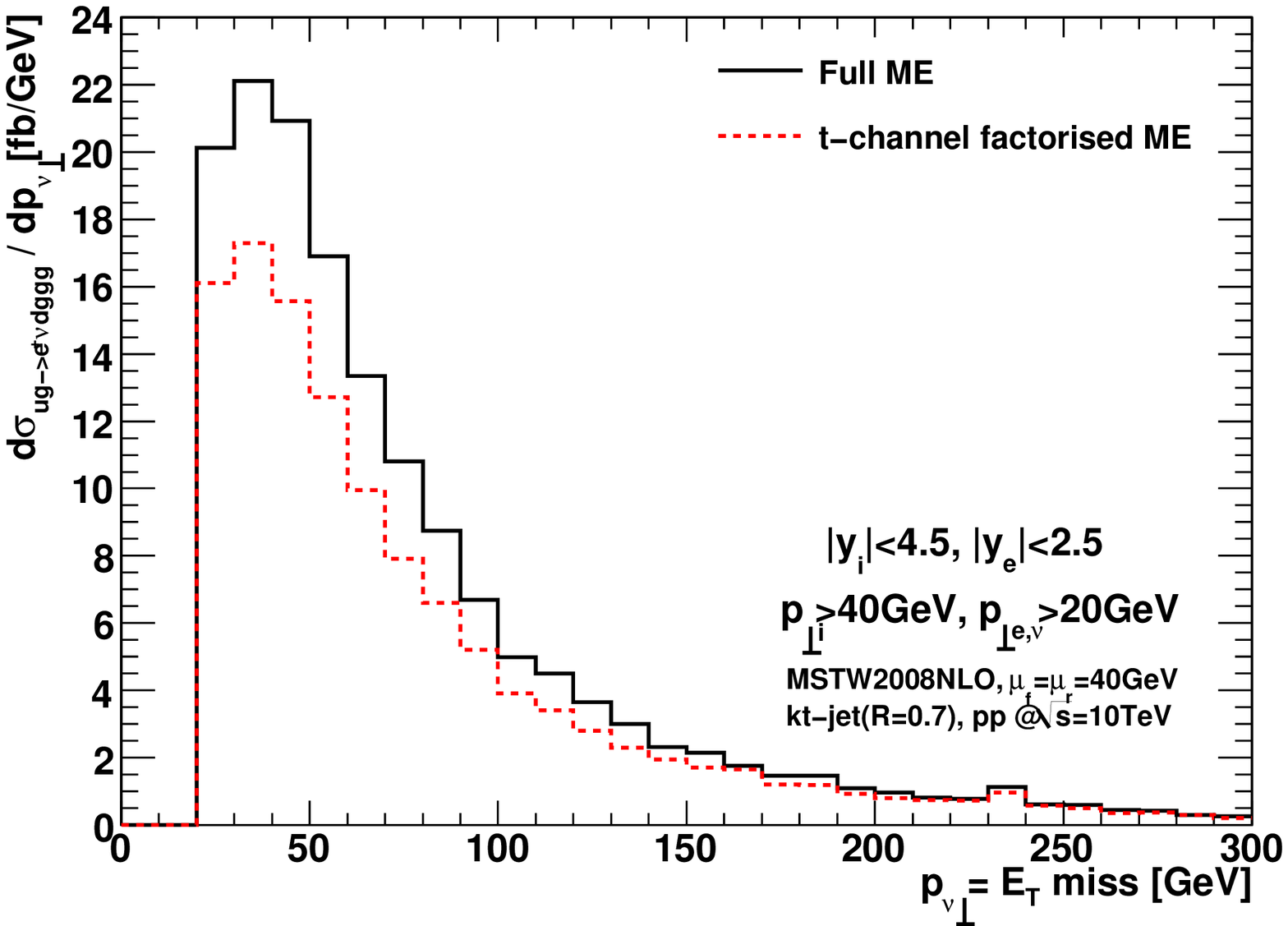}

  (c) \hspace{7.2cm}(d)\hspace{0.1cm}
  \caption{A comparison between the differential $W$ + 4-jet cross section obtained
    using lowest order matrix elements obtained from Madgraph and the
    $t$-channel factorised formalism for the process $ug\to e^+\nu dggg$ with the
    $u$-quark incoming in the positive $z$-direction. (a) the rapidity
    difference between the most forward and backward jet (b) the distribution
    on the azimuthal angle between the most forward and backward jet (c) the $p_\perp$
    distribution of the electron and (d) the $p_\perp$ distribution of the neutrino.}
  \label{fig:qg4jenu}
\end{figure}

Figure~\ref{fig:qg4jenu} repeats these distributions now for the production of a
$W$ boson in association with \emph{four} jets.  Again we see that
the shapes of the distributions are reproduced quite well.

\subsection{Z + Jets}
\label{sec:zjets}

\begin{figure}[tbp]
  \centering
  \epsfig{width=0.49\textwidth,file=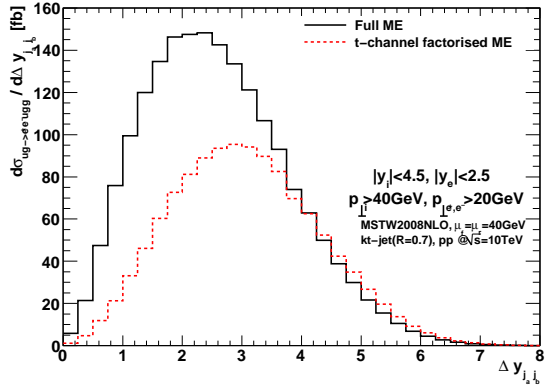}
  \epsfig{width=0.49\textwidth,file=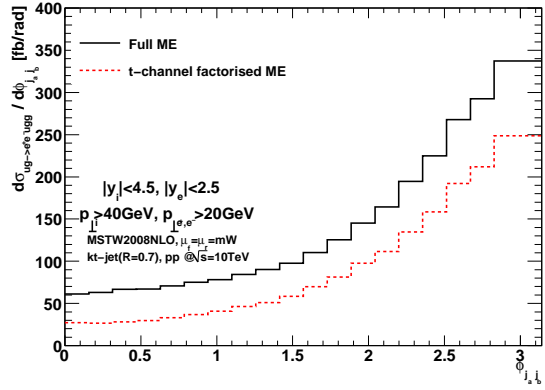}

  (a) \hspace{7.2cm}(b)\hspace{0.1cm}

  \epsfig{width=0.49\textwidth,file=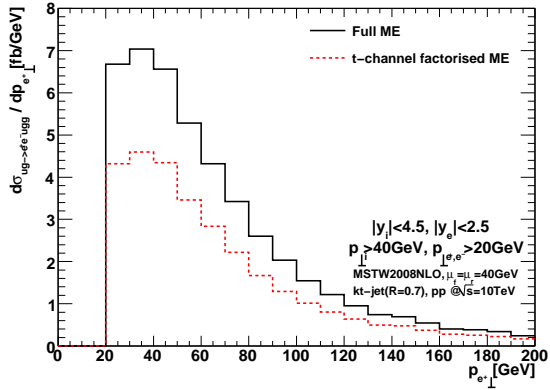}
  \epsfig{width=0.49\textwidth,file=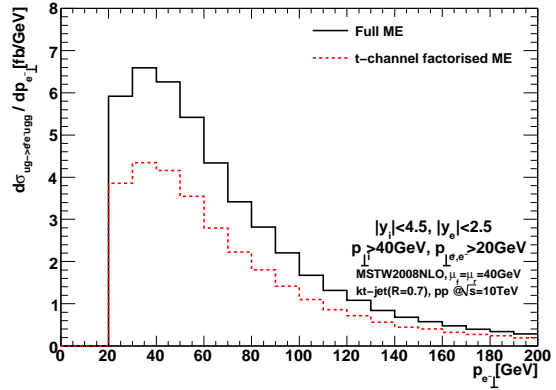}

  (c) \hspace{7.2cm}(d)\hspace{0.1cm}
  \caption{A comparison between the differential $Z$ + 3-jet cross section obtained
    using lowest order matrix elements obtained from Madgraph and the
    $t$-channel factorised formalism for the process $ug\to e^+e^- ugg$ with the
    $u$-quark incoming in the positive $z$-direction. (a) the rapidity
    difference between the most forward and backward jet (b) the distribution
    on the azimuthal angle between the most forward and backward jet (c) the $p_\perp$
    distribution of the $e^+$ and (d) the $p_\perp$ distribution of the $e^-$.}
  \label{fig:qg3jepem}
\end{figure}

We now present similar results for the production of a $Z$ boson (decaying to charged
leptons) in association with three (and four) jets, again for one of the dominant channels, $ug\to
e^+e^- ug(g)g$, with the following set of cuts:-
\begin{center}
  \begin{tabular}{|rl||rl|}
    \hline
    $p_{j_\perp}$ & $> 40$~GeV & $p_{e\perp}$ & $> 20$~GeV \\
    $|y_j|$ & $<$ 4.5 &  $|y_e|$& $< 2.5$ \\\hline
  \end{tabular}
\end{center}
The differential cross sections with respect to the rapidity difference, the
azimuthal angle between the most forward and most backward jets and the
transverse momentum of both the electron and anti-electron are all shown in
Fig.~\ref{fig:qg3jepem}.

The results resemble those for $W$ boson production.  The slight difference between
the $e^+$ and $\nu$ transverse momentum distributions, Fig.~\ref{fig:qg3jenu}, are not
seen in Fig.~\ref{fig:qg3jepem} because identical cuts are now applied to both leptons
here (the rapidity cut was not applied to the invisible neutrino for the $W$ events).  The
spin correlation of the $W$ to $e^+$ is also washed out when it is replaced by a $Z$
boson.

\begin{figure}[tbp]
  \centering
  \epsfig{width=0.49\textwidth,file=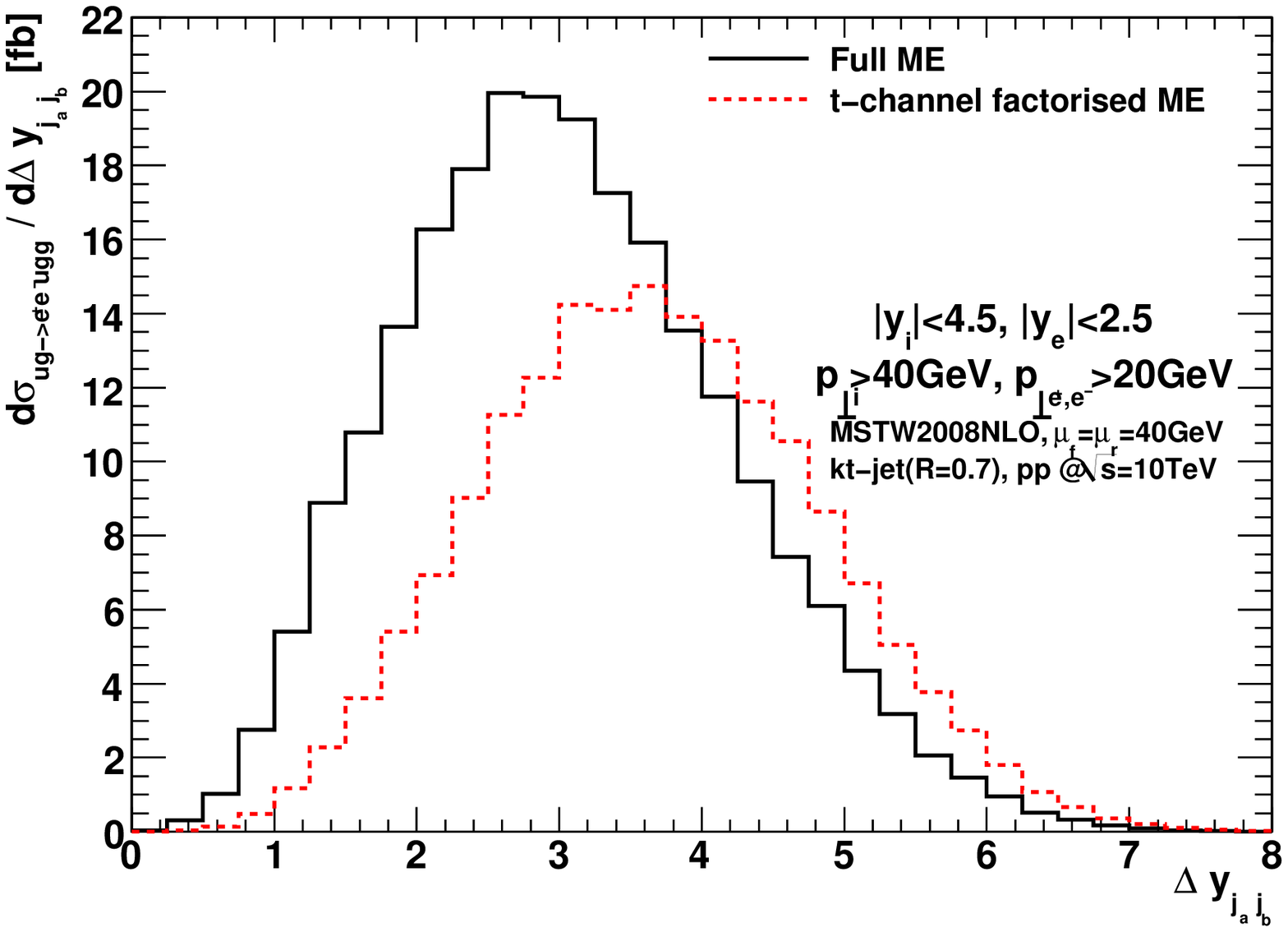}
  \epsfig{width=0.49\textwidth,file=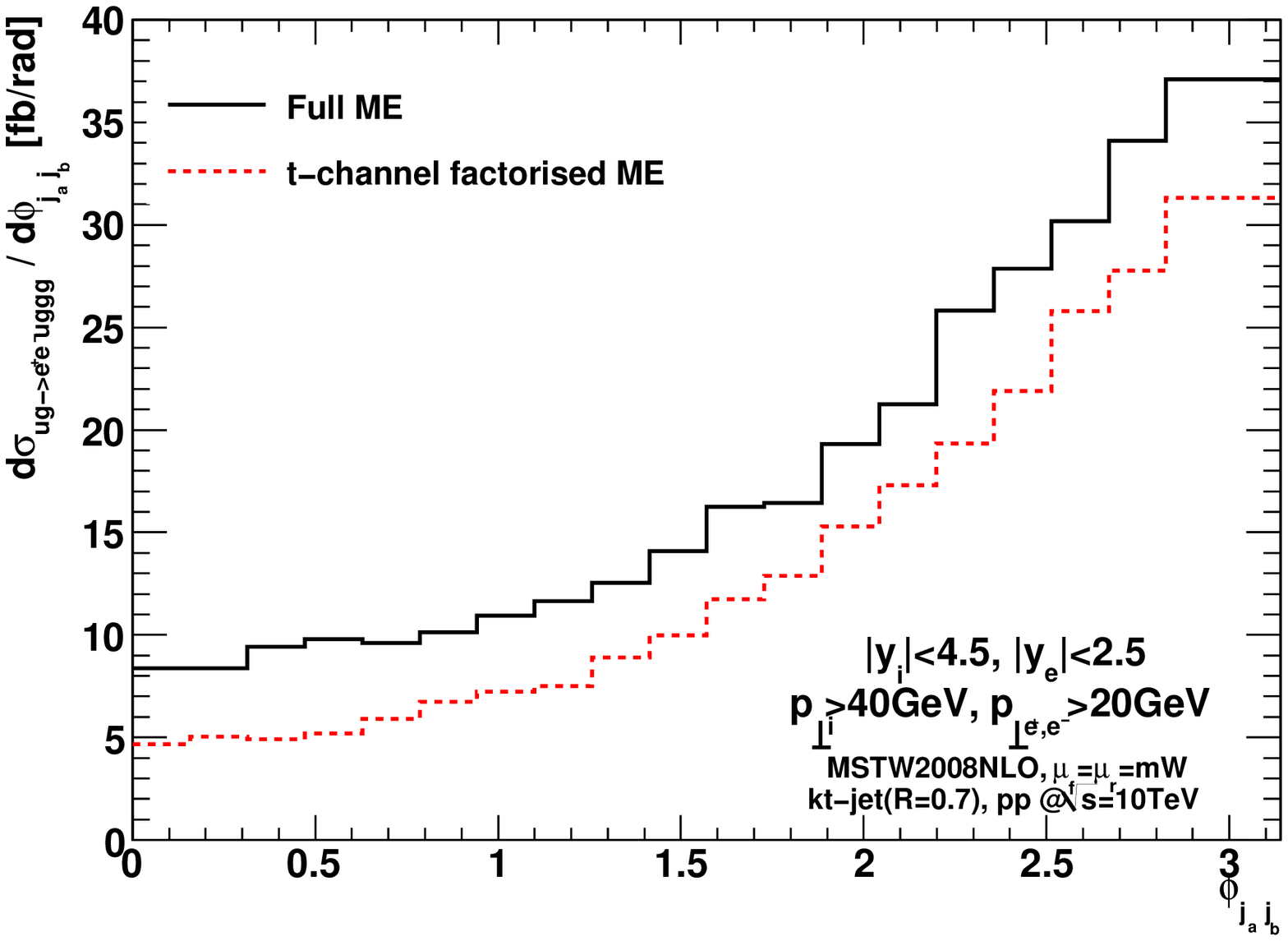}

  (a) \hspace{7.2cm}(b)\hspace{0.1cm}

  \epsfig{width=0.49\textwidth,file=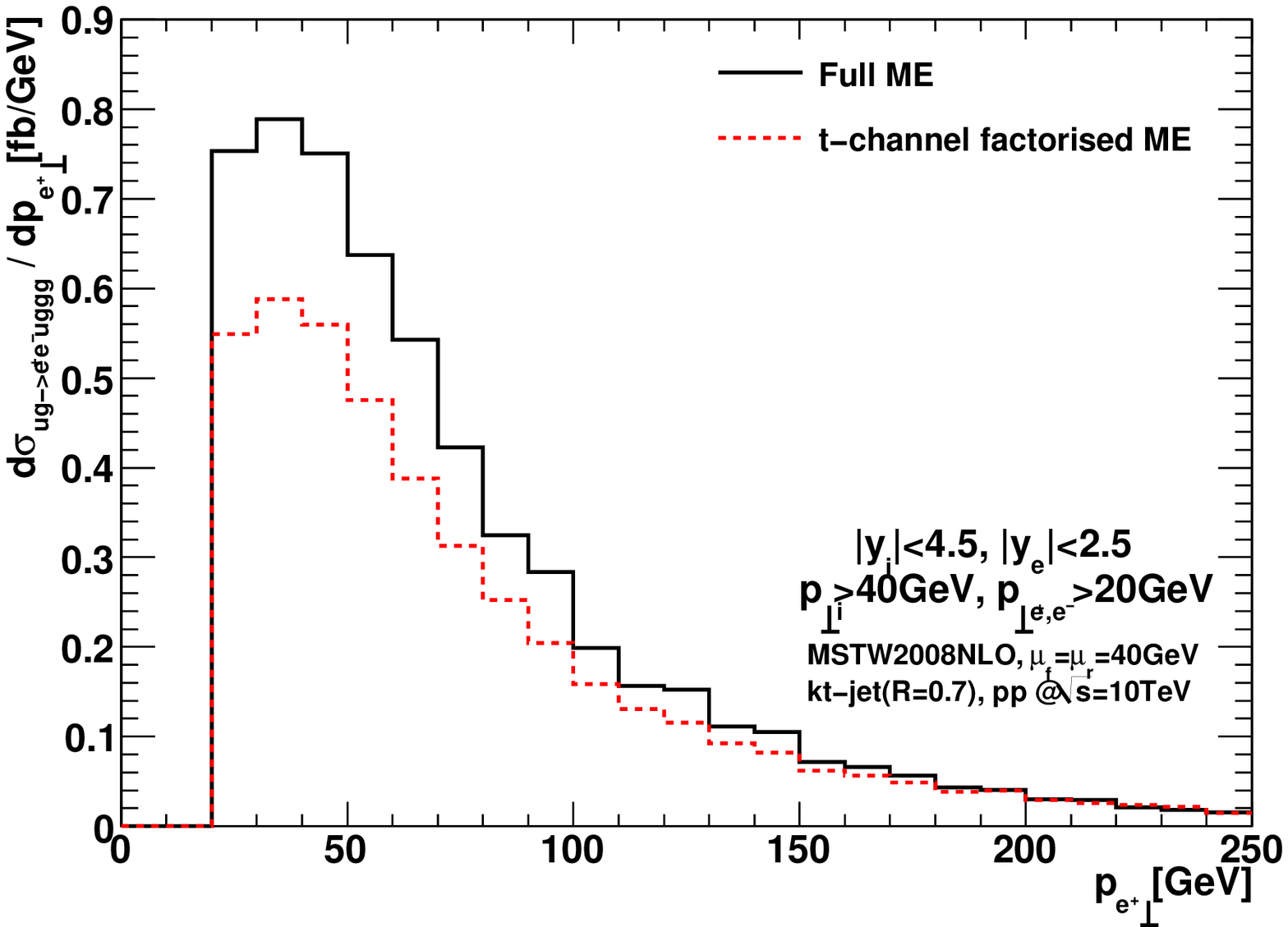}
  \epsfig{width=0.49\textwidth,file=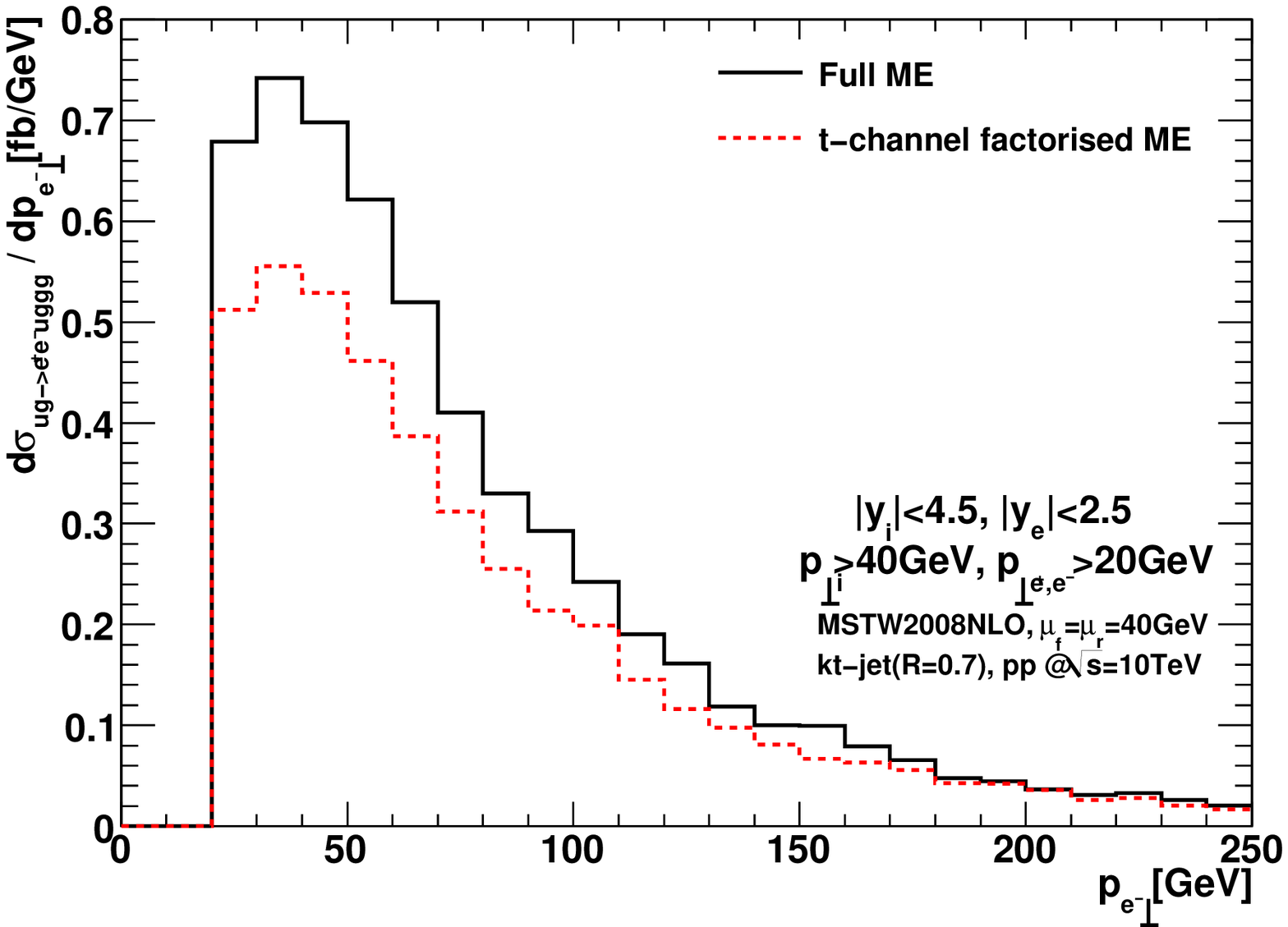}

  (c) \hspace{7.2cm}(d)\hspace{0.1cm}
  \caption{As in Fig.~\ref{fig:qg3jepem}, but now for 4 jets in the final state: $ug\to
    e^+ e^- uggg$.}
  \label{fig:qg4jepem}
\end{figure}

Figure~\ref{fig:qg4jepem} repeats these distributions now for the production of a
$Z$ boson in association with \emph{four} jets.  Again we see that
the shapes of the distributions are reproduced very well by the simple
approximations developed in Section~\ref{sec:w-boson-production}.

\subsection{Higgs Boson + Jets}
\label{sec:higgsjetsrevisited}

\begin{figure}[tbp]
  \centering
  \epsfig{width=0.49\textwidth,file=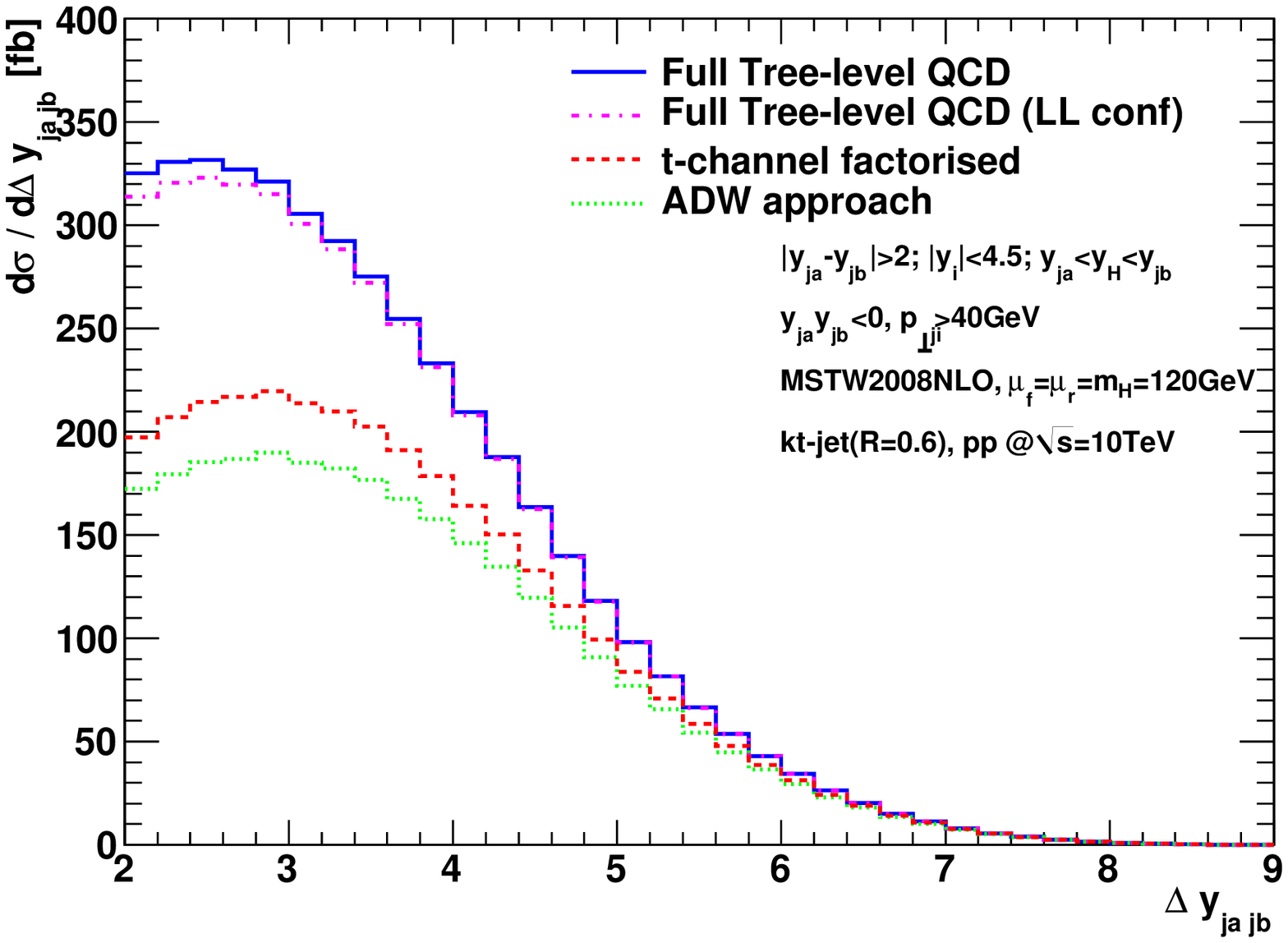}
  \epsfig{width=0.49\textwidth,file=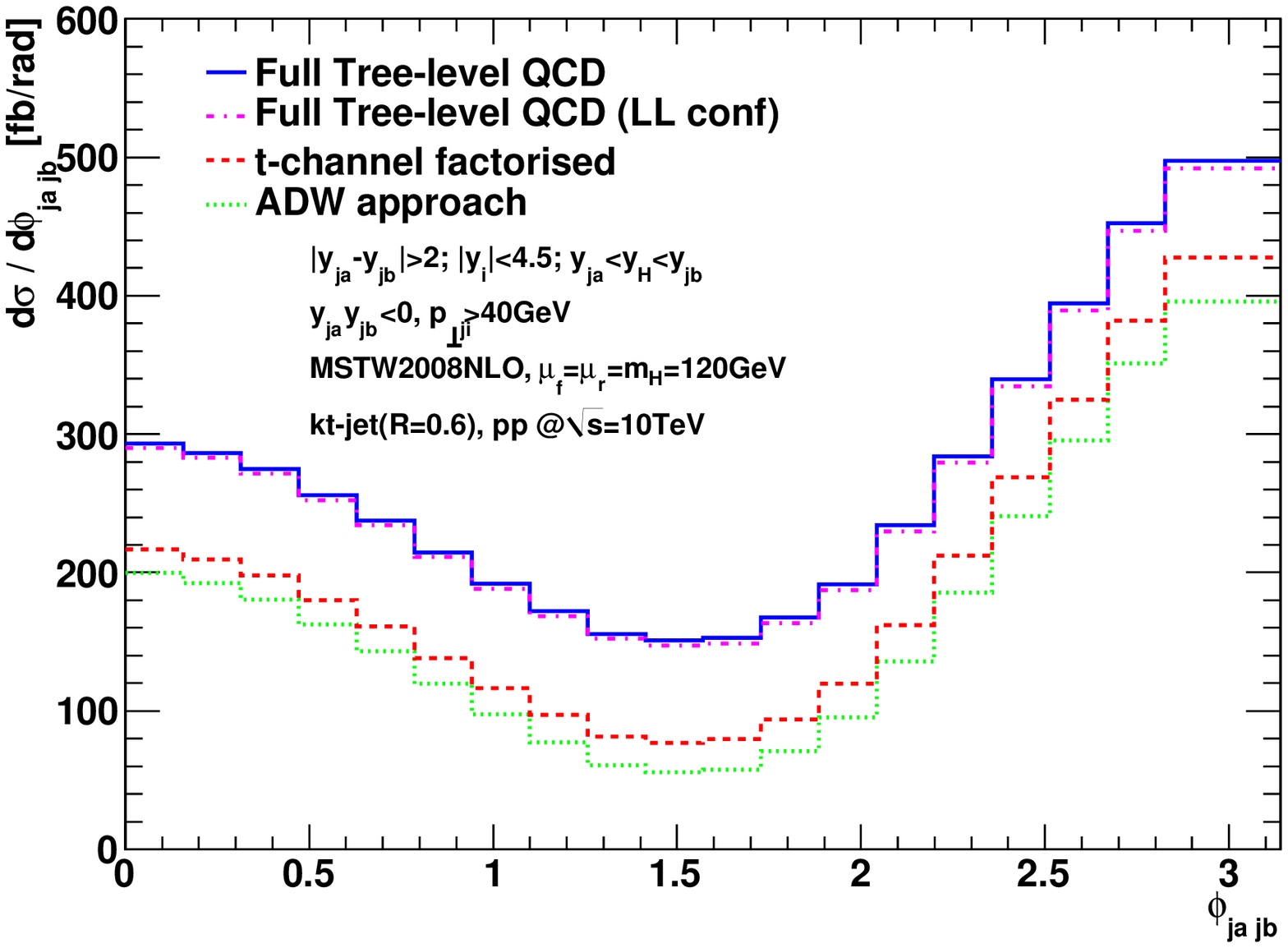}

  (a) \hspace{7.2cm}(b)\hspace{0.1cm}

  \epsfig{width=0.49\textwidth,file=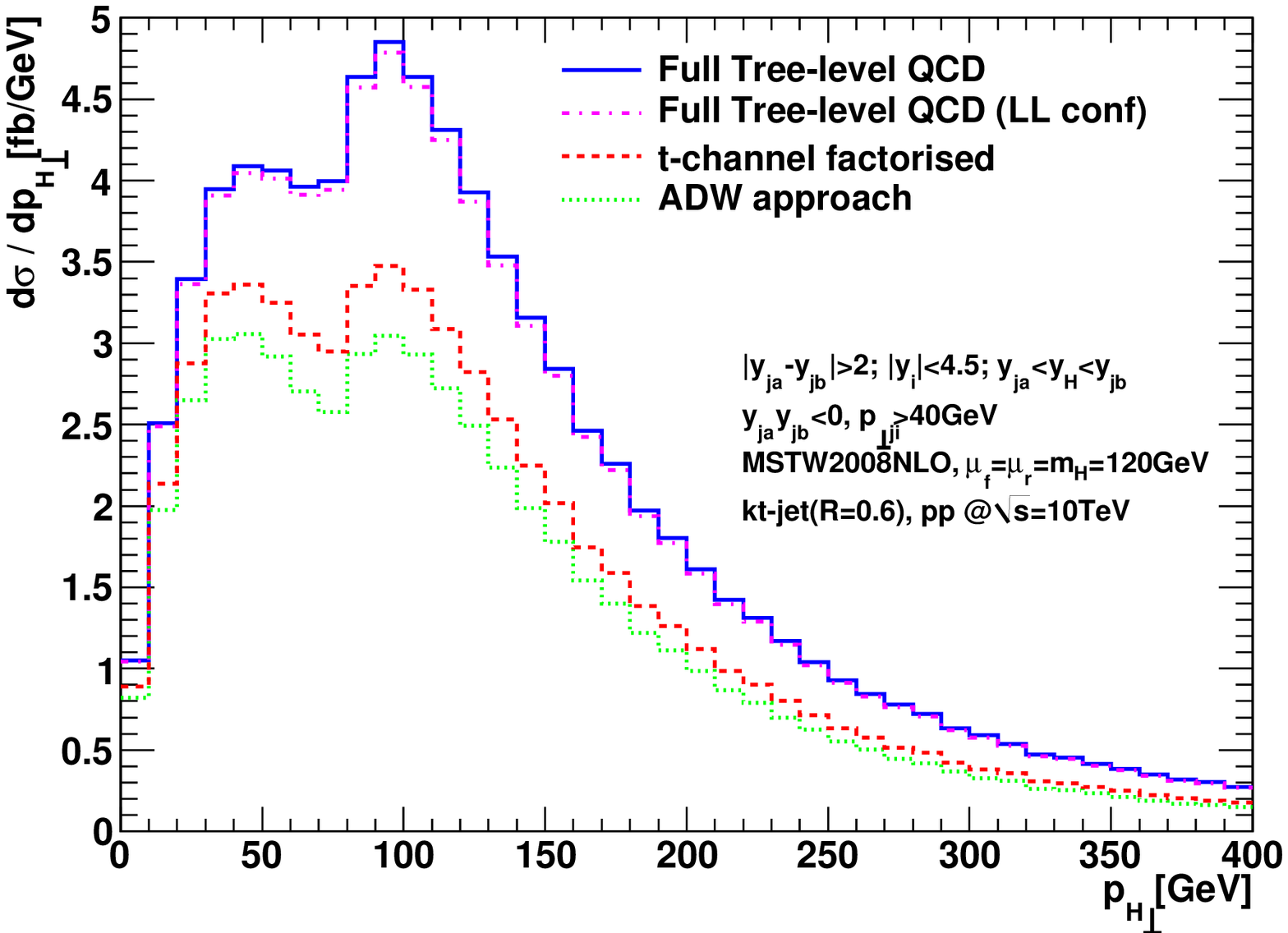}

  (c)
  \caption{A comparison between the different $H$+2 jet cross sections using
    lowest order matrix elements obtained from MadGraph, the factorised
    approximation of~\cite{Andersen:2008ue,Andersen:2008gc} and the formalism
    of section \ref{sec:higgs-boson-prod} for the sum over all relevant
    subprocesses $pp\to jHj$. For the results obtained using the full
    tree-level matrix elements, we have with the line marked ``LL conf''
    indicated the impact of taking into account only the flavour and phase
    space configurations (rapidity orderings) which are approximated in the
    (leading logarithmic) $t$-channel factorised picture. }
  \label{fig:Higgs2}
\end{figure}

In this section, we explore how well our new formalism (section
\ref{sec:higgs-boson-prod}) reproduces the results obtained using the full
matrix element (at tree-level, where the fixed order results can be readily
obtained). In the following analysis, we apply phase space cuts as follows:-
\begin{center}
  \begin{tabular}{|rl||rl|}
    \hline
    $p_{j_\perp}$ & $> 40$ GeV & $y_{ja}\cdot y_{jb}$& $<0$ \\
    $y_j$ & $<$ 4.5 & $\vert y_{ja}-y_{jb} \vert$ & $> 2$ \\
    $y_{ja}$&$\le y_h\!\le y_{jb}$ & &\\ \hline
  \end{tabular}
\end{center}
The merits of these cuts are discussed elsewhere\cite{Andersen:2008ue,Andersen:2008gc}.
In Figure~\ref{fig:Higgs2} we compare the results obtained for the
$hjj$-channel in four approximation:-
\begin{enumerate}
\item Lowest order QCD
\item Lowest order QCD, but including only the flavour and rapidity
  configurations which are taking into account in the $t$-channel factorised
  framework
\item The results obtained using the tree-level (lowest order predictions) of
  the framework of
  Ref.\cite{Andersen:2008ue,Andersen:2008gc} 
\item The results obtained using the tree-level (lowest order predictions) of
  the formalism of Section~\ref{sec:higgs-boson-prod}
\end{enumerate}

\begin{figure}[tbp]
  \centering
  \epsfig{width=0.49\textwidth,file=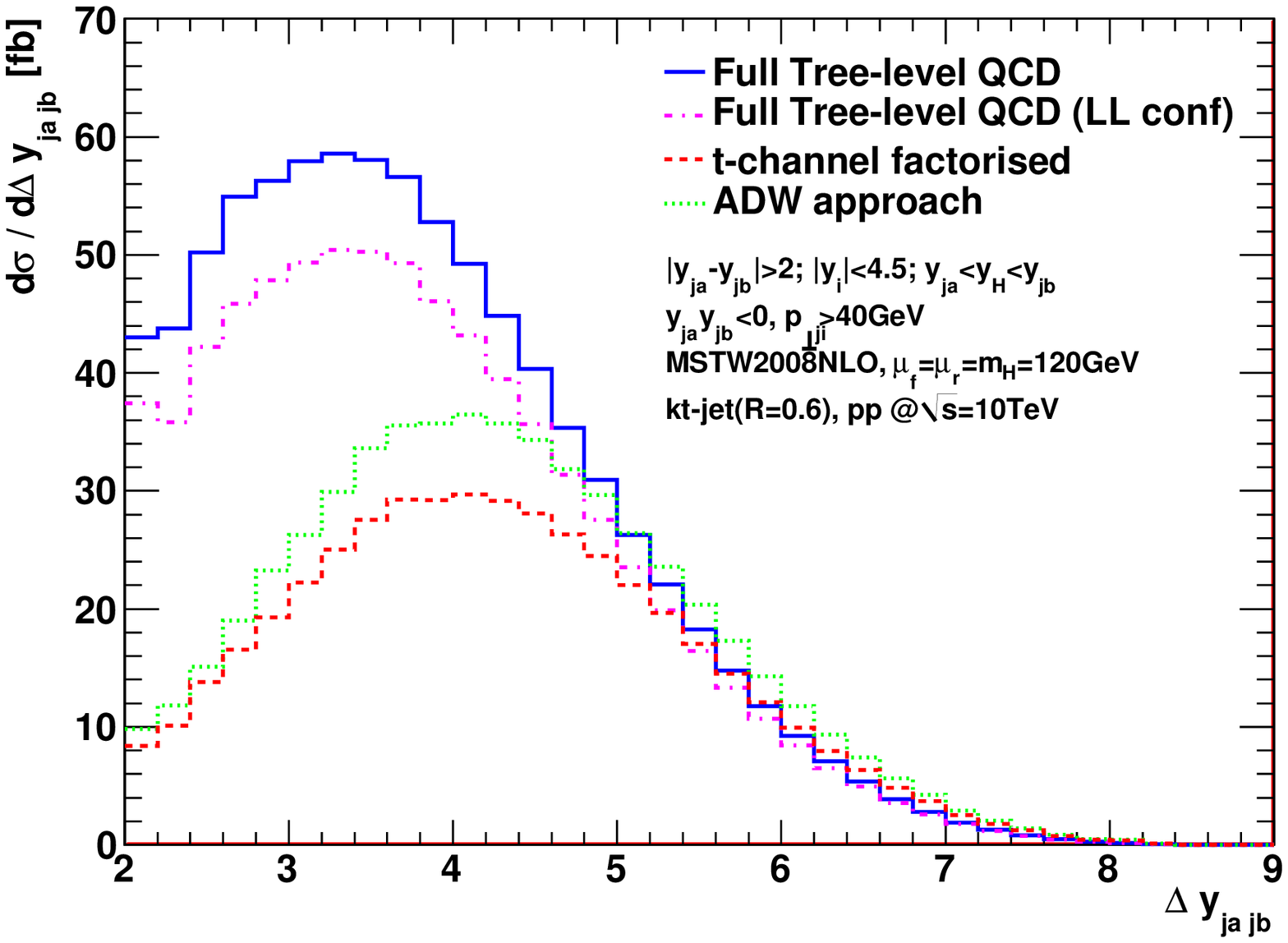}
  \epsfig{width=0.49\textwidth,file=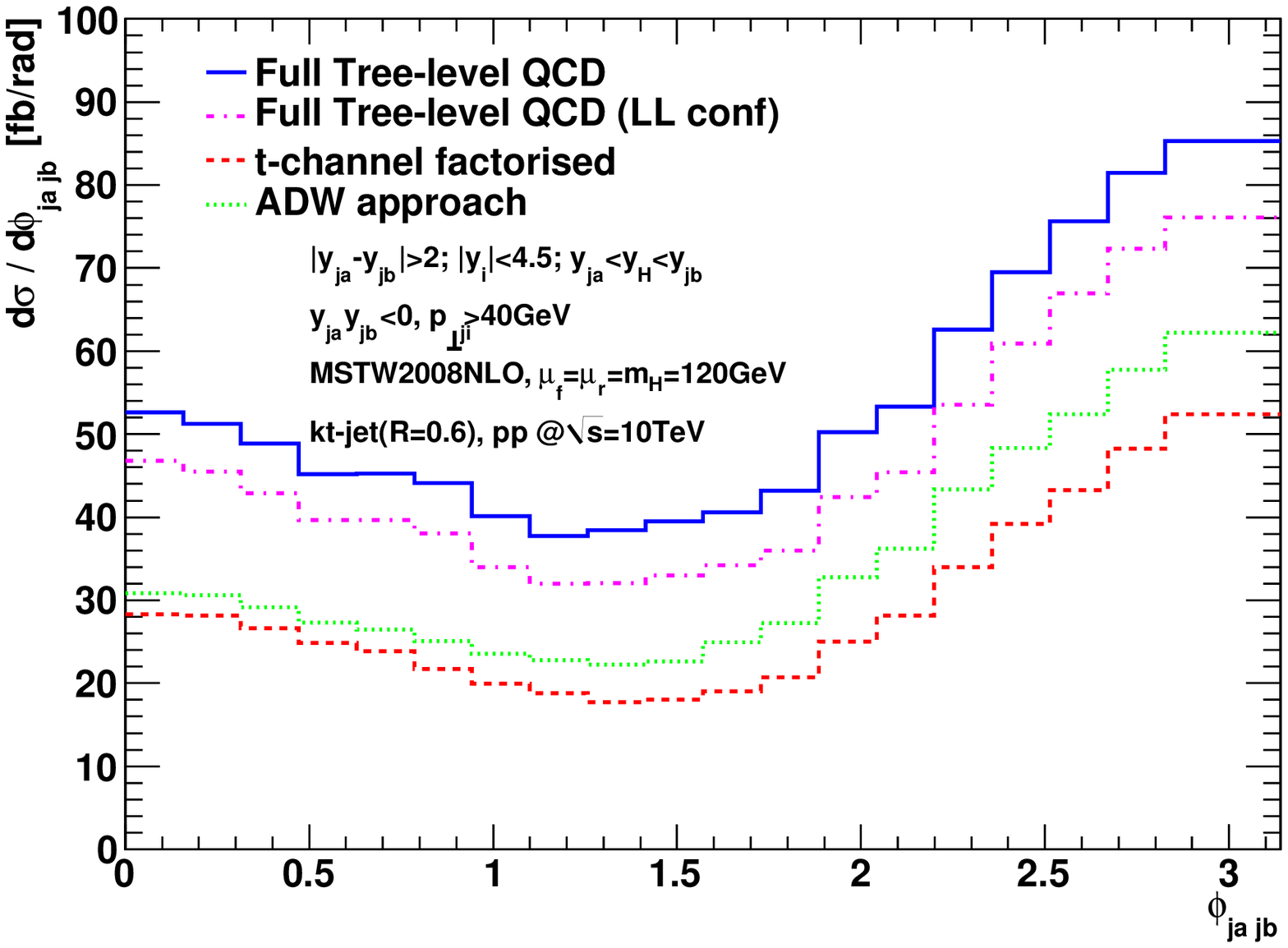}

  (a) \hspace{7.2cm}(b)\hspace{0.1cm}

  \epsfig{width=0.49\textwidth,file=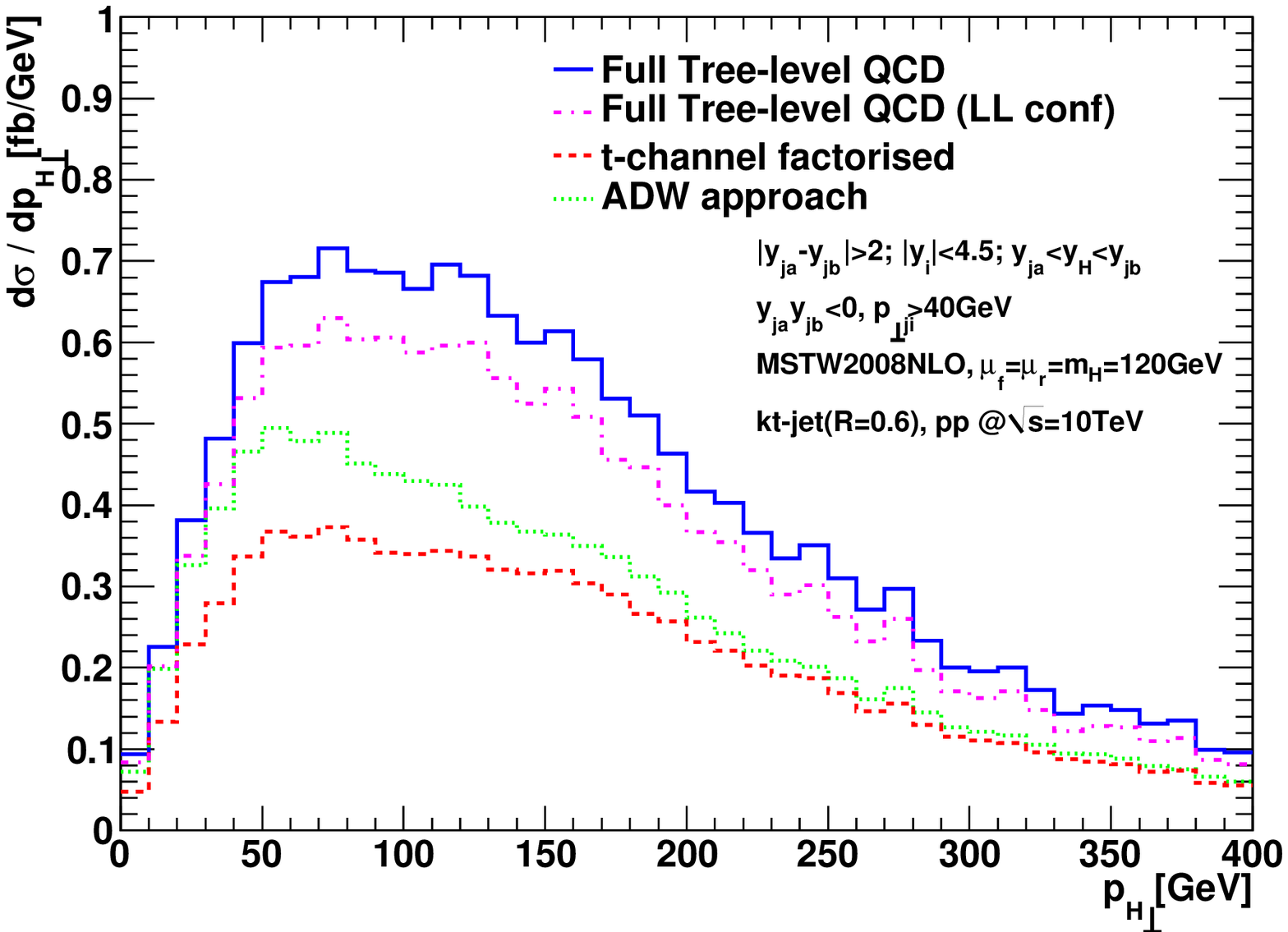}

  (c) 
  \caption{As in Fig.~\ref{fig:Higgs2}, but now for all subprocesses contributing to
    $pp\to Hjjj$.}
  \label{fig:Higgs3}
\end{figure}

We start by noting that distributions and normalisation is by far dominated
by the rapidity and flavour configurations of the final states which allow
for a colour octet exchange between every rapidity-neighbours (these are the
only contributions in the leading logarithmic approximation). The results
obtained by summing over all final state configurations are indicated by the
full lines, and the contribution from the ``leading logarithmic'' final
states are indicated by the dash-dotted line.

The difference between the results obtained with full QCD and in the
$t$-channel factorised is by far dominated by the channels consisting of all
gluons plus a Higgs boson. This is consistent with the observations made in
Figs.~\ref{fig:3j}--\ref{fig:4j}. Also, we note that using full tree-level
QCD, the distribution of the rapidity difference between the most forward and
backward hard jet is peaked around 2.5 for Higgs boson production in
association with two jets, and 3.5 for Higgs boson production in association
with three jets. The position of the peaks are similar to the situation of
pure jets, and highlights the universality of the QCD radiation pattern. The
rapidity distribution obtained with the $t$-channel factorised
matrix element peaks at slightly larger rapidities --- this is simply because
the gluon-gluon induced channel is underestimated at smaller rapidities; a
feature also observed for pure jets.

We note that the formalism developed in the present study is a slight
improvement over the earlier study in
Ref.\cite{Andersen:2008ue,Andersen:2008gc} in terms of reproducing the shape
and normalisation of results obtained with full fixed order QCD. It should be
noted that the results of the $t$-channel factorised approach are obtained
more than 2 orders of magnitude faster than the results relying on the
evaluation of the full matrix element with MadGraph\cite{Alwall:2007st}
(all other components of the calculation are identical: phase space
generation, pdfs etc.).

%%% Local Variables: 
%%% mode: latex
%%% TeX-master: "jetcurrents"
%%% End: 

\section{Conclusions}
\label{sec:conclusions}
We have demonstrated the universal analytical structure of scattering amplitudes in the
multi-Regge kinematic (MRK) limit of infinite rapidity separation between all produced
particles for a range of scattering processes which are important for the LHC
phenomenology (jets, $W$+jets, $Z$+jets and $H$+jets). The universal behaviour in this
limit is interpreted in terms of the scattering amplitudes being dominated by the poles in
the $t$-channel momenta.

In Section~\ref{sec:current-method} we developed a formalism which is exact
in the MRK limit, and fulfils the three requirement listed in
Section~\ref{sec:current-method} necessary for constructing a relevant
all-order summation:- Inclusiveness, Simplicity and Accuracy. The formalism
is inclusive in the sense that it captures the leading real and virtual
corrections in the MRK limit. It is sufficiently simple that the cancellation
of the infra-red poles can be organised explicitly to all orders (because of
the simple structure of radiative corrections in the MRK limit). The results
for the approximations to tree-level are obtained more than 2 orders of
magnitudes faster than using standard tools for evaluating the full
tree-level matrix element. This will allow for the all-order sum to be
computed by explicit evaluation of the exclusive $n$-parton final states, and
thus allow exclusive (or inclusive) analyses to be performed. Such an
implementation will allow for matching to full tree-level results, where such
are available. 

We have shown examples of the results obtained in the framework for all 4
processes mentioned above, and compared to the results of using full leading
order matrix elements. The results are very encouraging; the overall accuracy
is good even without phase space cuts to enhance the accuracy of the
approximation, and the discrepancies which do exists are limited to the
region of small rapidity separation, where the first few orders in the
perturbative series should be sufficient and the corrections can be included
by matching. This is obviously necessary in order for the resummation, which
will be built on the formalism, to be relevant for LHC phenomenology. The
phenomenological implications of the resummation will be the focus of future studies.

\subsection*{Acknowledgements}
\label{sec:acknowledgements}
JMS would like to thank the CERN theory group for kind hospitality at various
stages of this project and is supported by the UK Science and Technology Facilities
Council (STFC). This work was supported by the EC Marie-Curie
Research Training Network ``Tools and Precision Calculations for Physics
Discoveries at Colliders'' under contract MRTN-CT-2006-035505.

%%% Local Variables: 
%%% mode: latex
%%% TeX-master: "jetcurrents"
%%% End: 

\appendix

\section{Spinor Representation}
\label{sec:spin-repr}
We use the following representation for the spinors.  For outgoing particles with
4-momentum $p$, $p^\pm=E\pm p_z$ and $p_\perp=p_x+ip_y$, we use
\begin{align}
  \label{eq:outspin}
  u^+(p)=\left(
    \begin{array}{c} \sqrt{p^+} \\ \sqrt{p^-}\ \frac{p_\perp}{|p_\perp|} \\ 0 \\
      0 \end{array} \right) \qquad \mathrm{and} \qquad u^-(p)=\left( \begin{array}{c} 0 \\
      0 \\ \sqrt{p^-} \
        \frac{p_\perp^*}{|p_\perp|}\\ -\sqrt{p^+} \end{array} \right).
\end{align}
For incoming particles with 4-momentum $p$ moving in the + direction, we use:
\begin{align}
  \label{eq:inspinp}
  u^+(p)=\left(
    \begin{array}{c} \sqrt{p^+} \\ 0 \\ 0 \\ 0 \end{array}
    \right) \qquad \mathrm{and} \qquad u^-(p)=\left( \begin{array}{c} 0 \\ 0 \\ 0\\
        -\sqrt{p^+} \end{array} \right).
\end{align}
For incoming particles with 4-momentum $p$ moving in the - direction, we use:
\begin{align}
  \label{eq:inspinm}
  u^+(p)=\left(
    \begin{array}{c} 0 \\ -\sqrt{p^-} \\ 0 \\ 0 \end{array}
    \right) \qquad \mathrm{and} \qquad u^-(p)=\left( \begin{array}{c} 0 \\ 0 \\
        -\sqrt{p^-} \\ 0\end{array} \right).
\end{align}
We use the following representation for the gamma matrices:
\begin{align}
  \label{eq:gammas}
  \begin{split}
    &\gamma^0 = \left(
      \begin{array}{cccc}
        0 & 0 & 1 & 0 \\ 0 & 0 & 0 & 1 \\ 1 & 0 & 0 & 0 \\ 0 & 1 & 0 & 0 
      \end{array} \right),\quad
    \gamma^1 = \left(
      \begin{array}{cccc}
        0 & 0 & 0 & -1 \\ 0 & 0 & -1 & 0 \\ 0 & 1 & 0 & 0 \\ 1 & 0 & 0 & 0 
      \end{array} \right),\ \\
    &\gamma^2 = \left(
      \begin{array}{cccc}
        0 & 0 & 0 & i \\ 0 & 0 & -i &  \\ 0 & -i & 0 & 0 \\ i & 0 & 0 & 0 
      \end{array} \right),\quad
    \gamma^3 = \left(
      \begin{array}{cccc}
        0 & 0 & -1 & 0 \\ 0 & 0 & 0 & 1 \\ 1 & 0 & 0 & 0 \\ 0 & -1 & 0 & 0 
      \end{array} \right).
  \end{split}
\end{align}
We also use the shorthands
\begin{align}
  \label{eq:shorthands}
  \begin{split}
  &\langle ij \rangle = \bar u^-(p_i) u^+(p_j)\quad \mathrm{and} \quad [ ij ] = \bar
  u^+(p_i) u^-(p_j).
  \end{split}
\end{align}

\section{Momentum Configurations}
\label{sec:moment-conf}

Here we list, for completeness, the momentum configurations used for the plots of the
matrix elements.  Those for the pure jet events (Figs.~\ref{fig:M3j} and \ref{fig:2jMEs})
are given in the text.  For the production of either a $W$ or $Z$ (decaying to $\ell \bar
\ell$) in association with jets (Figs.~\ref{fig:WMRKlimit}, \ref{fig:ZMRKlimit},
\ref{fig:WMEs} and \ref{fig:ZMEs}), we use the following momentum configurations. For 2
jets in the final state :
\begin{align}
  \label{eq:2jmomWZ}
  \begin{split}
        p_i&=(k_{\perp i} \cosh(y_i), k_{\perp i} \cos(\phi_i), k_{\perp i} \sin(\phi_i) ,
    k_{\perp i} \sinh(y_i) ),\\
    k_{\perp 1}&=k_{\perp \bar \ell}=40\mathrm{GeV},\ k_{\perp \ell}=\frac{m_V^2}{2k_{\perp
        \bar \ell}(\cosh(y_{\bar \ell}-y_\ell)-\cos(\phi_{\bar \ell}-\phi_\ell))},\\ 
    \phi_1&=\pi,\ \phi_{\bar \ell}=\pi+0.2,\ \phi_\ell=-(\pi+0.2),\\
    y_1&=\Delta,\ y_2=-\Delta,\ y_{\bar \ell}=\Delta,\ y_\ell=\Delta-1.5,\\
    p_{2\perp}&=-p_{1\perp}-p_{\bar \ell \perp}-p_{\ell \perp}.
  \end{split}
\end{align}
For 3 jets in the final state:
\begin{align}
  \label{eq:3jmomWZ}
  \begin{split}
        p_i&=(k_{\perp i} \cosh(y_i), k_{\perp i} \cos(\phi_i), k_{\perp i} \sin(\phi_i) ,
    k_{\perp i} \sinh(y_i) ),\\
    k_{\perp 1}&=k_{\perp 2}=k_{\perp \bar \ell}=40\mathrm{GeV},\ k_{\perp
      \ell}=\frac{m_V^2}{2k_{\perp \bar \ell}(\cosh(y_{\bar \ell}-y_\ell)-\cos(\phi_{\bar
        \ell}-\phi_\ell))},\\
    \phi_1&=2\pi/3,\ \phi_2=0,\ \phi_{\bar \ell}=\pi/2,\ \phi_\ell=-\pi/2,\\
    y_1&=\Delta,\ y_2=0,\ y_3=-\Delta,\ y_{\bar \ell}=\Delta,\ y_\ell=\Delta-1.5,\\
    p_{3\perp}&=-p_{1\perp}-p_{2\perp}-p_{\bar \ell \perp}-p_{\ell \perp},
  \end{split}
\end{align}
while the $4j$ final state events use:
\begin{align}
  \label{eq:4jmomWZ}
  \begin{split}
        p_i&=(k_{\perp i} \cosh(y_i), k_{\perp i} \cos(\phi_i), k_{\perp i} \sin(\phi_i) ,
    k_{\perp i} \sinh(y_i) ),\\
    k_{\perp 1}&=k_{\perp 2}=k_{\perp 3}=k_{\perp \bar \ell}=40\mathrm{GeV},\ k_{\perp
      \ell}=\frac{m_V^2}{2k_{\perp \bar \ell}(\cosh(y_{\bar \ell}-y_\ell)-\cos(\phi_{\bar \ell}-\phi_\ell))},\\ 
    \phi_1&=\pi,\ \phi_2=\pi/2,\ \phi_3=-\pi/3,\ \phi_{\bar \ell}=\pi/4,\ \phi_\ell=-\pi/4,\\
    y_1&=\Delta,\ y_2=\Delta/3,\ y_3=-\Delta/3,\ y_4=-\Delta,\ y_{\bar \ell}=\Delta,\
    y_\ell=\Delta,\\ p_{4\perp}&=-p_{1\perp}-p_{2\perp}-p_{3\perp}-p_{\bar
      \ell\perp}-p_{\ell \perp}.
  \end{split}
\end{align}
For the production of a Higgs boson in association with jets (Figs.~\ref{fig:HMRKlimit}
and \ref{fig:HME}) we use the following momentum configurations.  For 2 jets in the final
state:
\begin{align}
  \label{eq:2jHmoms}
  \begin{split}
        p_1&=(40\sqrt{2} \cosh(\Delta),-40,40,40\sqrt{2}\sinh(\Delta),\\
        p_H&=(\sqrt{40^2+m_H^2},0,-40,0)\ \mathrm{GeV}, \\
        p_2&=(40 \cosh(-\Delta), 40, 0, 40 \sinh(-\Delta) )\ \mathrm{GeV},
  \end{split}
\end{align}
and for 3 jets in the final state:
\begin{align}
  \label{eq:3jHmoms}
  \begin{split}
        p_1&=(40 \cosh(\Delta),-40,0,40\sinh(\Delta),\\
        p_H&=(\sqrt{40^2+m_H^2}\cosh(\Delta/3),0,-40,\sqrt{40^2+m_H^2}\sinh(\Delta/3))\
        \mathrm{GeV}, \\
        p_2&=(40 \cosh(-\Delta/3), 0, 40, 40 \sinh(-\Delta/3) )\ \mathrm{GeV}, \\
        p_3&=(40 \cosh(-\Delta), 40, 0, 40 \sinh(-\Delta) )\ \mathrm{GeV}.
  \end{split}
\end{align}

%%% Local Variables: 
%%% mode: latex
%%% TeX-master: "jetcurrents"
%%% End: 

\bibliographystyle{JHEP}
\bibliography{jetpapers}

\end{document}